\documentclass[runningheads]{llncs}

\usepackage{eccv}

\usepackage{eccvabbrv}

\usepackage{graphicx}
\usepackage{booktabs}
\usepackage{multirow}
\usepackage{bm}
\usepackage{appendix}

\usepackage[accsupp]{axessibility}  % Improves PDF readability for those with disabilities.

\usepackage{hyperref}

\usepackage{orcidlink}

\begin{document}

\title{Bidirectional Stereo Image Compression with Cross-Dimensional Entropy Model} 

\titlerunning{BiSIC: Bidirectional Stereo Image Compression}

\author{Zhening Liu\orcidlink{0009-0001-6502-368X} \and
Xinjie Zhang\orcidlink{0000-0002-3194-7518} \and
Jiawei Shao\orcidlink{0000-0001-8836-1430} \and \\
Zehong Lin\thanks{Corresponding author}\orcidlink{0000-0002-9503-2464} \and 
Jun Zhang\orcidlink{0000-0002-5222-1898}
}

\authorrunning{Z.~Liu et al.}

\institute{Hong Kong University of Science and Technology, Clear Water Bay, Hong Kong SAR\\
\email{\{zhening.liu,xzhangga,jiawei.shao\}@connect.ust.hk, \\ \{eezhlin,eejzhang\}@ust.hk}}

\maketitle

\begin{abstract}
  With the rapid advancement of stereo vision technologies, stereo image compression has emerged as a crucial field that continues to draw significant attention. Previous approaches have primarily employed a unidirectional paradigm, where the compression of one view is dependent on the other, resulting in imbalanced compression. To address this issue, we introduce a symmetric bidirectional stereo image compression architecture, named BiSIC. Specifically, we propose a 3D convolution based codec backbone to capture local features and incorporate bidirectional attention blocks to exploit global features. Moreover, we design a novel cross-dimensional entropy model that integrates various conditioning factors, including the spatial context, channel context, and stereo dependency, to effectively estimate the distribution of latent representations for entropy coding. Extensive experiments demonstrate that our proposed BiSIC outperforms conventional image/video compression standards, as well as state-of-the-art learning-based methods, in terms of both PSNR and MS-SSIM.
  The code is available at \url{https://github.com/LIUZhening111/BiSIC}.
\end{abstract}

% \footnotetext{The code is available at \url{https://github.com/LIUZhening111/BiSIC}.}

\section{Introduction}
\label{sec:intro}
Stereo vision, which mimics human binocular vision, has emerged as an important area in computer vision. It enables a wide range of applications, such as 3D movies, remote sensing \cite{song2020edgestereo}, autonomous driving \cite{yin20203d}, and augmented/virtual reality (AR/VR) \cite{arena2022overview}. 
With the rapid development of stereo cameras, stereo images have become essential for providing immersive and convincing visual presentations that enhance user experience and benefit downstream applications. 
Meanwhile, the proliferation of stereo cameras leads to a significant surge in the volume of stereo images, resulting in substantial storage costs and posing challenges to transmission efficiency. Consequently, stereo image compression plays a pivotal role in ensuring the practicality and effectiveness of stereo vision.

In recent years, numerous methods have been developed for single image compression, including traditional codecs \cite{wallace1991jpeg, BPGWeb} and learning-based methods \cite{balle2016end, balle2018variational, minnen2018joint, balle2020nonlinear, Cheng_2020_CVPR}. These methods, however, focus on extracting individual features for compression and fail to capture the cross-view correlation. Consequently, they are less effective for stereo image compression that requires the extraction of both intra-view information and inter-view information. To exploit the inherent shared information in stereo images, traditional methods for stereo image coding follow a predictive compression procedure of video standards \cite{tech2015overview, bross2021overview}. In this process, one view is compressed first and then used to predict the other view. Subsequently, the disparity between the predicted and actual views is compressed. Nonetheless, these methods typically rely on hand-crafted modules and lack end-to-end optimization, resulting in suboptimal performance.

Inspired by the success of learning-based single image compression, researchers have proposed novel stereo image compression methods using deep neural networks (DNNs) to enhance compression performance \cite{Liu_2019_ICCV, Deng_2021_CVPR, Wodlinger_2022_CVPR, deng2023masic, zhai2022disparity, Lei_2022_CVPR, zhang2023ldmic, wodlinger2024ecsic}. Initially, extensive research focuses on unidirectional learned stereo image compression methods \cite{Liu_2019_ICCV, Deng_2021_CVPR, Wodlinger_2022_CVPR, deng2023masic, zhai2022disparity, wodlinger2024ecsic}, where one view is compressed and used as a reference to compress the other view. Despite achieving high overall rate-distortion (RD) performance, these unidirectional learned compression methods yield a noticeable imbalance in compression quality between stereo views, similar to traditional predictive compression methods. This imbalance is unfavorable for human vision and undermines downstream machine vision tasks \cite{liu2020visually}. 
Moreover, some of these methods \cite{Deng_2021_CVPR, Wodlinger_2022_CVPR, deng2023masic, zhai2022disparity} heavily rely on sequential coding order and accurate estimation of the explicit warping relationship between views.
To avoid these issues, a recent study \cite{Lei_2022_CVPR} proposes a bidirectional compression method that leverages 2D convolution to extract individual features and facilitate information sharing between views. Nonetheless, the compression performance is limited due to the separate processing of convolutions and the plain spatial context based entropy model.

Based on the above discussion, we propose a novel bidirectional method for stereo image compression, named BiSIC, to address the aforementioned issues. BiSIC achieves improvements in four key aspects: encoder/decoder structure, mutual stereo learning, entropy model, and coding speed. Firstly, regarding the codec structure, although 2D convolutions have proven effective in various computer vision tasks, they fail to capture the aligned morphs and features between stereo views. To overcome this limitation, we propose using 3D convolution as the backbone for the encoder/decoder, which takes both views as input to capture the inter-view dependencies that are inaccessible in previous works based on 2D convolution. Secondly, for mutual stereo learning, we propose a mutual attention block to facilitate the transfer of features between views.
This module enhances the integration and cooperation of the stereo views, leading to improved compression performance. Thirdly, for the entropy model, prior methods primarily use hyperprior and spatial context as conditions, overlooking the rich inter-view dependencies across other dimensions. 
To exploit these dependencies, we construct a cross-dimensional entropy model that aggregates hyperprior, spatial context, channel context, and stereo dependency.

Specifically, we introduce a masked 3D convolution to capture the stereo spatial context and utilize an attention block to process the stereo channel context. 
Finally, to improve coding speed and reduce auto-regressive steps, we design a fast variant based on a stereo-checkerboard structure. This design converts auto-regressive inference into a two-fold operation, improving the coding efficiency. The contributions of this paper are summarized as follows: 
\begin{itemize}
    \item We propose a novel 3D convolution based codec backbone that processes stereo images as a whole, enabling the learning of local stereo features and effective capture of shared patterns across views.
    \item We design a bidirectional mutual attention block that facilitates the exchange of global features between stereo views. This block produces compact attention maps and is robust to input images of different sizes. 
    \item We develop a powerful symmetric cross-dimensional entropy model, which incorporates hyperpriors, spatial context, channel context, and stereo dependency as conditioning factors, to enhance the probability estimation.
    \item To accelerate coding, we propose a stereo-checkerboard design that transforms the spatial auto-regressive inference into a two-fold process. This design strikes a balance between time consumption and compression quality.
\end{itemize}

\begin{figure}[tb]
    \centering
    \includegraphics[height=4.265cm]{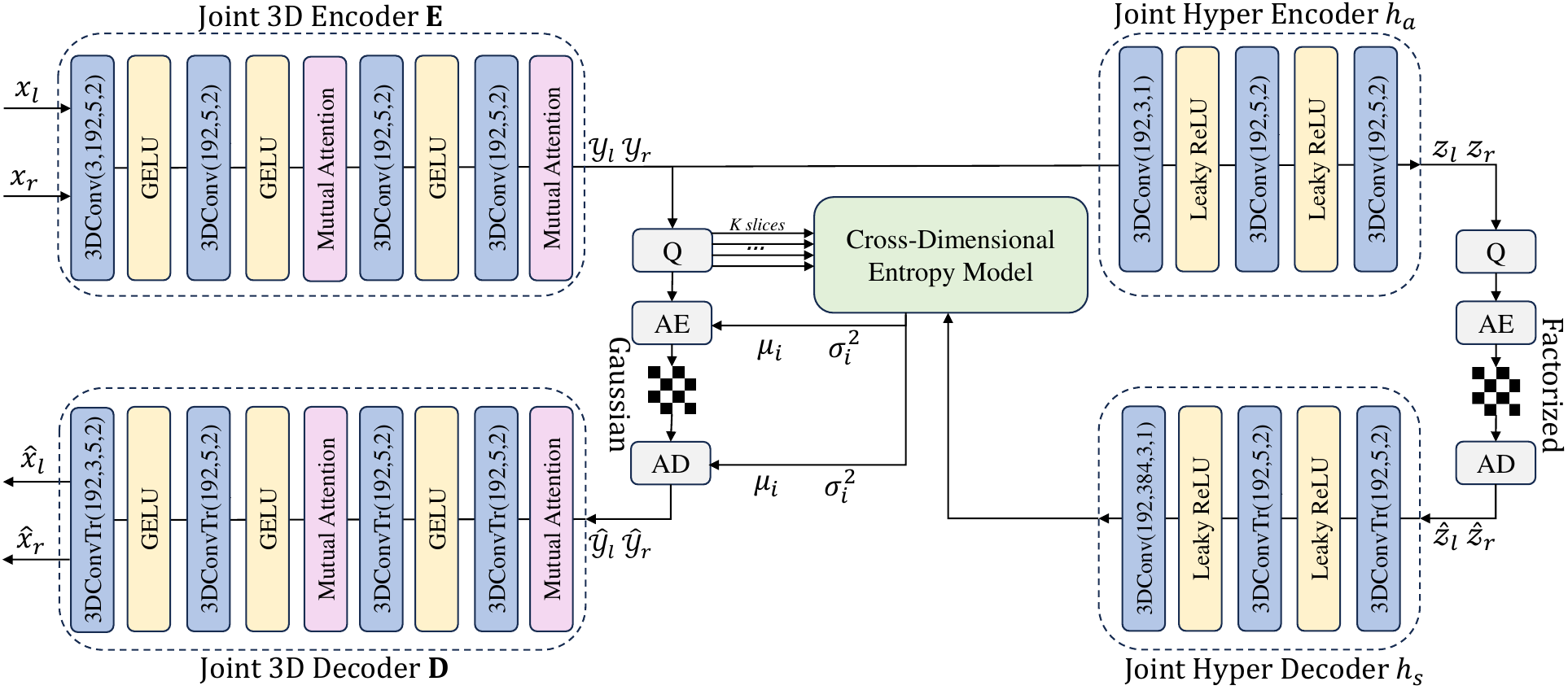}
    \caption{The proposed bidirectional stereo image compression architecture. AE and AD are arithmetic coder for entropy coding. The backbones of encoder $\mathbf{E}$, decoder $\mathbf{D}$, hyper encoder $h_a$, and hyper decoder $h_s$ are constructed with 3D convolution to model local features. \textit{3DConvTr} denotes transposed 3D convolution. To enhance global feature extraction, bidirectional mutual attention blocks are inserted between 3D convolutional layers. Moreover, a novel cross-dimensional entropy model is utilized to capture complex inter-view dependencies.}
    \label{fig:framework}
\end{figure}

\section{Related Works}
\label{sec:related}

\subsection{Single Image Compression}
Image compression is an important topic that continuously attracts significant research interest. With the increasing demand for vision products and applications, steady progress has been made in traditional image compression standards, evolving from JPEG \cite{wallace1991jpeg, skodras2001jpeg} and BPG \cite{BPGWeb} to video compression standards such as HEVC \cite{sullivan2012overview} and VVC \cite{bross2021overview}. These traditional methods establish a standard compression pipeline comprising transform, quantization, and entropy coding. 
Over the past five years, deep learning has revolutionized various vision applications. As demonstrated in the seminal work \cite{balle2016end}, DNNs have the potential to replace components in traditional compression methods and enable end-to-end compression. This work introduces uniform noise to mimic the distortion caused by quantization, thereby paving the way for end-to-end optimization in image compression.

In learned image compression models \cite{balle2016end,balle2018variational,minnen2018joint,minnen2020channel,Cheng_2020_CVPR,hu2020coarse,he2021checkerboard,he2022elic,pan2022content,zhu2021transformer,jiang2023mlic,liu2023learned}, neural networks are employed to parameterize the probability distribution of transformed latents. Accurate estimation of this distribution is crucial for achieving compact compression via entropy coding. To enhance probability estimation, various conditions have been integrated into entropy models \cite{balle2020nonlinear}, including hyperpriors \cite{balle2018variational}, spatial context \cite{minnen2018joint}, and channel-wise context \cite{minnen2020channel}, to capture a broader range of dependencies. To be specific, hyperprior is a lightweight subsidiarity information transmitted alongside the encoded latents. Both spatial context and channel-wise context leverage auto-regressive inference, which is conditioned on the preceding segments of data by iteratively interleaving the inference.
Checkerboard methods \cite{he2021checkerboard}, on the other hand, avoid the spatial auto-regressive process and accelerate the coding speed by constraining the spatial context to the near neighbors of each pixel instead of spatial causal components. These dependencies operate on diverse scales and can be progressively combined to further improve effectiveness, as demonstrated in \cite{he2022elic, jiang2023mlic}.

\subsection{Stereo Image Compression}
Stereo image compression is an extension of single image compression that considers capturing both intra-view features and inter-view correlations. Traditional approaches employ key-frame based  predictive coding, akin to video coding \cite{vetro2011overview, kadaikar2018joint}.
However, these methods rely on hand-crafted feature extractors and classical optimization techniques. As these modules are optimized individually, they often fail to achieve satisfactory RD performance. Moreover, the predictive compression pipeline also leads to imbalanced compression. 

In recent years, learned stereo image compression methods have gained increasing attention \cite{Liu_2019_ICCV, Deng_2021_CVPR, Lei_2022_CVPR, deng2023masic, Wodlinger_2022_CVPR, zhai2022disparity, wodlinger2024ecsic, zhang2023ldmic}. These methods can be broadly categorized into unidirectional and bidirectional approaches. Unidirectional methods first compress one view image and then use it as a reference to compress the other view, significantly reducing the required bits. The pioneering work \cite{Liu_2019_ICCV} introduces a skip module to warp the features from one view to encode the other. Subsequent works have utilized various techniques to represent the warping relationship between stereo views, such as homography matrices \cite{Deng_2021_CVPR, deng2023masic}, disparity maps \cite{zhai2022disparity}, and horizontal shifts \cite{Wodlinger_2022_CVPR}.

These unidirectional methods heavily rely on the precise estimation of the spatial transformations. A very recent work, ECSIC \cite{wodlinger2024ecsic}, avoids explicit parametric transformation in favor of constructing a unidirectional entropy model that conditions the right image on the known left image. However, the decompressed results of these unidirectional methods often have imbalanced quality due to the one-way dependency between stereo views.
On the other hand, bidirectional methods, while remaining less explored, produce more balanced quality by extracting shared information from both views and compressing stereo images simultaneously. One such approach \cite{Lei_2022_CVPR} introduces a contextual transform module for feature transfer between views. However, it employs separate downsampling networks and only takes spatial context into account, which limits the compression performance. In contrast, to fully exploit the shared information and establish stronger dependencies, we employ a joint downsampling codec and a cross-dimensional entropy model.

\section{Proposed Methods}
% \label{sec:method}
% \subsection{System Overview}
The proposed architecture for stereo image compression, as depicted in \cref{fig:framework}, consists of two main components: a joint codec and an entropy model. The codec employs 3D convolutions to transform the original stereo images into compact latent representations, effectively exploiting the correlation between the two views. Subsequently, the entropy model estimates the probability distribution of these latent representations, enabling the generation of a compact bitstream through entropy coding. In the following subsections, we elaborate on the codec and the entropy model.

\subsection{Joint Codec}
\noindent \textbf{3D Convolution Backbone.}
The codec acts as a non-linear transformation that downsamples the raw images into compact latent representations and generates hyperpriors for the entropy model. Different from previous approaches to stereo image compression that employ 2D convolutional layers, we adopt 3D convolutions as the backbone of our codec. Unlike 2D convolution, which operates separately on each view, 3D convolution concurrently processes both stereo images. Consequently, it inherently possesses the capability to extract inter-view correlations. 

Our encoder, denoted by $\mathbf{E}$, comprises four 3D convolutional layers, with bidirectional mutual attention blocks inserted after the second and the fourth layers to facilitate global feature communication between the two views. 
The encoder transforms a pair of stereo images, i.e., the left and right images denoted by $\textbf{\textit{x}}_l, \textbf{\textit{x}}_r \in \mathbb{R}^{B\times 3 \times H\times W}$, into compact latent representations $\textbf{\textit{y}}_l, \textbf{\textit{y}}_r  \in \mathbb{R}^{B\times N \times \frac{H}{16} \times \frac{W}{16}}$ and quantizes them into $\hat{\textbf{\textit{y}}}:=\{\hat{\textbf{\textit{y}}}_l,\hat{\textbf{\textit{y}}}_r\}$ for efficient transmission. 
Conversely, the decoder $\mathbf{D}$ transforms the quantized latents $\hat{\textbf{\textit{y}}}$ back into images $\hat{\textbf{\textit{x}}}_l$ and $\hat{\textbf{\textit{x}}}_r$.
The encoding and decoding processes are expressed as:
\begin{align}
    \textbf{\textit{y}}_l, \textbf{\textit{y}}_r &=\mathbf{E}(\textbf{\textit{x}}_l,\textbf{\textit{x}}_r), \\
    % \hat{\textbf{\textit{y}}}_l =\mathit{Q}&(\textbf{\textit{y}}_l), \; \hat{\textbf{\textit{y}}}_r =\mathit{Q}(\textbf{\textit{y}}_r) \\
    \hat{\textbf{\textit{x}}}_l, \hat{\textbf{\textit{x}}}_r &=\mathbf{D}(\hat{\textbf{\textit{y}}}_l, \hat{\textbf{\textit{y}}}_r).
\end{align} 

To capture the dependencies among elements in the latents $\hat{\textbf{\textit{y}}}$ for efficient entropy coding \cite{bishop1998latent, balle2018variational}, hyperpriors are generated through another 3D convolution based hyper encoder $h_a$.
The generated hyperpriors $\textbf{\textit{z}}_l, \textbf{\textit{z}}_r  \in \mathbb{R}^{B\times M \times \frac{H}{64} \times \frac{W}{64}}$ are quantized into $\hat{\textbf{\textit{z}}}_l,\hat{\textbf{\textit{z}}}_r$ and then transmitted alongside the quantized latents $\hat{\textbf{\textit{y}}}$. At the decoder side, a 3D convolution based hyper decoder $h_s$ upscales the quantized hyperpriors $\hat{\textbf{\textit{z}}}_l, \hat{\textbf{\textit{z}}}_r$ into $\tilde{\textbf{\textit{z}}}_l$, $\tilde{\textbf{\textit{z}}}_r$, which are subsequently utilized in the entropy model, as elaborated in \cref{sec:entropy}. The processes of the hyper encoder and hyper decoder are expressed as:
\begin{align}
    \textbf{\textit{z}}_l, \textbf{\textit{z}}_r&={h}_a(\textbf{\textit{y}}_l, \textbf{\textit{y}}_r), \\
    \tilde{\textbf{\textit{z}}}_l, \tilde{\textbf{\textit{z}}}_r&={h}_s(\hat{\textbf{\textit{z}}}_l, \hat{\textbf{\textit{z}}}_r).
    \label{eq:hs}
\end{align}

\begin{figure}[tb]
    \centering
    \includegraphics[height=4cm]{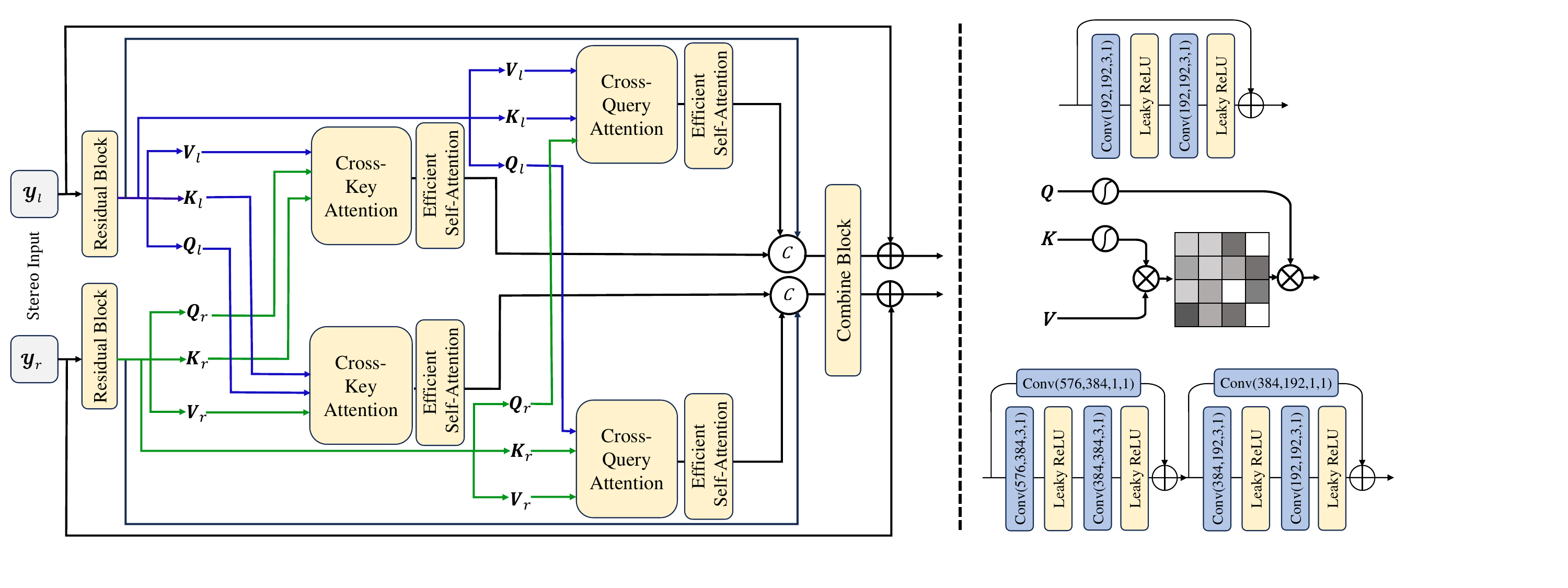}
    \caption{Overview of the proposed bidirectional mutual attention block. The blue and green lines represent features extracted from the left view and right view, respectively. The network structures of the residual block, the basic efficient attention unit, and the combine block are illustrated on the right, from top to bottom.}
    \label{fig:mutualatten}
\end{figure}

\noindent \textbf{Bidirectional Mutual Attention Block.}
Convolutional layers have an excellent ability for local modeling, but their ability for long-range feature modeling is relatively limited. On the other hand, attention modules \cite{vaswani2017attention} excel at extracting long-range features. Therefore, we propose combining the attention mechanism with convolutional layers to compensate for the shortcomings of the latter and leverage the advantages of both convolution and attention. To this end, we design a novel bidirectional mutual attention block, as depicted in \cref{fig:mutualatten}, to be inserted between convolutional layers as the pink blocks shown in \cref{fig:framework}.

As illustrated in \cref{fig:mutualatten}, the bidirectional mutual attention block contains two cross-attention stages, each followed by a self-attention layer. The two cross-attention stages differ in how they concentrate on the feature map and are referred to as cross-key attention and cross-query attention. Before each cross-attention stage, the input stereo data $\textbf{\textit{y}}_{l}, \textbf{\textit{y}}_{r}$ first pass through a residual block, which are then embedded into low-dimensional query, key, and value tensors $\textbf{\textit{Q}}, \textbf{\textit{K}}, \textbf{\textit{V}} \in \mathbb{R}^{B\times C \times (H \times W)}$.
To avoid the prohibitive computation of classical attention, which yields an $(H \times W) \times (H \times W)$ sized attention map, we adopt the efficient attention approach proposed in \cite{shen2021efficient}. This approach computes the product of the key and the transposed value, thereby reducing the attention map size to $ C_{\textbf{\textit{K}}} \times  C_{\textbf{\textit{V}}}$. 

In the cross-key attention stage, for each view, we generate an attention map using its key and the value from the other view, which is then queried by the current view. By doing so, we can extract the cross-key feature $\mathbf{\Phi}$ from the stereo views as:
\begin{equation}
     \mathbf{\Phi}_{r \to l} =({\sigma(\textbf{\textit{K}}_{r}) \times {\textbf{\textit{V}}_{l}}^T})^T \times \sigma(\textbf{\textit{Q}}_{r}), \  
    \mathbf{\Phi}_{l \to r} =({\sigma(\textbf{\textit{K}}_{l}) \times {\textbf{\textit{V}}_{r}}^T})^T \times \sigma(\textbf{\textit{Q}}_{l}),
\end{equation}
where $\sigma$ is the softmax function. This approach calculates the attention map using the key and value from different views, thereby facilitating inter-view alignment and the identification of common patterns. When a feature in one view is correlated with the corresponding feature in the other view, it is effectively captured through the attention map. 

In the cross-query attention stage, the attention map is obtained using the key and value from a single view, and then queried by the other view, yielding the cross-query feature $\mathbf{\Psi}$ as: 
\begin{equation}
    \mathbf{\Psi}_{r \to l} =({\sigma(\textbf{\textit{K}}_{l}) \times {\textbf{\textit{V}}_{l}}^T})^T \times \sigma(\textbf{\textit{Q}}_{r}), \ 
    \mathbf{\Psi}_{l \to r} =({\sigma(\textbf{\textit{K}}_{r}) \times {\textbf{\textit{V}}_{r}}^T})^T \times \sigma(\textbf{\textit{Q}}_{l}).
\end{equation}
The attention map in this stage is generated within one view, thus maintaining an inner-focused relationship. This process preserves the individual self-features of each view, while the other view serves as a reference for determining the allocation of attention weights.

After obtaining the cross-key attention and cross-query attention features $\mathbf{\Phi},\mathbf{\Psi}$, we employ self-attention to augment these features as $\tilde{\mathbf{\Phi}},\tilde{\mathbf{\Psi}}$ and use a shared-parameter combine block to get the desired mutual attention output, as shown in the bottom right of \cref{fig:mutualatten}. Specifically, the combine block concatenates the augmented features $\tilde{\mathbf{\Phi}},\tilde{\mathbf{\Psi}}$ and the residual component, and shrinks the channel number to be consistent with the block input.

This bidirectional mutual attention block captures diverse features and provides a global receptive field for the codec. It is suitable for any stereo information transfer scenarios, provided that the inputs are two features of equal size, making it a generalized plug-and-play block. Moreover, the size of the attention maps is determined only by the embedding channel $C$ and does not vary with the size of input images.

\begin{figure}[tb]
    \centering
    \includegraphics[height=4.2cm]{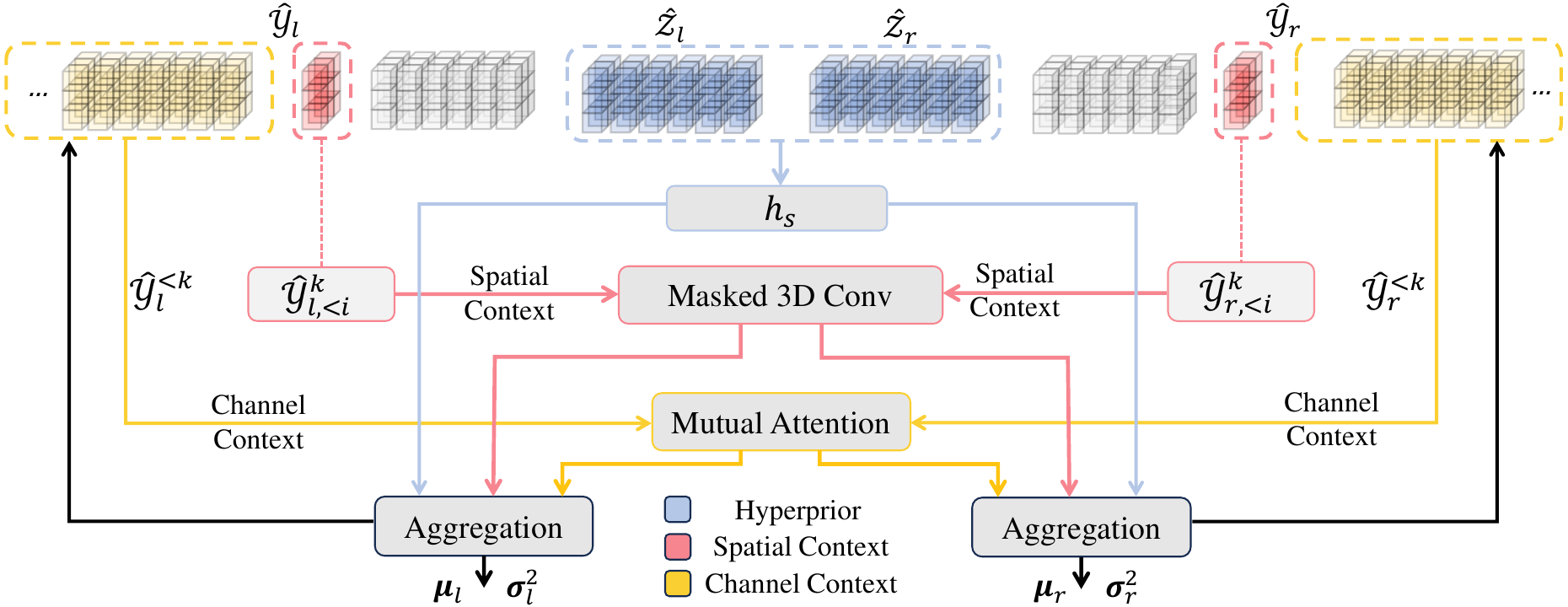}
    \caption{Illustration of the proposed symmetric cross-dimensional entropy model. It jointly aggregates the hyperprior (blue), stereo spatial context (red), and stereo channel context (yellow) as conditions for an effective probability distribution estimation. The hyper decoder ${h}_s$ for hyperprior is shown in \cref{fig:framework}. The masked 3D convolution for spatial context is provided in \cref{fig:Masked3D}. Moreover, the mutual attention block for channel context is detailed in \cref{fig:mutualatten}.}
    \label{fig:entropy}
\end{figure}

\subsection{Cross-Dimensional Entropy Model }
\label{sec:entropy}
The proposed codec effectively transforms stereo images into small-sized latent representations. To achieve efficient compression, the next crucial step is to encode these latents into compact bitstreams using accurate probability distribution and entropy coding \cite{balle2018variational, minnen2018joint}. 
Thus, an entropy model is critical for parameterizing and estimating probabilities. In learned image compression, the probability distribution is commonly modeled as Gaussian distributions \cite{minnen2018joint}, with the mean $\mathbf{\mu}$ and variance $\mathbf{\sigma}^2$ being estimated. To achieve accurate estimation, it is critical to provide appropriate references. Apart from hyperpriors, previous stereo image compression methods primarily use spatial context \cite{Lei_2022_CVPR, Deng_2021_CVPR, zhang2023ldmic} and unidirectional stereo dependency as references \cite{Liu_2019_ICCV,wodlinger2024ecsic}, while abundant features on the channel dimension are overlooked. In this work, we propose a symmetric cross-dimensional entropy model that integrates hyperprior, spatial context, channel context, and stereo dependency in a fully symmetric bidirectional way. 

The architecture of the proposed cross-dimensional entropy model is depicted in \cref{fig:entropy}. We use $\hat{\textbf{\textit{y}}}_{l, i}^k$ to denote the $i$-th $(i=1,2,...,n)$ entry of the $k$-th $(k=1,2,...,K)$ channel slice from the received left view latent $\hat{\textbf{\textit{y}}}_{l}$, and likewise for the others. The cross-dimensional condition formulation is expressed as:
\begin{align}
    \label{eq:entropyA}
    p_{\hat{\textbf{\textit{y}}}_{l}}(\hat{\textbf{\textit{y}}}_{l,i}^k | \hat{\textbf{\textit{z}}}_{l}, \hat{\textbf{\textit{y}}}_{l}^{<k}, \hat{\textbf{\textit{y}}}^k_{l,<i}, \hat{\textbf{\textit{y}}}_{r}^{<k}, \hat{\textbf{\textit{y}}}^k_{r,<i}) \sim \mathcal{N}&(\bm{\mu}_{l},\bm{\sigma}^2_{l}), \\
    p_{\hat{\textbf{\textit{y}}}_{r}}(\hat{\textbf{\textit{y}}}_{r,i}^k | \hat{\textbf{\textit{z}}}_{r}, \hat{\textbf{\textit{y}}}_{r}^{<k}, \hat{\textbf{\textit{y}}}^k_{r,<i}, \hat{\textbf{\textit{y}}}_{l}^{<k}, \hat{\textbf{\textit{y}}}^k_{l,<i}) \sim \mathcal{N}&(\bm{\mu}_{r},\bm{\sigma}^2_{r}).
    \label{eq:entropyB}
\end{align}
For the hyperprior component, we generate the dependency $\tilde{\textbf{\textit{z}}}_l, \tilde{\textbf{\textit{z}}}_r$ from the hyper decoder using \cref{eq:hs}.
To obtain the channel context, we evenly split the stereo latents on the channel dimension into $K$ slices. These latent slices share a great similarity, providing abundant conditions \cite{minnen2020channel}. To exploit this similarity, the probability estimation is performed slice by slice, using the previous slices as references for the following one. In addition, the decoded slices of the other view also serve as references. Therefore, we propose to use the mutual attention block introduced in the previous subsection to aggregate the channel-wise features $\mathbf{\Theta}_{l},\mathbf{\Theta}_{r}$ from both views, conditioning on $\hat{\textbf{\textit{y}}}_{l}^{<k}$ and  $\hat{\textbf{\textit{y}}}_{r}^{<k}$, as depicted by the yellow lines in \cref{fig:entropy}.

\begin{figure}[t]
  \centering
\begin{subfigure}{0.18\linewidth}
    \includegraphics[height=1.5cm]{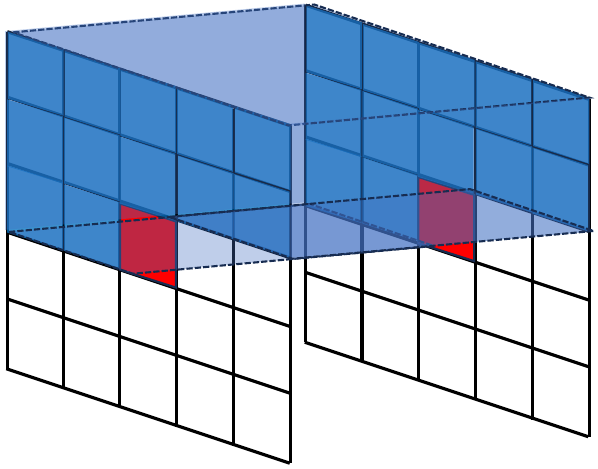}
    \caption{Masked 3D convolution.}
    \label{fig:Masked3D}
  \end{subfigure}
  \hfill
  \begin{subfigure}{0.35\linewidth}
    \includegraphics[height=2cm]{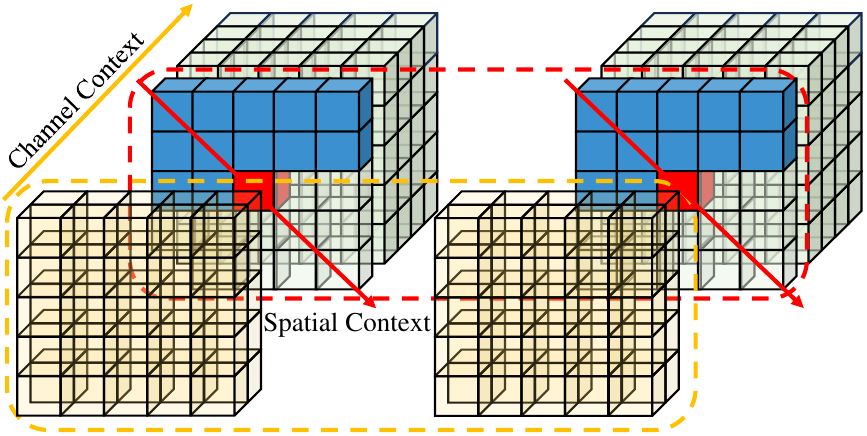}
    \caption{Auto-regressive pattern in BiSIC entropy model.}
    \label{fig:entropycube}
  \end{subfigure}
  \hfill
    \begin{subfigure}{0.35\linewidth}
    \includegraphics[height=2cm]{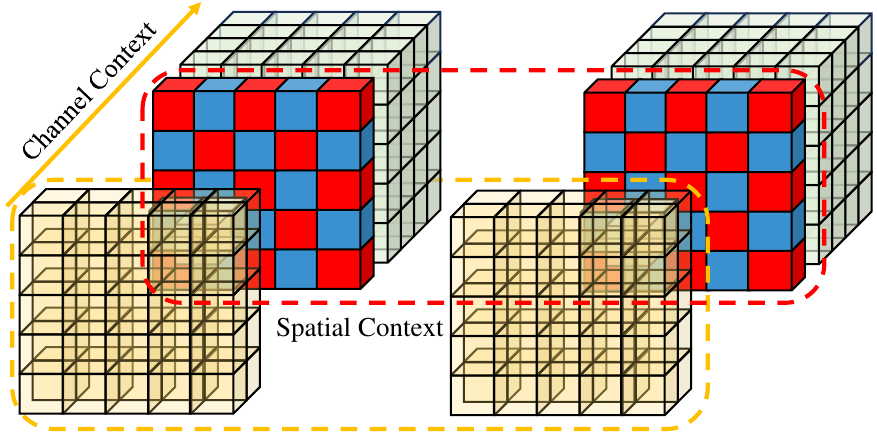}
    \caption{Stereo-checkerboard pattern in BiSIC-Fast.}
    \label{fig:CKBDcube}
  \end{subfigure}
  \hfill
  \caption{(Left) Illustration of masked 3D convolution. The blue region is used as the condition for the probability estimation of the current target depicted in red. The weights of the convolution kernel in blue are valid, while the white region is masked to maintain the causal spatial context. (Middle) Demonstration of the auto-regressive process in the cross-dimensional entropy model. The yellow and red arrows represent the channel context and spatial context, respectively. The spatial context is processed entry by entry. (Right) Demonstration of the stereo-checkerboard in the fast variant, where the blue and red parts are each processed only once.}
  \label{fig:CNNs}
\end{figure}

To capture the spatial context within each channel slice, we consistently employ 3D convolution to jointly learn spatial features from the causal areas $\hat{\textbf{\textit{y}}}^k_{<i}$. Specifically, during the coding process, the current compression/decompression only accesses the areas that have already been encoded/decoded. To maintain the causality of the compression, we mask out the uncoded part, which is referred to as masked 3D convolution, as shown in \cref{fig:Masked3D}. In this approach, for the $i$-th entry $\hat{\textbf{\textit{y}}}^k_i$, the weights of uncoded part are set to zero, and we obtain spatial context features $\mathbf{\Upsilon}_{l,i},\mathbf{\Upsilon}_{r,i}$ by considering the causal context $\hat{\textbf{\textit{y}}}^k_{l,<i}$ and $\hat{\textbf{\textit{y}}}^k_{r,<i}$ from both views simultaneously, as depicted by the red lines in \cref{fig:entropy}.

Upon extracting various dependency features, we employ an aggregation network $\mathbf{G}_{\textrm{ag}}$ to fuse the conditions and obtain the probability estimations as:
\begin{equation}
    \bm{\mu}_{l},\bm{\sigma}^2_{l}=\mathbf{G}_{\textrm{ag}}(\tilde{\textbf{\textit{z}}}_l, \mathbf{\Theta}_{l},\mathbf{\Upsilon}_{l}), \ 
    \bm{\mu}_{r},\bm{\sigma}^2_{r}=\mathbf{G}_{\textrm{ag}}(\tilde{\textbf{\textit{z}}}_r, \mathbf{\Theta}_{r},\mathbf{\Upsilon}_{r}).
\end{equation}
The entire procedure in the entropy model is designed to be fully symmetric and bidirectional, which aligns with our motivation for a bidirectional design.

\subsection{Loss Function}
The proposed stereo image compression network is trained using the rate-distortion loss. The rate is defined by bit-per-pixel (BPP), which includes bits for transmitting latents $\hat{\textbf{\textit{y}}}_{l}, \hat{\textbf{\textit{y}}}_{r}$ and hyperpriors $\hat{\textbf{\textit{z}}}_{l}, \hat{\textbf{\textit{z}}}_{r}$. The distortion is measured using the mean squared error (MSE) and the multi-scale structural similarity index measure (MS-SSIM) \cite{wang2003multiscale}. The overall loss function is given by:
\begin{equation}
     \mathcal{L}:=\lambda D + R=\lambda \sum_{l,r} d(\textbf{\textit{x}}, \hat{\textbf{\textit{x}}}) + \sum_{l,r}(r(\hat{\textbf{\textit{y}}})+r(\hat{\textbf{\textit{z}}})),
\end{equation}
where $d(\cdot)$ denotes the distortion function with MSE or MS-SSIM,  $r(\cdot)$ calculates the bitrate using the estimated probability of $\hat{\textbf{\textit{y}}}$ and $\hat{\textbf{\textit{z}}}$ \cite{balle2018variational}, and $\lambda$ is a hyperparameter that balances the rate and distortion.

\begin{figure}[tb]
    \centering
    \includegraphics[height=3.7cm]{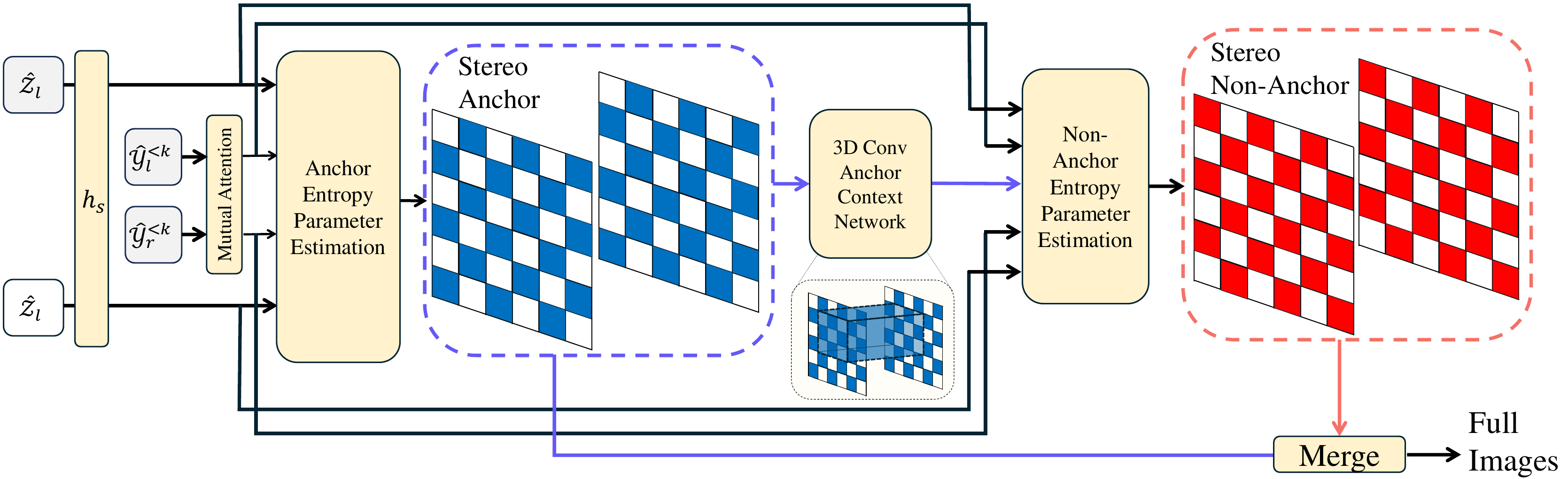}
    \caption{Illustration of the entropy model for the proposed fast variant based on stereo-checkerboard. The blue and red parts represent the stereo anchor and stereo non-anchor, respectively. By concurrently processing two views, the 3D convolution based anchor context network effectively captures the dependency from the stereo anchor.}
    \label{fig:CKBD}
\end{figure}

\subsection{Fast Variant}
The proposed entropy model effectively captures multiple references and estimates the probability distribution. However, the auto-regressive iterations in the spatial context are time-consuming. To strike a better balance between compression runtime and performance, we need to reduce the auto-regressive executions while maintaining a strong condition for the entropy model. To this end, we propose a fast variant of our model, named BiSIC-Fast, by extending the checkerboard structure \cite{he2021checkerboard} into stereo-checkerboard and utilizing 3D convolutions to jointly model the spatial context of two views with only two iterations. 

This fast variant maintains the codec as BiSIC, while the entropy model network is modified as illustrated in \cref{fig:CKBD}. In this approach, we split the latents ${\textbf{\textit{y}}}$ into two parts, stereo anchor part and stereo non-anchor part, as represented in blue and red in \cref{fig:CKBD}, respectively. The entropy parameter estimation for the stereo anchor part relies on the hyperprior condition and channel context. The stereo anchor is decompressed first, and then the stereo non-anchor utilizes the decompressed stereo anchor as a dependency. Consequently, the entropy parameter estimation for the stereo non-anchor part is based on the hyperprior, channel context, and stereo anchor context.

By adopting this approach, we avoid the spatial auto-regressive process in favor of a two-fold operation, as demonstrated in \cref{fig:entropycube} and \cref{fig:CKBDcube}. Although the spatial auto-regressive process is omitted to save runtime, the neighbors of each pixel in the latents are still maintained, preserving a good perceptive field.

\section{Experiments}
\subsection{Experimental Implementation}
\noindent \textbf{Datasets.}
We evaluate the performance of our methods on two benchmark datasets: InStereo2K \cite{bao2020instereo2k} and Cityscapes \cite{cordts2016cityscapes}. 
InStereo2K contains 2,060 indoor scene stereo image pairs of size $1,080 \times 860$. It is split into 2,010 image pairs for training and 50 for testing. Cityscapes contains 5,000 outdoor scene stereo image pairs of size $2,048 \times 1,024$, where 2,975 pairs are allocated for training and 1,525 for testing. 

\noindent \textbf{Benchmarks.}
We compare our methods with traditional codecs, such as BPG \cite{BPGWeb}, H.265/HEVC \cite{sullivan2012overview}, MV-HEVC \cite{tech2015overview}, and H.266/VVC \cite{bross2021overview}.
BPG encodes each view of the stereo pair independently, while H.265/HEVC and H.266/VVC compress the left and right view images jointly. 
In addition, we show the results of learning-based stereo image compression schemes, including HESIC+ \cite{Deng_2021_CVPR}, SASIC \cite{Wodlinger_2022_CVPR}, BCSIC \cite{Lei_2022_CVPR}, LDMIC \cite{zhang2023ldmic}, and ECSIC \cite{wodlinger2024ecsic}.

\noindent \textbf{Metrics.} The quality of the reconstructed images is evaluated by the peak signal-to-noise ratio (PSNR) and MS-SSIM \cite{wang2003multiscale}. The rate is measured by BPP. Besides, we use the widely employed Bjøntegaard Delta bitrate (BDBR) and Bjøntegaard Delta PSNR (BD-PSNR) \cite{bjontegaard2001calculation} to compare the rate-distortion (RD) performance among the methods. Specifically, BDBR measures the average bit saving at the same distortion level, while BD-PSNR reflects the relative PSNR gap at an equivalent bitrate. A lower value of BDBR and a higher value of BD-PSNR indicate better performance.

\noindent \textbf{Implementation Details.}
During the training process, we optimize the model using the MSE loss over 1.5 million steps. We employ the Adam optimizer \cite{kingma2014adam} and set the batch size to 8. The learning rate is initialized at $10^{-4}$ and is reduced by half after every 300,000 steps. The hyperparameter $\lambda$ is chosen from the set $\{256, 512, 1024, 2048, 3072, 4096\}$ for different bitrates.
To achieve better MS-SSIM performance, we fine-tune the pre-trained models for an additional 500,000 steps.
More implementation details are provided in the supplementary material.

\subsection{Results}
\begin{figure}[tb]
  \centering
  \begin{subfigure}{0.45\linewidth}
    \includegraphics[height=4cm]{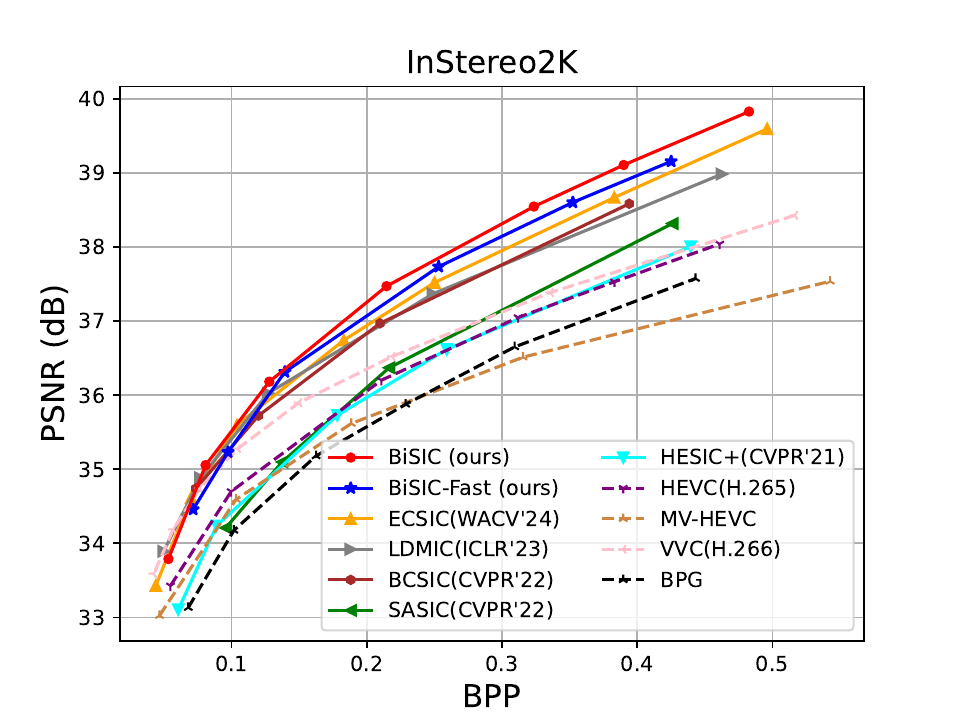}
    % \caption{}
    \label{fig:PSNRIns}
  \end{subfigure}
  \hfill
  \begin{subfigure}{0.45\linewidth}
    \includegraphics[height=4cm]{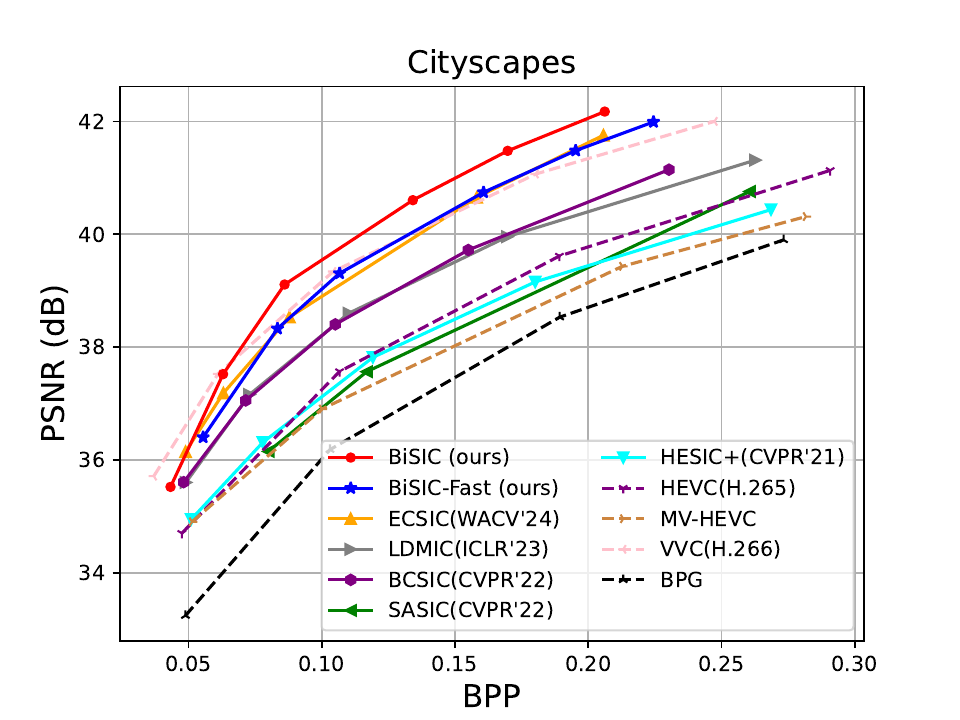}
    % \caption{}
    \label{fig:PSNRCity}
  \end{subfigure}
  \\
    \begin{subfigure}{0.45\linewidth}
    \includegraphics[height=4cm]{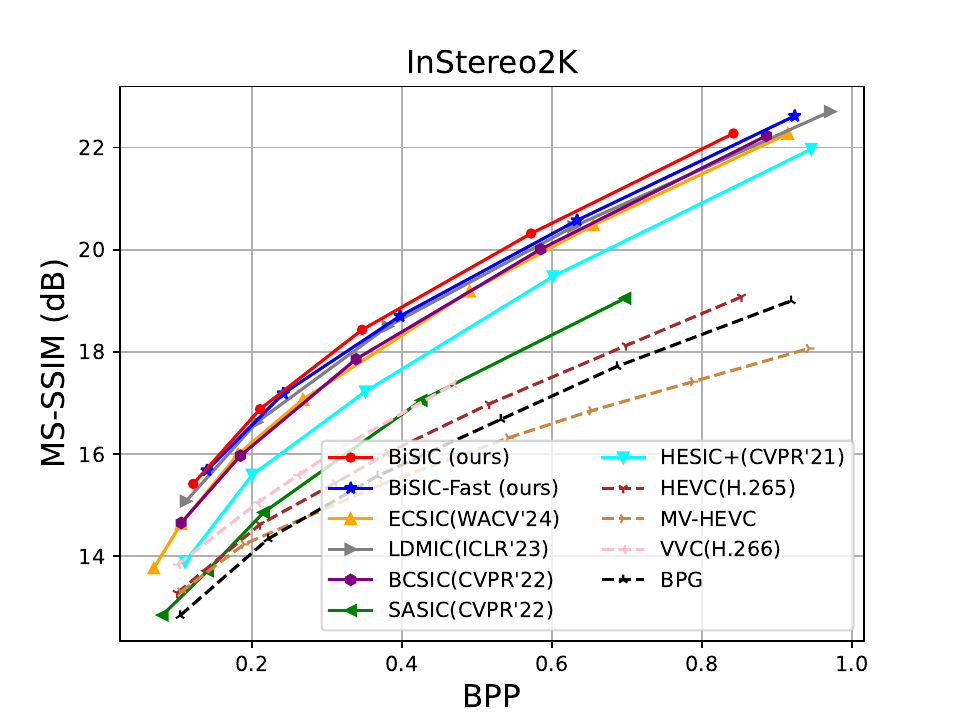}
    % \caption{}
    \label{fig:SSIMIns}
  \end{subfigure}
  \hfill
   \begin{subfigure}{0.45\linewidth}
    \includegraphics[height=4cm]{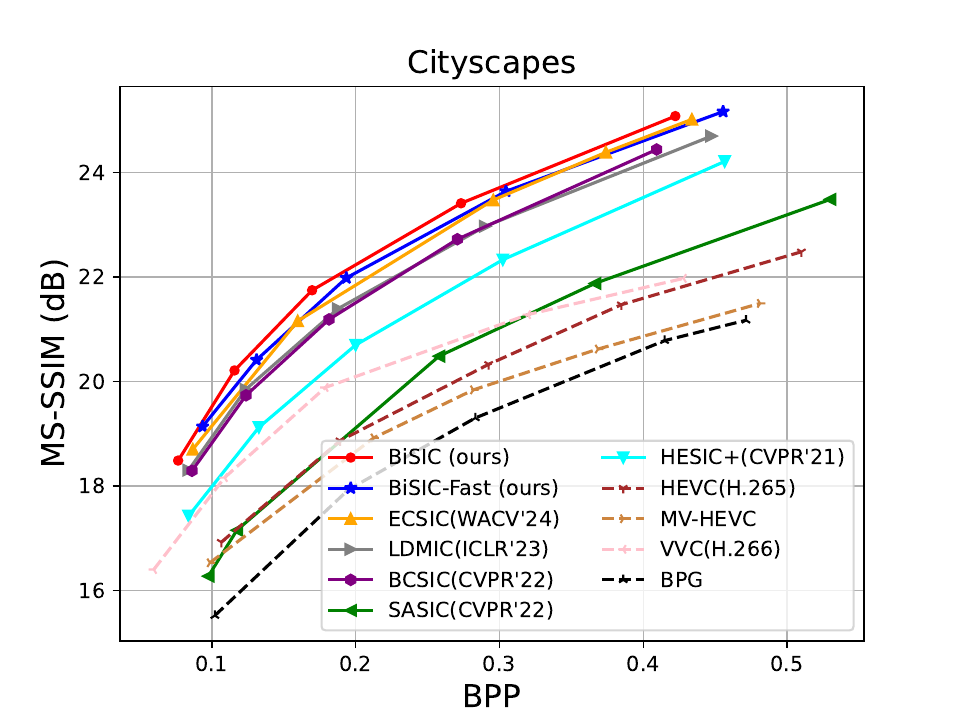}
    % \caption{}
    \label{fig:SSIMCity}
  \end{subfigure}
  \caption{Rate-distortion curves in terms of PSNR and MS-SSIM on the InStereo2K and Cityscapes datasets. }
  \label{fig:RD2x2}
\end{figure}
\noindent \textbf{RD Performance.}
\cref{fig:RD2x2} shows the RD curves of different compression methods on the InStereo2K and Cityscapes datasets.
In terms of the MS-SSIM metric, our proposed method BiSIC outperforms all the other compression methods on both datasets.
Besides, BiSIC achieves higher PSNR in the high BPP range compared to the baselines, while maintaining competitive PSNR values at low bitrates.
The BDBR results are presented in \cref{tab:BDBR}, where BPG is set as the baseline. 
Our BiSIC achieves more than 40\% and 50\% BDBR reduction on InStereo2K and Cityscapes, respectively.
Notably, compared to the state-of-the-art video codec VVC, BiSIC achieves an additional 12.76\% BDBR reduction for PSNR and an additional 30.08\% BDBR reduction for MS-SSIM on InStereo2K.
Compared with unidirectional methods, BiSIC achieves significant bit savings by observing a holistic view and mutually sharing features, which helps remove redundancies in each view.
Moreover, compared with the state-of-the-art bidirectional coding scheme BCSIC \cite{Lei_2022_CVPR}, our method achieves an additional 6.5\% to 15\% BDBR reduction. 
% ######
This is because the employed 3D convolution captures more correlations than the separately deployed 2D convolution used in BCSIC \cite{Lei_2022_CVPR}, and our cross-dimensional entropy model effectively aggregates various abundant dependencies from different dimensions, providing accurate probability estimation.
In addition, the proposed fast variant also demonstrates satisfactory performance, with a marginal degradation in compression performance compared to our BiSIC.

\begin{figure}[t]
    \centering
    \includegraphics[width= 0.99\linewidth ]{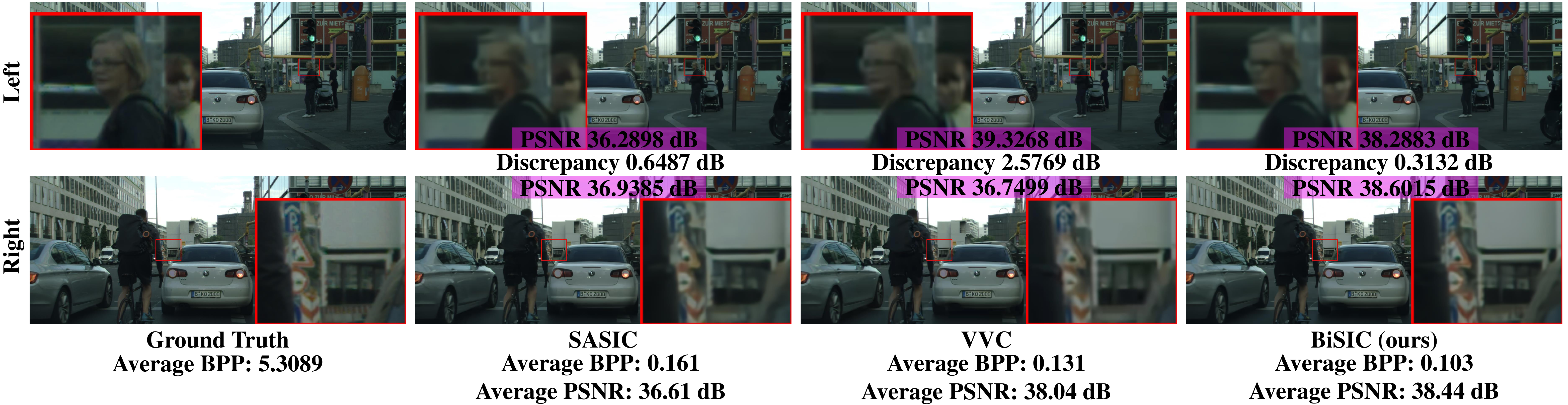}
    \caption{Qualitative comparison of image reconstruction quality across various codecs.}
    \label{fig:visua_city}
\end{figure}

\noindent \textbf{Qualitative Results.}
\cref{fig:visua_city} visualizes the reconstructed images of different methods on the Cityscapes dataset.
For a fair comparison, all images are compressed to similar bitrates.
We observe that the proposed method not only achieves higher PSNR for the reconstructed images at reduced bitrates but also maintains consistent PSNR across stereo views, thanks to its bidirectional architecture. 
In comparison, the SASIC method displays a PSNR discrepancy of 0.6487 dB between views, as it compresses the right image based on the left one.
The VVC codec compresses images in a predictive manner, resulting in an even larger PSNR gap of 2.5769 dB between views.
We leave additional qualitative analysis to the supplementary material.

\begin{table}[t]
  \centering
  \setlength{\tabcolsep}{12pt}
  \caption{BDBR values of different compression methods. \textcolor[rgb]{ 1,  0,  0}{\textbf{Bold}} indicates best results, and \textcolor[rgb]{ 0,  0,  1}{\underline{underlined}} values are the second-best ones.}
    \begin{tabular}{ccccc}
    \toprule
    \multirow{2}[4]{*}{Method} & \multicolumn{2}{c}{InStereo2K} & \multicolumn{2}{c}{Cityscapes} \\
\cmidrule{2-5}          & PSNR  & MS-SSIM & PSNR  & MS-SSIM \\
    \midrule
    HEVC \cite{sullivan2012overview}  & -18.80\% & -12.59\% & -27.46\% & -24.36\% \\
    MV-HEVC \cite{tech2015overview} & -7.69\% & 2.14\% & -18.02\% & -17.13\% \\
    VVC \cite{bross2021overview}  & -35.31\% & -31.05\% & \textcolor[rgb]{ 0, 0,  1}{\underline{-56.25\%}} & -44.04\% \\
    HESIC+ \cite{Deng_2021_CVPR} & -14.96\% & -43.22\% & -23.83\% & -50.79\% \\
    SASIC \cite{Wodlinger_2022_CVPR} & -18.40\% & -23.87\% & -21.47\% & -29.78\% \\
    BCSIC \cite{Lei_2022_CVPR} & -41.22\% & -54.67\% & -42.62\% & -60.72\% \\
    LDMIC \cite{zhang2023ldmic} & -41.95\% & -58.98\% & -41.92\% & -61.90\% \\
    ECSIC \cite{wodlinger2024ecsic} & -43.71\% & -55.65\% & -52.06\% & -64.96\% \\
    BiSIC (ours)  & \textcolor[rgb]{ 1,  0,  0}{\textbf{-48.07\%}} & \textcolor[rgb]{ 1,  0,  0}{\textbf{-61.13\%}} & \textcolor[rgb]{ 1,  0,  0}{\textbf{-57.49\%}} & \textcolor[rgb]{ 1,  0,  0}{\textbf{-67.98\%}} \\
    BiSIC-Fast (ours) & \textcolor[rgb]{ 0,  0,  1}{\underline{-45.35\%}} & \textcolor[rgb]{ 0,  0,  1}{\underline{-59.36\%}} & -51.96\% & \textcolor[rgb]{ 0,  0,  1}{\underline{-65.56\%}} \\
    \bottomrule
    \end{tabular}%
  \label{tab:BDBR}%
\end{table}%

\begin{table}[t]
  \centering
  \caption{Average runtime of different methods.}
  \resizebox{0.99\textwidth}{!}{
    \begin{tabular}{c|cccccccc}
    \toprule
    Codec & BPG & HEVC  & VVC   & SASIC & BCSIC & LDMIC & BiSIC (ours) & BiSIC-Fast (ours) \\
    \midrule
    Runtime & 16.17s & 28.16s & 190.27s & 20.24s & 89.44s & 81.35s & 167.25s & 22.82s \\
    \bottomrule
    \end{tabular}%
    }
  \label{tab:runtime}%
\end{table}%

\subsection{Runtime Comparison}

In this part, we evaluate the computational complexity of various compression algorithms. 
To ensure a fair comparison, each method is executed on an Intel Xeon Platinum 8370C CPU.
\cref{tab:runtime} reports the average runtime on the InStereo2K test set, which includes the encoding and decoding latency. 
We observe that our BiSIC-Fast significantly reduces the runtime compared with the proposed BiSIC.
This is because we replace the spatial auto-regressive interweaving in the entropy model with a two-fold stereo-checkerboard, which simplifies the dependency structure within the latent representations. Note that BiSIC-Fast still maintains competitive performance, as demonstrated in \cref{tab:BDBR}. Therefore, it achieves a better trade-off between the compression runtime and performance. 

\subsection{Ablation Study}
\begin{figure}[t]
    \begin{minipage}[b]{0.55\linewidth}
        \centering
        \includegraphics[height=3.85cm]{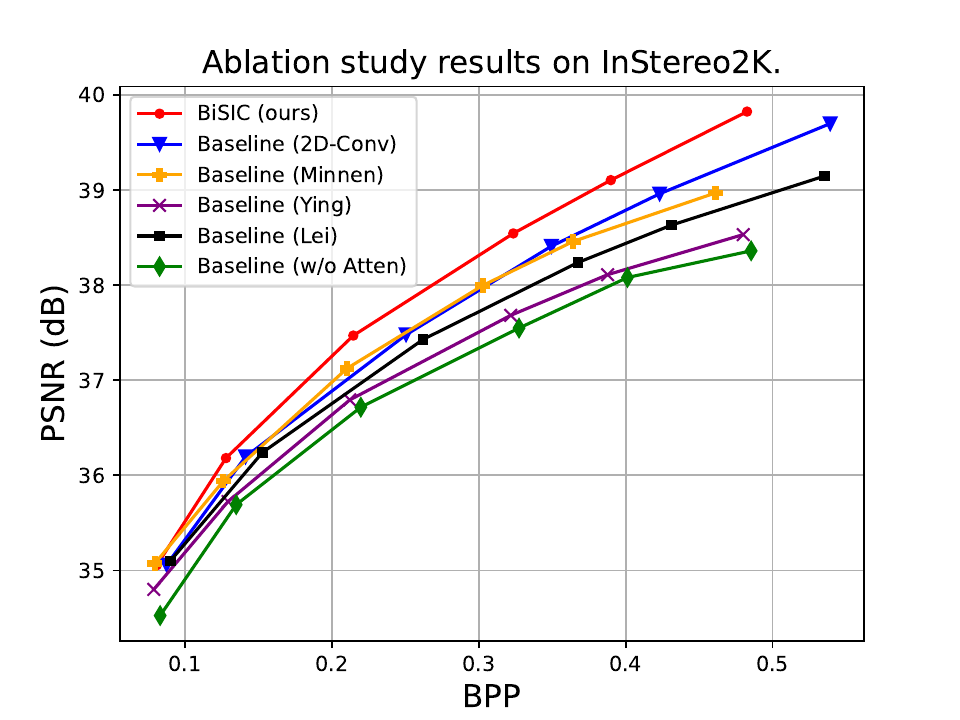}
        \caption{Rate-distortion curves of different ablation baselines on InStereo2K.}
        \label{fig:ablation}
    \end{minipage}
    \hspace{4 pt}
    \begin{minipage}[b]{0.4\linewidth}
        \centering
        \setlength{\tabcolsep}{4pt}
        \begin{tabular}{cc}
            \toprule
            Method & \multicolumn{1}{l}{BD-PSNR} \\
            \midrule
            BiSIC (ours) & 0 \\
            Baseline (2D-Conv) & -0.3218 dB \\
            Baseline (Minnen) & -0.3475 dB \\
            Baseline (Ying) & -0.6554 dB \\
            Baseline (Lei) &  -0.4956 dB \\
            Baseline (w/o Atten) & -0.7882 dB \\
            \bottomrule
        \end{tabular}
        % \vspace{14pt}
        \captionof{table}{BD-PSNR \cite{bjontegaard2001calculation} results relative to the proposed BiSIC.}
        \label{tab:ablation}
    \end{minipage}
\end{figure}

We conduct ablation experiments on the InStereo2K dataset to analyze the contribution of the proposed modules. Our model comprises several key components, including the 3D convolution backbone, cross-dimensional entropy model, and mutual attention block. 
To evaluate their impact, we replace each module with a baseline counterpart or remove it from the model, and compare the performance in terms of PSNR.
The experimental results are shown in \cref{fig:ablation} and \cref{tab:ablation}.

\noindent \textbf{Effectiveness of 3D Convolution Backbone.} 
To assess the capability of 3D convolutional layers in learning inter-view features, we replace 3D convolutional layers in codec with traditional 2D convolutional layers. The resultant model, referred to as Baseline (2D-Conv), experiences a PSNR degradation of $0.3218$ dB at the same bitrate, as shown in \cref{fig:ablation} and \cref{tab:ablation}.
This indicates that the 3D convolutional layers are more effective at capturing the spatio-stereo information inherent in the data and reducing the redundancies between views.

\noindent \textbf{Effectiveness of Cross-Dimensional Entropy Model.} 
To validate the performance improvement offered by our proposed cross-dimensional entropy model, we compare it with the joint auto-regressive and hyperprior entropy model \cite{minnen2018joint}, which we refer to as Baseline (Minnen).
As shown in \cref{fig:ablation} and \cref{tab:ablation}, the proposed cross-dimensional entropy model achieves better rate-distortion performance than Baseline (Minnen).
This suggests that our model provides more accurate probability estimations, which in turn minimizes the coding overhead.

\noindent \textbf{Effectiveness of Mutual Attention Block.} 
To evaluate the effectiveness of the proposed mutual attention block, we consider three baselines for comparisons, namely Baseline (Ying), Baseline (Lei), and Baseline (w/o Atten).
Baseline (Ying) and Baseline (Lei) replace the mutual attention block with the stereo attention module \cite{ying2020stereo} and the contextual transfer module \cite{Lei_2022_CVPR}, respectively. In particular, Baseline (w/o Atten) represents a variant without the attention module.
As shown in \cref{fig:ablation} and \cref{tab:ablation}, our proposed model outperforms all baselines by a large margin.
Notably, removing the proposed mutual attention block results in a dramatic degradation, which demonstrates its significance.  

\section{Conclusion}
In this work, we introduce a bidirectional stereo image compression network that employs 3D convolution to encode local stereo features and propose a mutual attention block to capture global features. Besides, we design a symmetric cross-dimensional entropy model that integrates hyperprior, spatial context, channel context, and stereo dependency. Our bidirectional design mitigates the issue of imbalanced compression quality intrinsic to unidirectional methods and eliminates the requirement for sequential coding. As demonstrated in the experimental results, our BiSIC achieves greater bit savings at the same level of compression quality compared to both traditional coding standards and existing learning-based methods. Moreover, we propose a fast variant that significantly reduces the compression runtime while maintaining competitive performance.

\section*{Acknowledgements}
This work was supported by the General Research Fund (Project No. 16209622) from the Hong Kong Research Grants Council. 

\bibliographystyle{splncs04}
\bibliography{main}

\newpage
\section*{Appendix}

\appendix
The supplementary material provides additional visualization results in \cref{sec:visualization}, extra ablation studies in \cref{sec:ab}, more experimental details in \cref{sec:exp}, and discussion about potential extensions in \cref{sec:exten}.

\section{Additional Visualization Results} 
\label{sec:visualization}
\subsection{Qualitative Results}
\begin{figure}[t!] % \setcounter{figure}{8}
\centering

    \begin{subfigure}{0.3\textwidth}
    \centering
    \includegraphics[width=\linewidth]{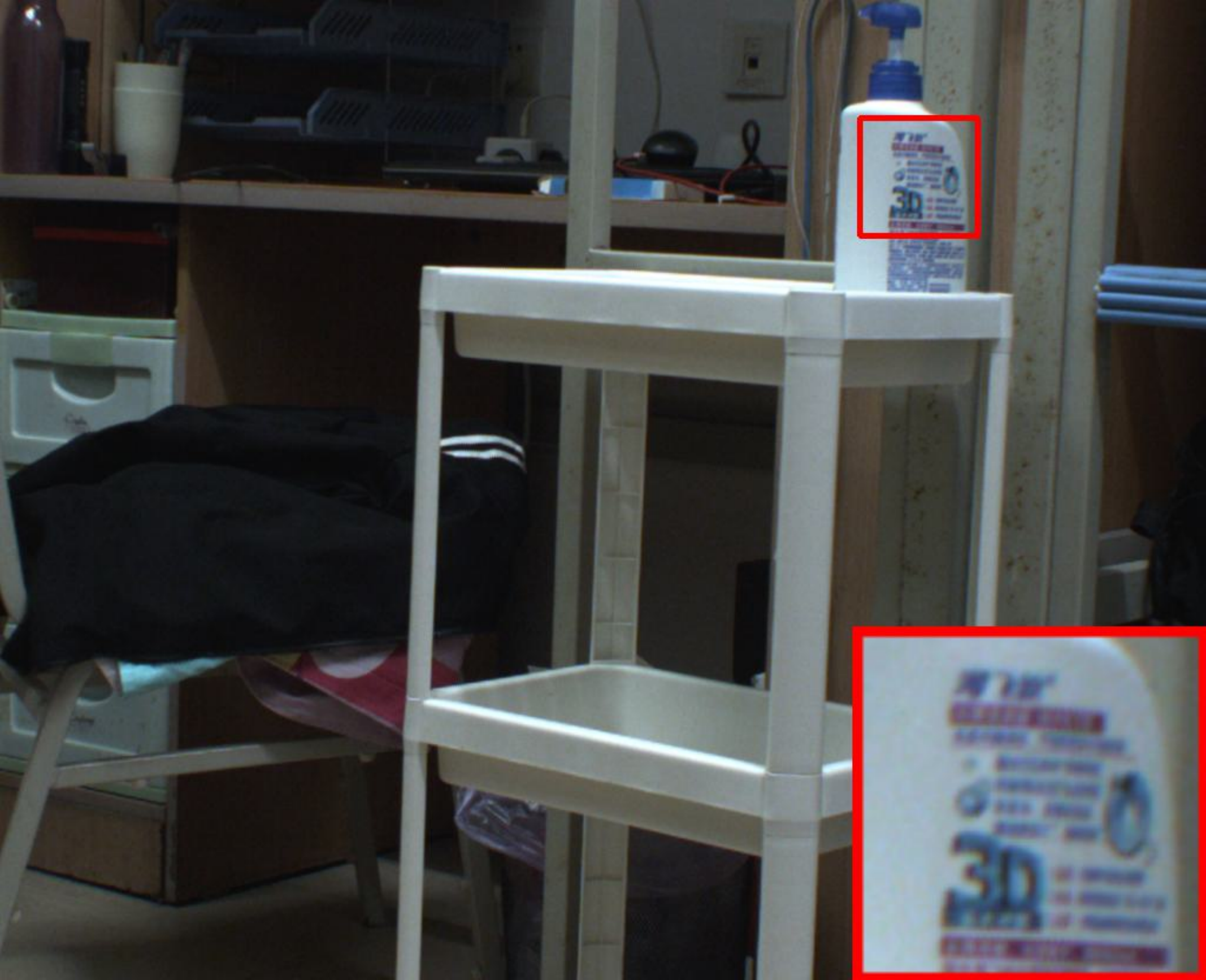}
    \caption{Ground truth left. \\ \quad}
    \label{fig:ins1GT}
    \end{subfigure}
    ~
    \begin{subfigure}{0.3\textwidth}
    \centering
    \includegraphics[width=\linewidth]{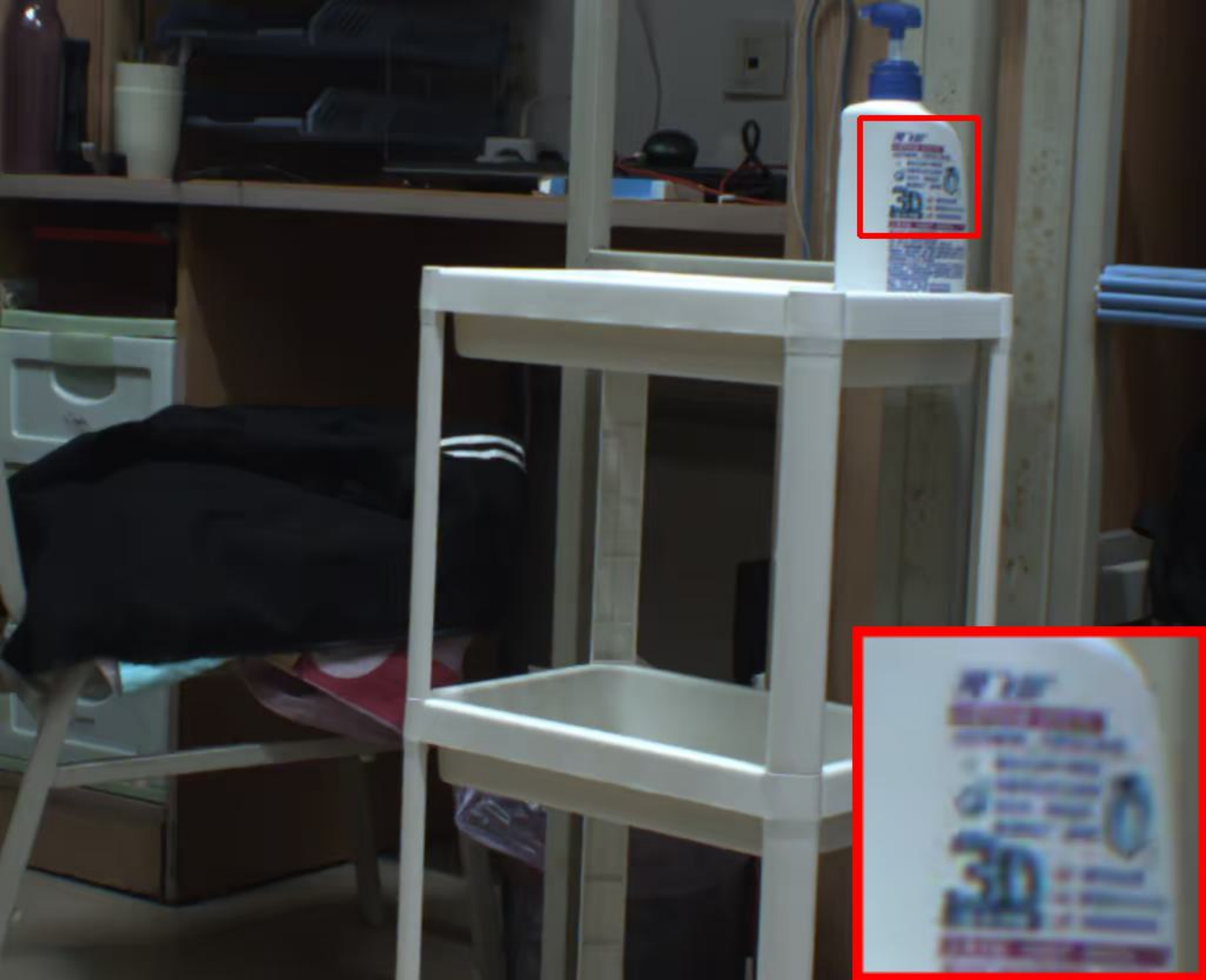}
    \caption{BPG left. \\ BPP: 0.1337 PSNR: 37.496 dB}
    \label{fig:ins1A}
    \end{subfigure}
    ~
    \begin{subfigure}{0.3\textwidth}
    \centering
    \includegraphics[width=\linewidth]{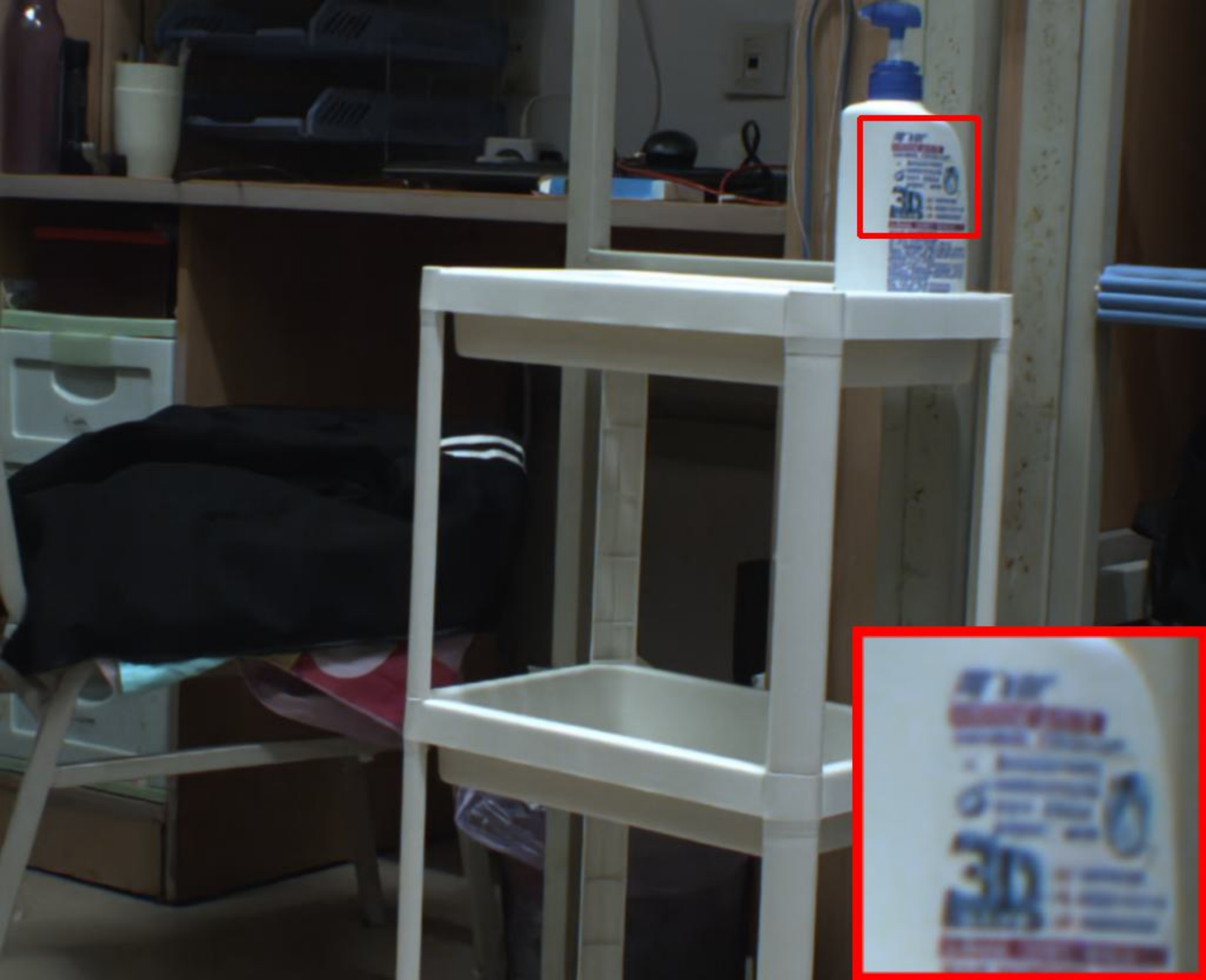}
    \caption{Proposed BiSIC left. \\ BPP: 0.1330 PSNR: 39.288 dB}
    \label{fig:ins1B}
    \end{subfigure}
    
    % \bigskip
    
    \begin{subfigure}{0.3\textwidth}
    \centering
    \includegraphics[width=\linewidth]{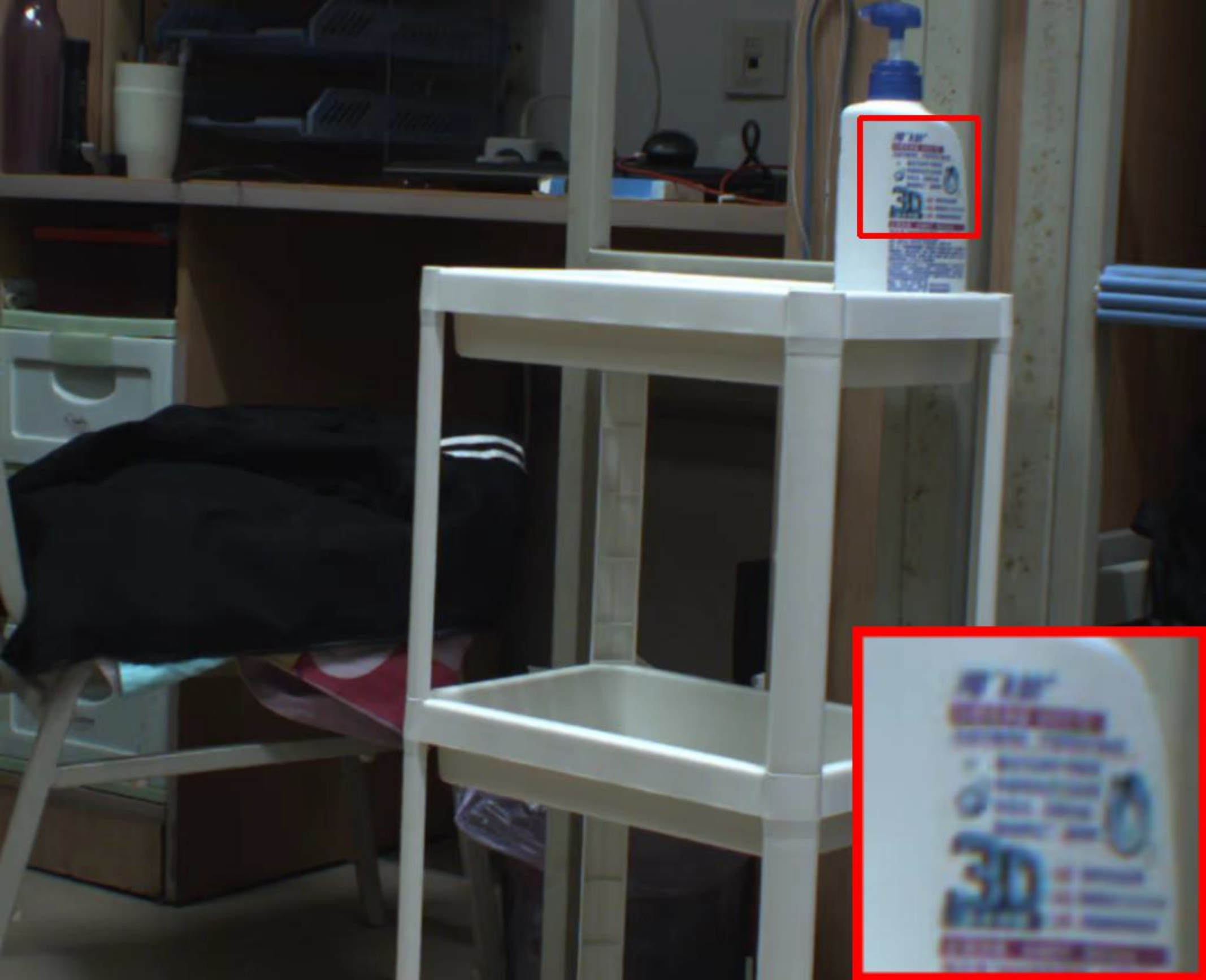}
    \caption{HEVC left. \\ BPP: 0.1448 PSNR: 38.661 dB}
    \label{fig:ins1C}
    \end{subfigure}
    ~
    \begin{subfigure}{0.3\textwidth}
    \centering
    \includegraphics[width=\linewidth]{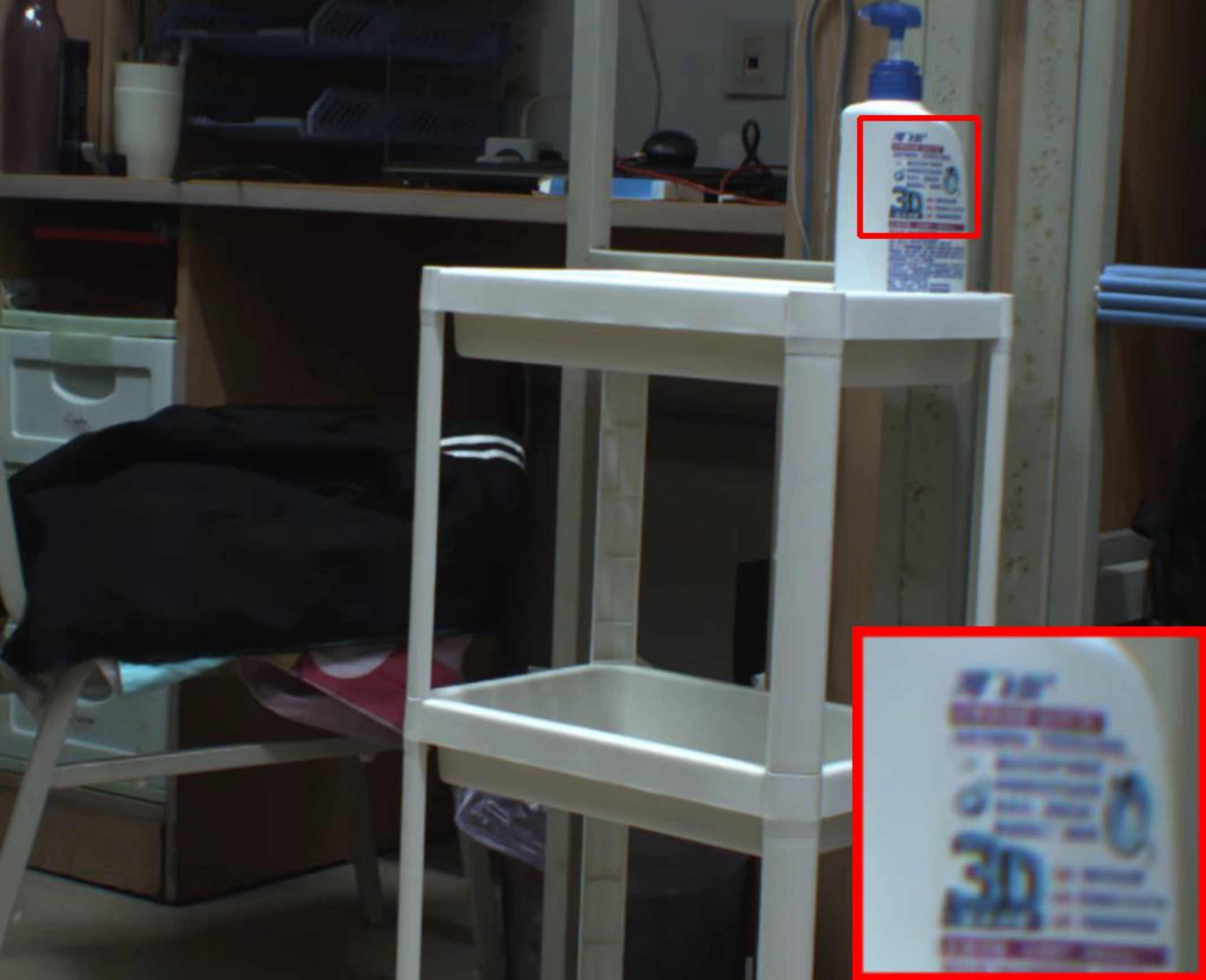}
    \caption{VVC left. \\ BPP: 0.1344 PSNR: 39.081 dB}
    \label{fig:ins1D}
    \end{subfigure}
    ~
    \begin{subfigure}{0.3\textwidth}
    \centering
    \includegraphics[width=\linewidth]{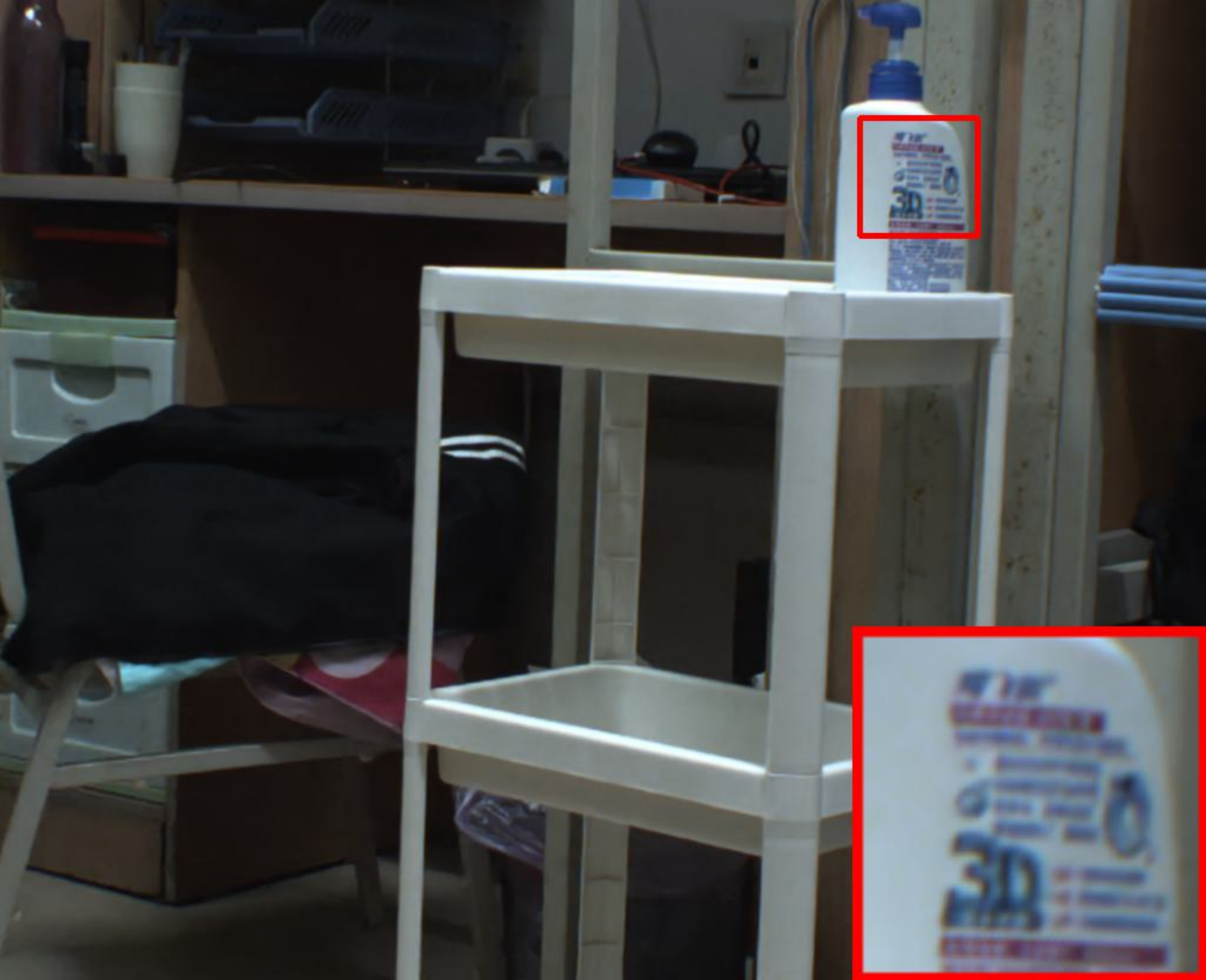}
    \caption{ECSIC left. \\ BPP: 0.1820 PSNR: 38.979 dB}
    \label{fig:ins1F}
    \end{subfigure}

    % \bigskip
    
    \begin{subfigure}{0.3\textwidth}
    \centering
    \includegraphics[width=\linewidth]{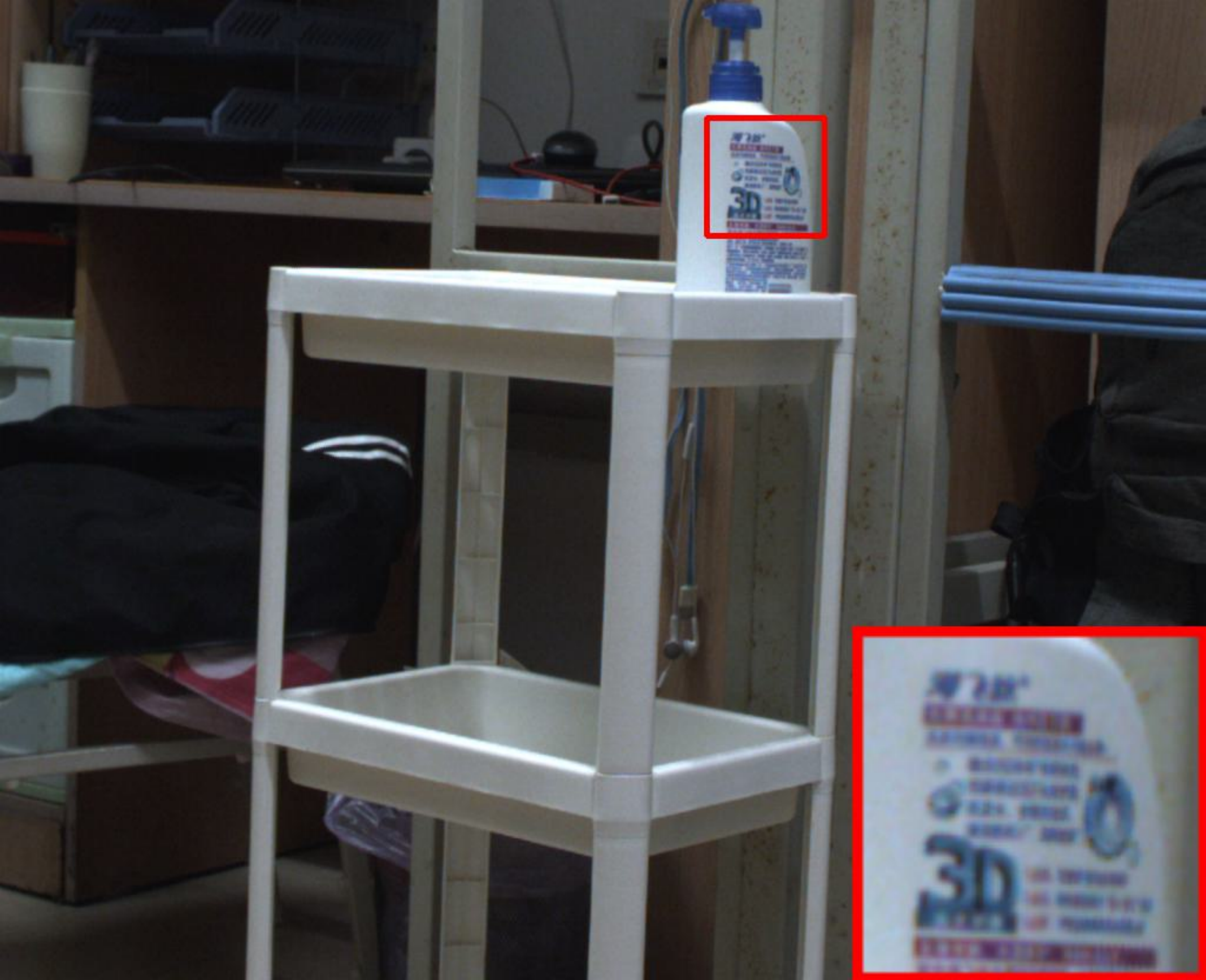}
    \caption{Ground truth right. \\ \quad}
    \label{fig:ins2GT}
    \end{subfigure}
    ~
    \begin{subfigure}{0.3\textwidth}
    \centering
    \includegraphics[width=\linewidth]{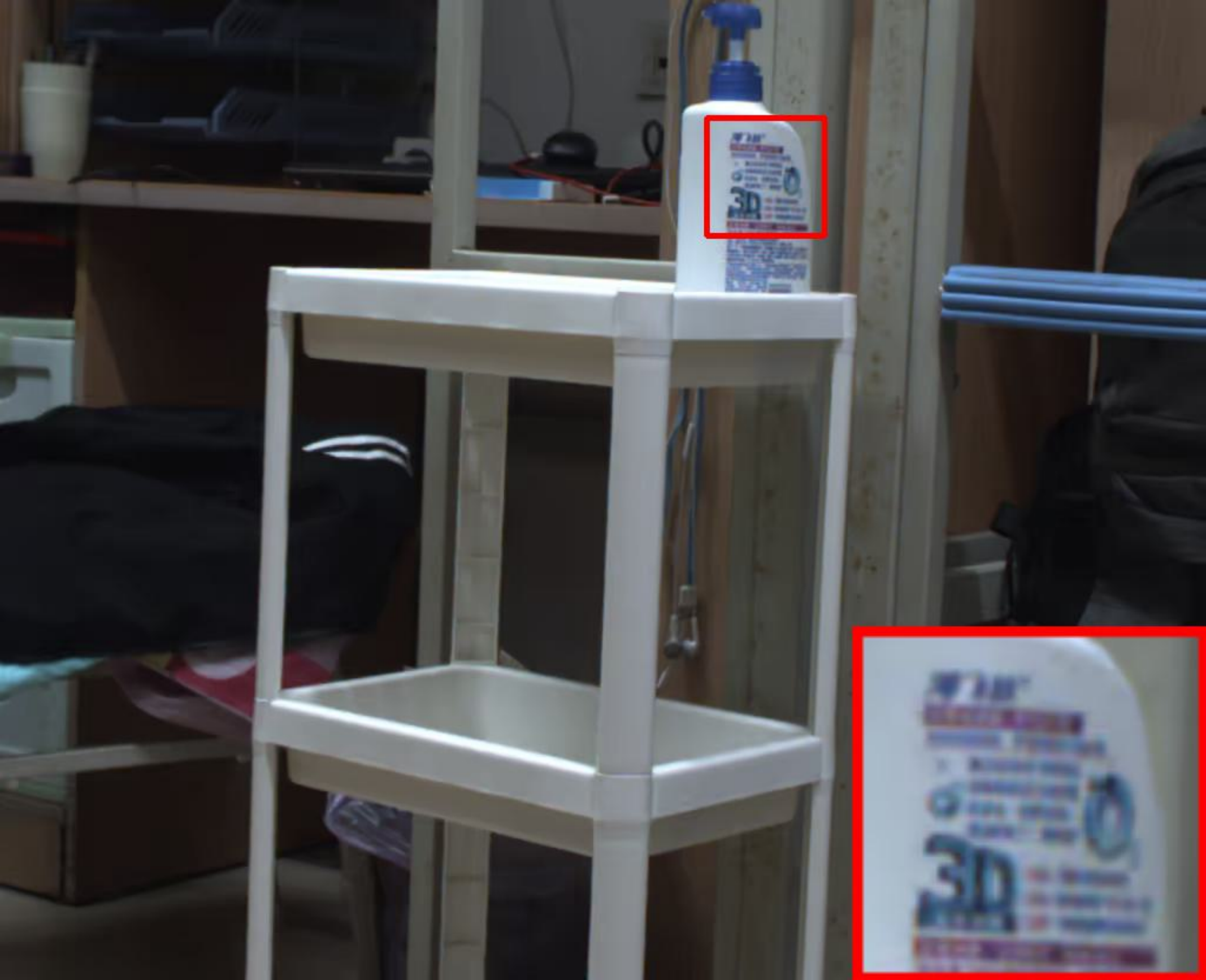}
    \caption{BPG right. \\ BPP: 0.1322 PSNR: 37.580 dB}
    \label{fig:ins2A}
    \end{subfigure}
    ~
    \begin{subfigure}{0.3\textwidth}
    \centering
    \includegraphics[width=\linewidth]{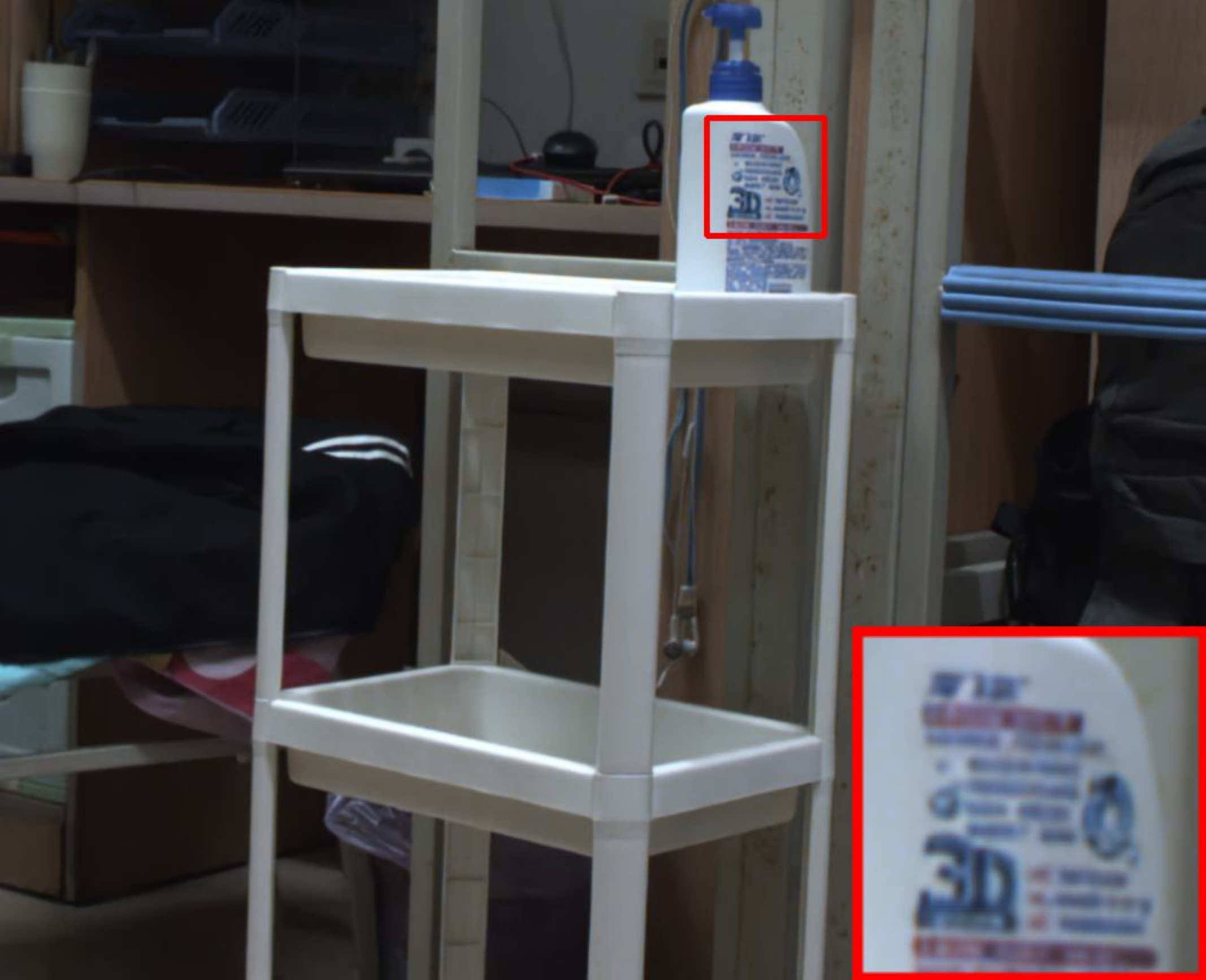}
    \caption{Proposed BiSIC right. \\ BPP: 0.1292 PSNR: 39.422 dB}
    \label{fig:ins2B}
    \end{subfigure}
    
    % \bigskip
    
    \begin{subfigure}{0.3\textwidth}
    \centering
    \includegraphics[width=\linewidth]{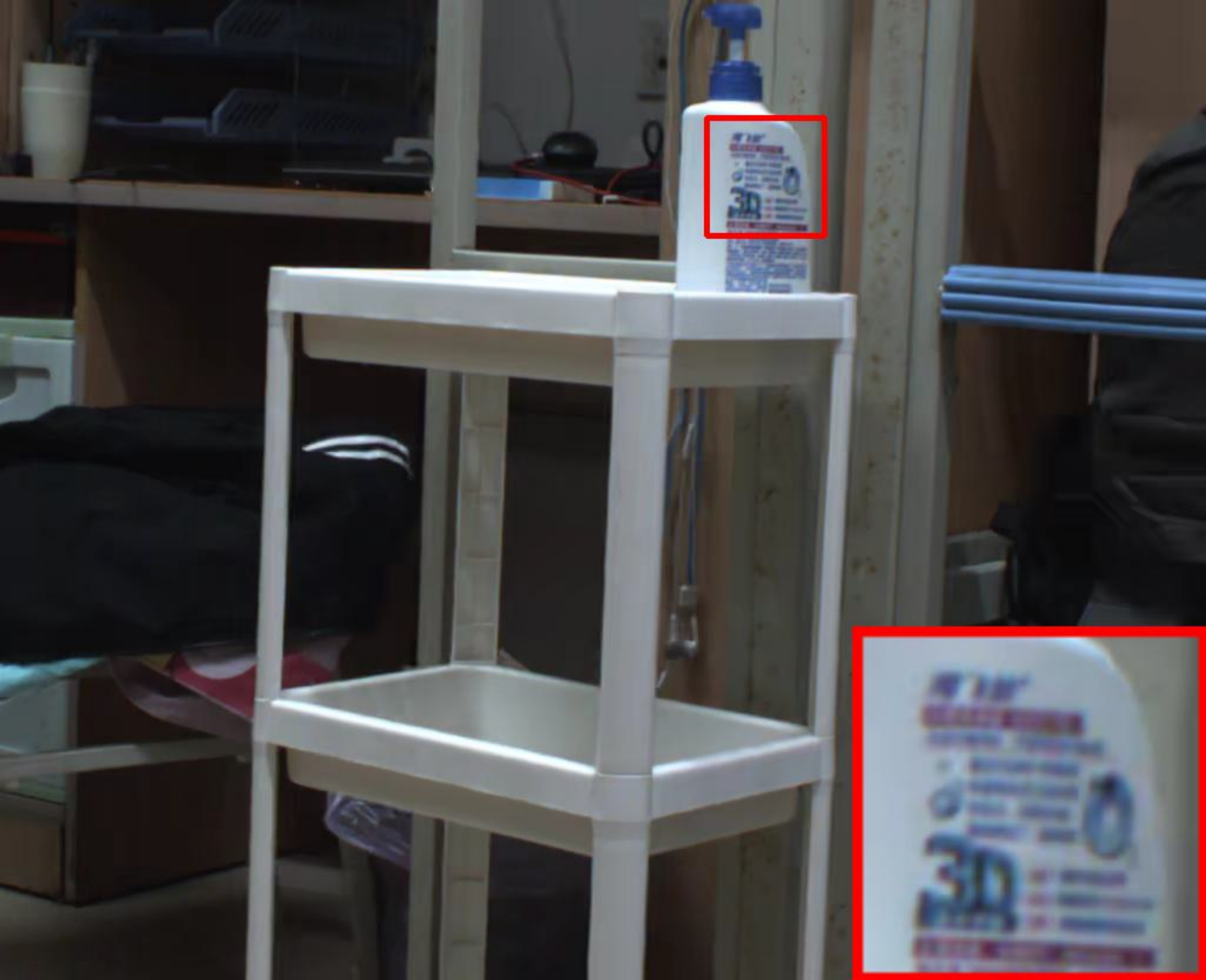}
    \caption{HEVC right. \\ BPP: 0.1448 PSNR: 36.396 dB}
    \label{fig:ins2C}
    \end{subfigure}
    ~
    \begin{subfigure}{0.3\textwidth}
    \centering
    \includegraphics[width=\linewidth]{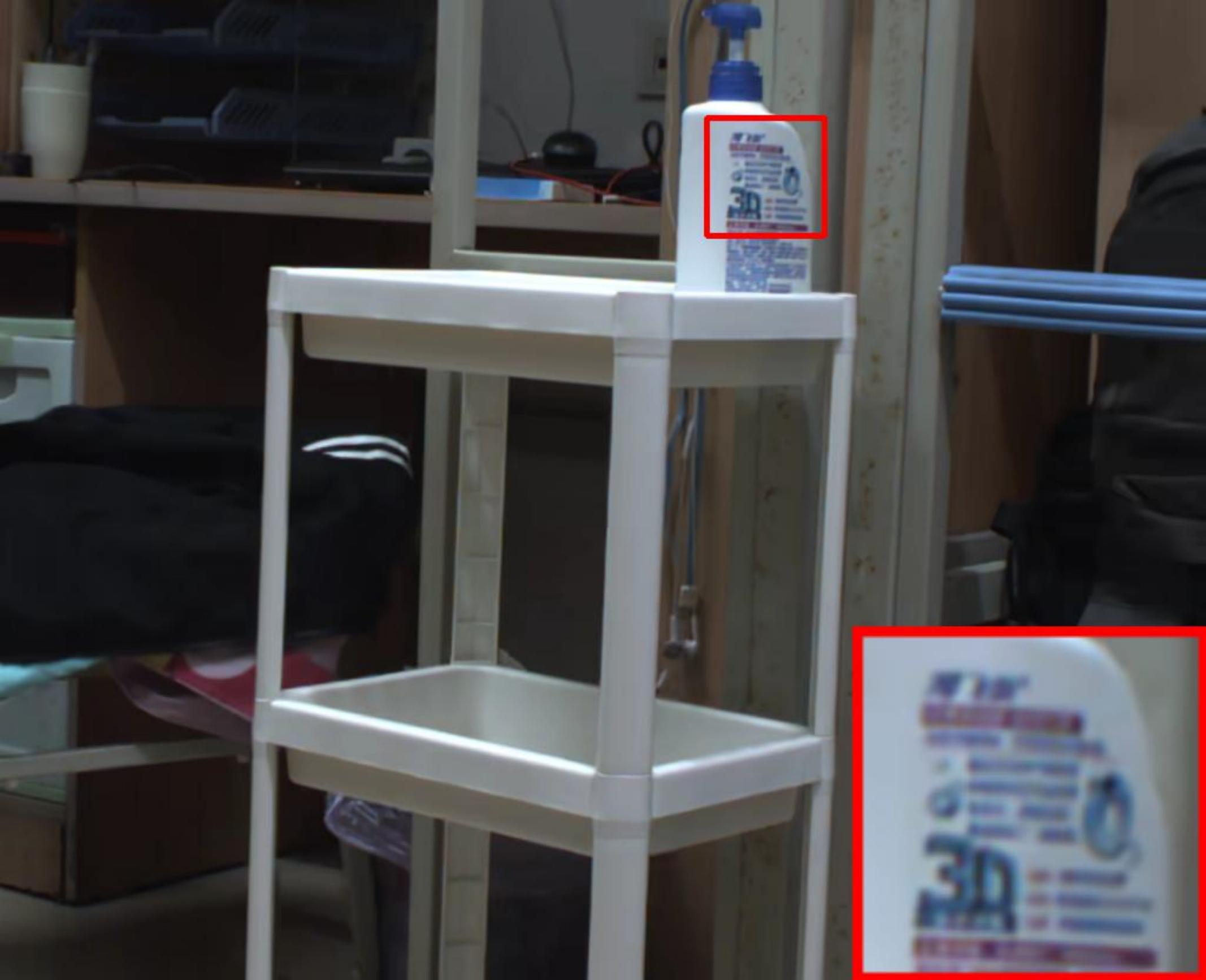}
    \caption{VVC right. \\ BPP: 0.1344 PSNR: 37.237 dB}
    \label{fig:ins2D}
    \end{subfigure}
    ~
    \begin{subfigure}{0.3\textwidth}
    \centering
    \includegraphics[width=\linewidth]{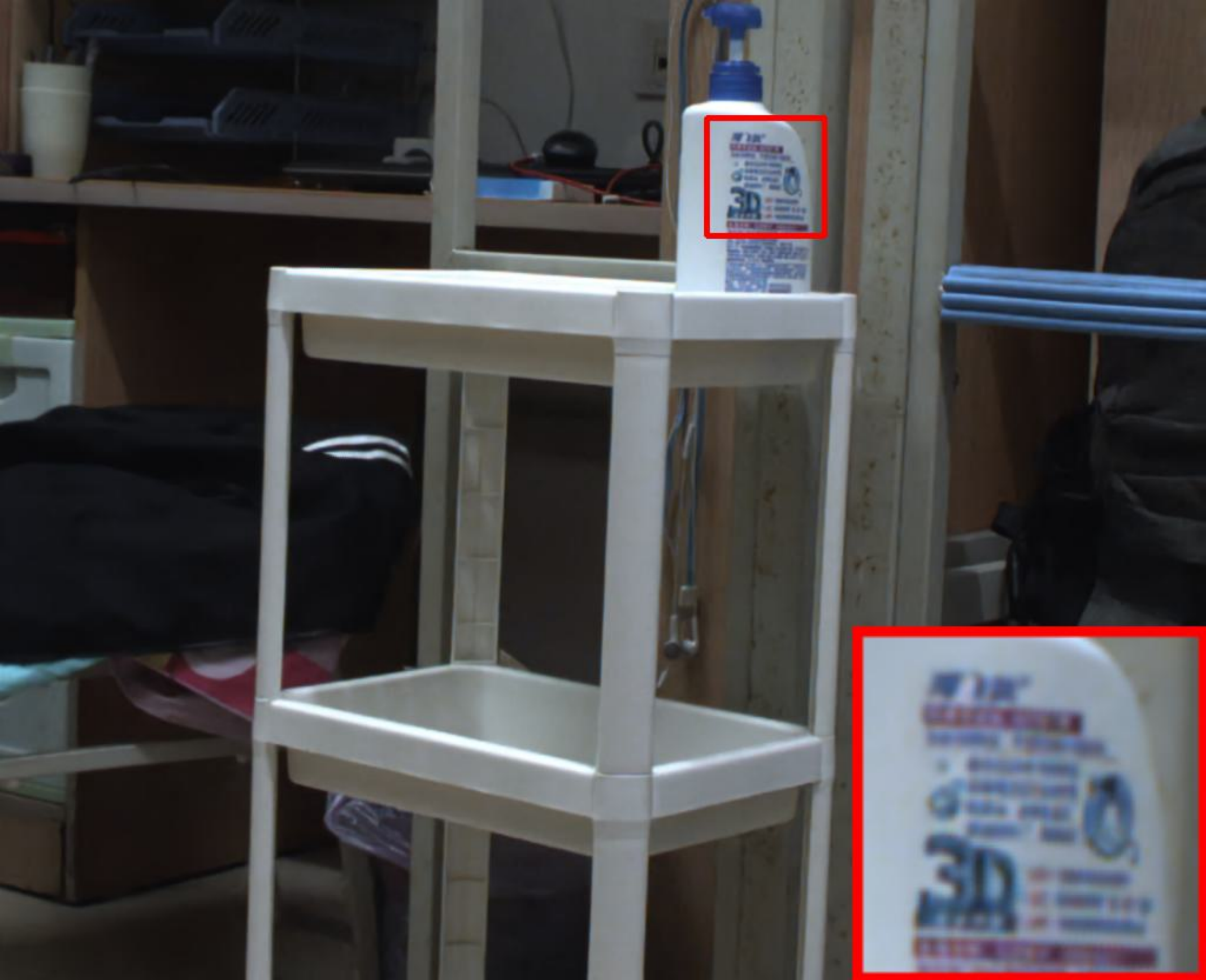}
    \caption{ECSIC right. \\ BPP: 0.1681 PSNR: 39.157 dB}
    \label{fig:ins2F}
    \end{subfigure}
    
\caption{Visualization of the reconstructed images.
For classical video coding methods, such as HEVC and VVC, BPP is calculated as an average across two views.}
\label{fig:ins6}
\end{figure}

We visualize the qualitative results in \cref{fig:ins6}, \cref{fig:ins34}, and \cref{fig:City1459} to show the effectiveness of the proposed method, compared with baselines BPG \cite{BPGWeb}, HEVC \cite{sullivan2012overview}, VVC \cite{bross2021overview}, SASIC \cite{Wodlinger_2022_CVPR}, and ECSIC \cite{wodlinger2024ecsic}. 
As shown in \cref{fig:ins6}, our proposed BiSIC achieves higher PSNR quality with a lower BPP for both left and right views, compared with other methods. Besides, the reconstruction details and texture of BiSIC are closer to the ground truth. 
Moreover, the image qualities of the left and right views in our bidirectional design remain close, mitigating the imbalance issue in unidirectional methods. In contrast, HEVC and VVC adopt a predictive compression pipeline where one view is compressed normally, and the other view is generated through the disparity between the prediction and the real view. The unidirectional compression results in a 2.265~dB PSNR gap for HEVC and a 1.844~dB PSNR gap for VVC between stereo views, as seen in \cref{fig:ins6}.
ECSIC utilizes the spatial context from the left image to compress the right one, resulting in a higher compression quality of the right image. In \cref{fig:ins34}, we illustrate another example on InStereo2K, where we can visually observe that the same area appears differently in the left and right views between \cref{fig:34LHEVC} and \cref{fig:34RH}, as well as \cref{fig:34LVVC} and \cref{fig:34RVVC}, due to the unidirectional compression. Another group of visualization comparisons on Cityscapes is shown in \cref{fig:City1459}.

\begin{figure}[t!]
  \centering
  \begin{subfigure}[b]{0.48\textwidth}
    \includegraphics[width=\textwidth]{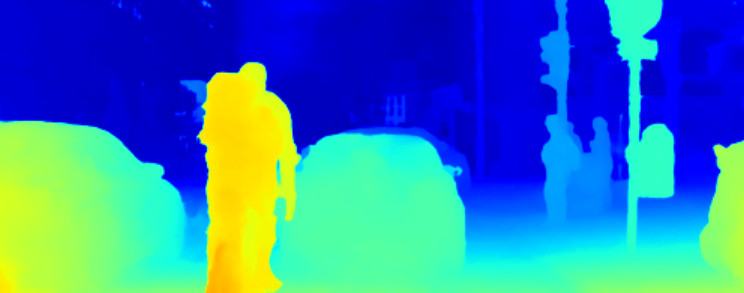}
    \caption{Estimation from the original images.}
    \label{depth_GT}
  \end{subfigure}
    \begin{subfigure}[b]{0.48\textwidth}
    \includegraphics[width=\textwidth]{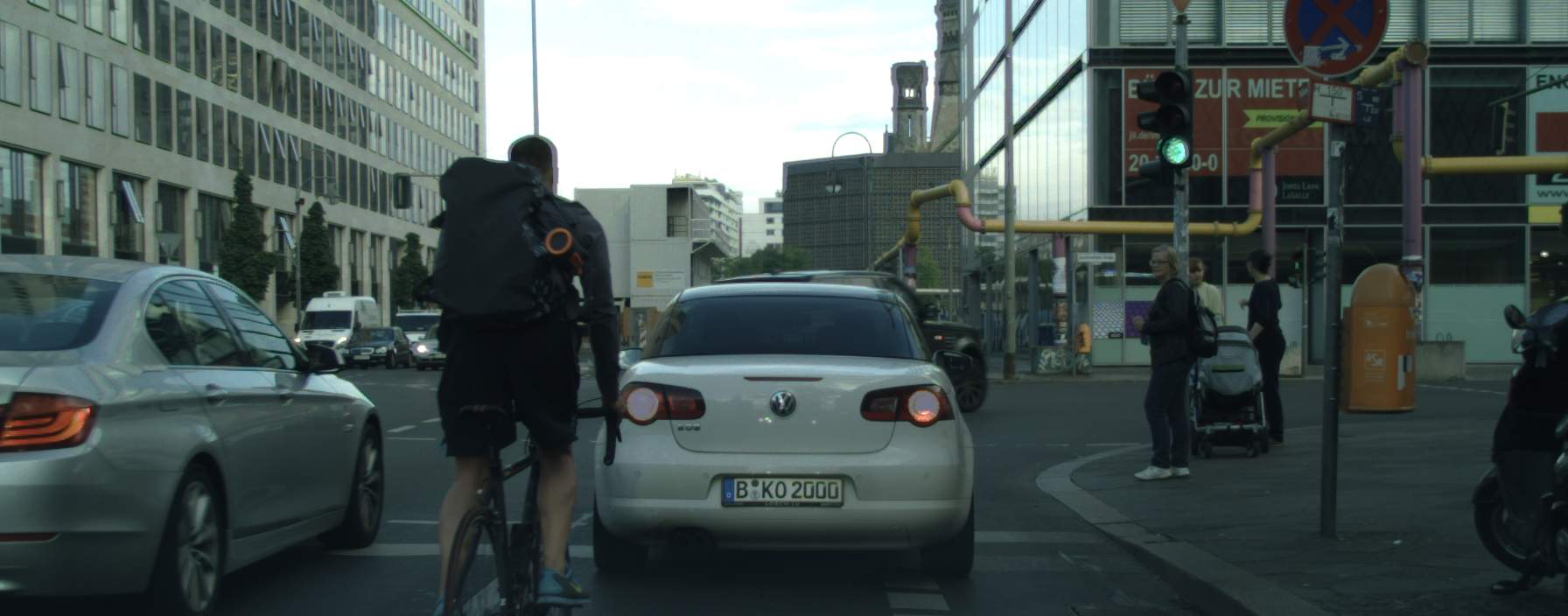}
    \caption{Original image.}
  \end{subfigure}
  \begin{subfigure}[b]{0.48\textwidth}
    \includegraphics[width=\textwidth]{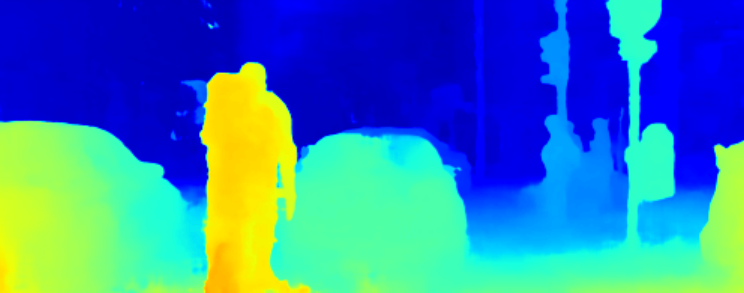}
    \caption{Estimation from the proposed BiSIC.}
    \label{depth_My}
  \end{subfigure}
    \begin{subfigure}[b]{0.48\textwidth}
    \includegraphics[width=\textwidth]{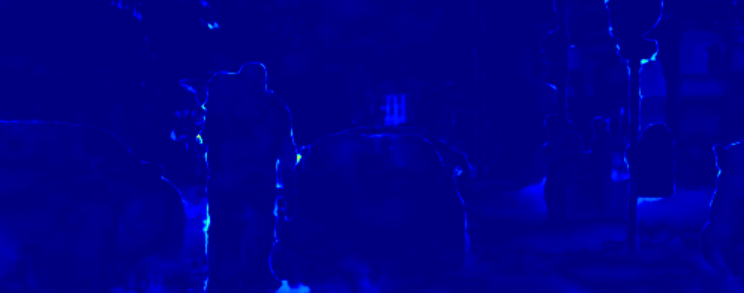}
    \caption{Disparity map of BiSIC. RMSE = 2.9838.}
    \label{disp_My}
  \end{subfigure}
  \begin{subfigure}[b]{0.48\textwidth}
    \includegraphics[width=\textwidth]{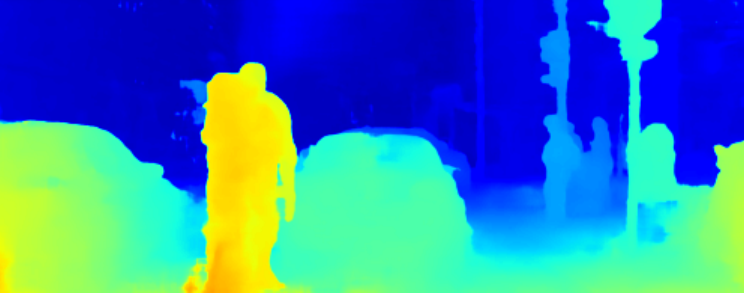}
    \caption{Estimation from ECSIC.}
    \label{depth_ECSIC}
  \end{subfigure}
    \begin{subfigure}[b]{0.48\textwidth}
    \includegraphics[width=\textwidth]{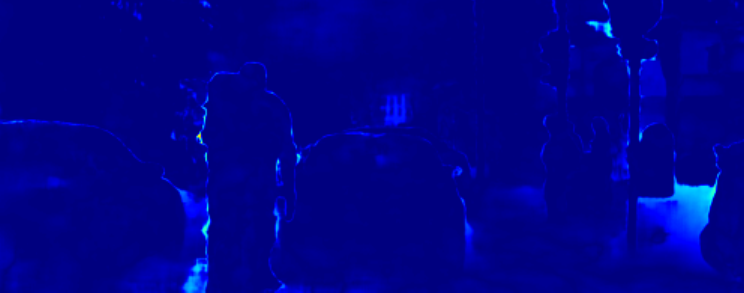}
    \caption{Disparity map of ECSIC. RMSE = 4.2445.}
    \label{disp_ECSIC}
  \end{subfigure}
  \begin{subfigure}[b]{0.48\textwidth}
    \includegraphics[width=\textwidth]{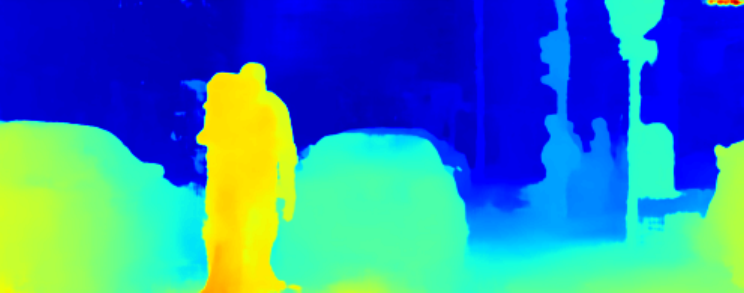}
    \caption{Estimation from VVC.}
    \label{depth_VVC}
  \end{subfigure}
  \begin{subfigure}[b]{0.48\textwidth}
    \includegraphics[width=\textwidth]{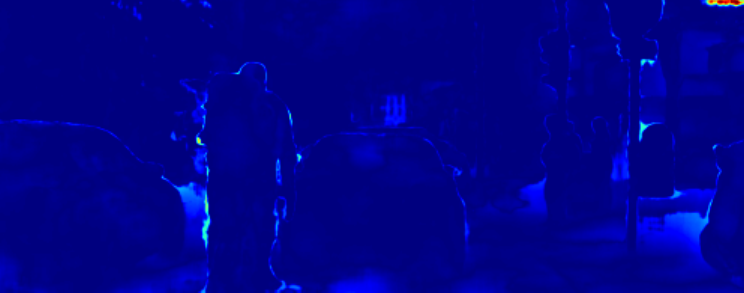}
    \caption{Disparity map of VVC. RMSE = 4.5740.}
    \label{disp_VVC}
  \end{subfigure}
  \\
  \begin{subfigure}[b]{0.48\textwidth}
    \includegraphics[width=\textwidth]{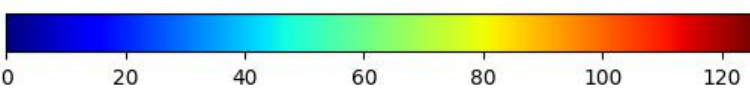}
    % \caption{}
  \end{subfigure}
  ~
    \begin{subfigure}[b]{0.48\textwidth}
    \includegraphics[width=\textwidth]{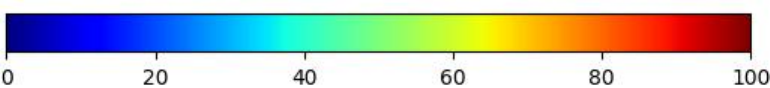}
    % \caption{}
  \end{subfigure}
  \caption{Stereo matching results of different compression methods. The original image, the estimation from the original images, the estimations from reconstruction results, and their disparity maps are provided. All stereo matching results are estimated using the method in \cite{chang2018pyramid}, and RMSE is calculated to reflect the accuracy \cite{scharstein2002taxonomy}. 
  The corresponding BPPs for BiSIC, ECSIC, and VVC are 0.103, 0.116, and 0.131, respectively.}
  % The average BPP and PSNR after various compression methods are listed here. BiSIC: BPP 0.103, PSNR 38.44 dB, VVC: BPP 0.131, PSNR 38.04 dB, ECSIC: BPP 0.1160, PSNR 37.954 dB. 
  \label{fig:matching}
\end{figure}

\subsection{Downstream Task Verification}
The previous subsection provides a visual comparison between our method and the baselines. Note that imbalanced stereo quality is unfavorable for machine vision and downstream tasks \cite{liu2020visually}. Therefore, it is interesting to investigate the degradation caused by different compression methods. In this subsection, we compare their performance on the stereo matching task. We employ a benchmark stereo matching method \cite{chang2018pyramid} on both ground truth stereo image pairs and reconstructed stereo image pairs from various compression methods to illustrate the degradation effect brought by compression.

The stereo matching results are visualized in \cref{fig:matching}. 
We calculate the root-mean-square-error (RMSE) \cite{scharstein2002taxonomy} to quantify the disparity between the estimations from original images and reconstructed images.
% each method and the ground truth.
% between the estimations from the ground truth and each of the methods to measure their disparity. 
As illustrated in \cref{depth_GT} and \cref{depth_My}, the estimation from the reconstructed results of our proposed BiSIC achieves a nearly identical estimation to the one from the ground truth. Moreover, it achieves the lowest RMSE among others, while requiring the lowest BPP. This finding demonstrates that BiSIC preserves most of the features and information in the stereo images after compression. In contrast, the decompression result of ECSIC (see \cref{depth_ECSIC}) fails to accurately estimate the right part of the image, where objects at different depths are confused.
Although VVC (38.04 dB) achieves approximately the same average PSNR level as our BiSIC (38.44 dB), there exists a discrepancy of 2.5769 dB in VVC between the left view (39.3268 dB) and the right view (36.7499 dB). 
Moreover, the estimation from VVC (see \cref{depth_VVC}) produces severe artifacts on the top right region and incorrect estimations in the right part. These observations suggest that balanced stereo image quality, which can be achieved through bidirectional compression, is beneficial to both visual perception and downstream tasks.

\subsection{Bit Allocation Visualization}
\begin{figure}
    \centering
    \includegraphics[width=0.99\linewidth]{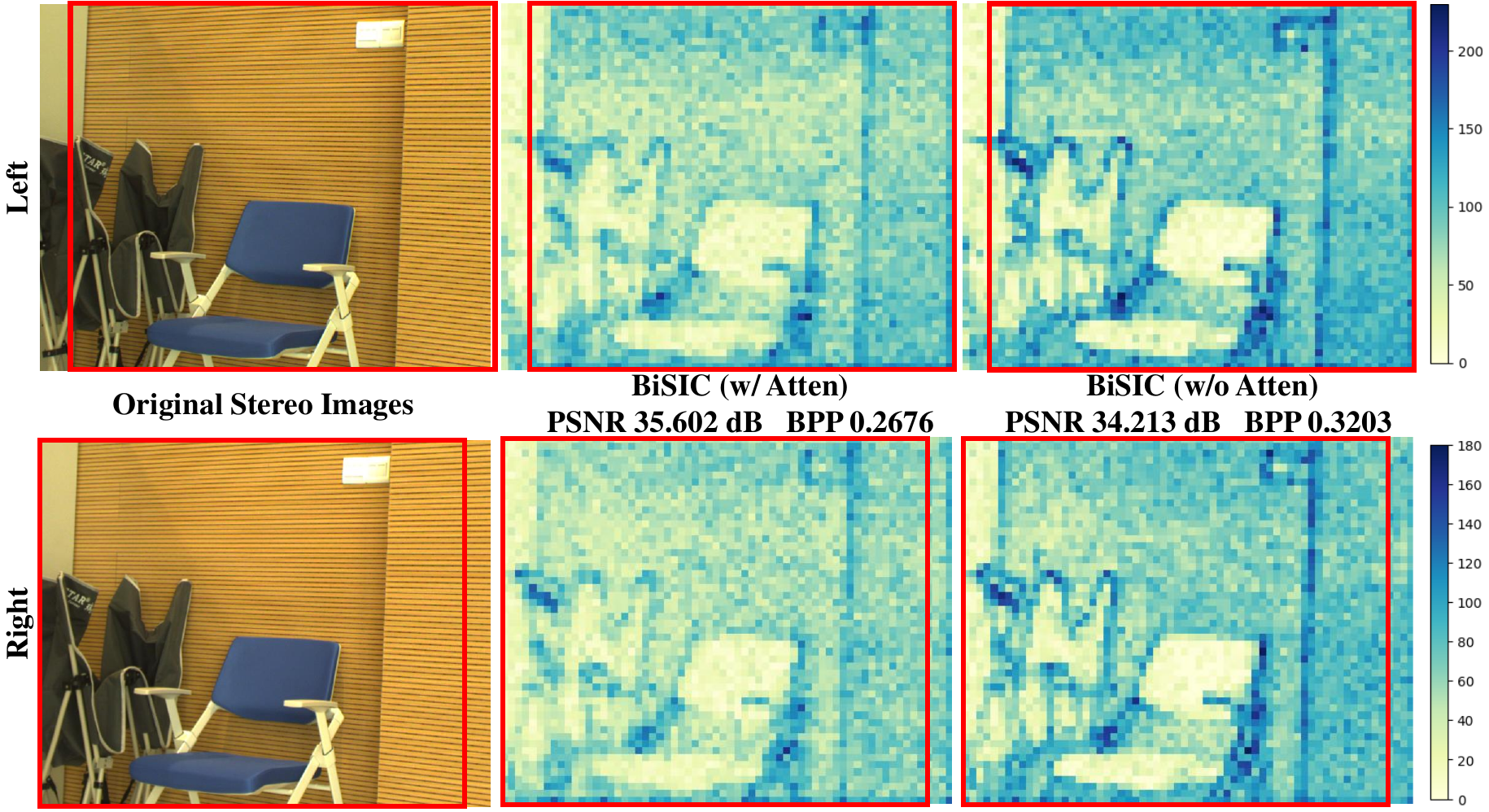}
    \caption{Visualization of the bit allocation. 
    The regions within red boundaries present the common areas of stereo images. 
    (Left) Original images. (Middle) Bit allocation of BiSIC. (Right) Bit allocation of BiSIC without mutual attention blocks. BiSIC with mutual attention blocks achieves a higher average PSNR of 35.602 dB with a lower average BPP, compared to the baseline without attention (34.213 dB).}
    \label{fig:attentionmapswb}
\end{figure}

In this subsection, we examine the effectiveness of the proposed mutual attention blocks on compression performance.
\cref{fig:attentionmapswb} visualizes the bit allocation of latents $\hat{\textbf{\textit{y}}}_l$ and $\hat{\textbf{\textit{y}}}_r$ in BiSIC with and without the mutual attention blocks.
In particular, darker-colored regions indicate a greater number of allocated bits for image encoding (i.e., higher BPP), while regions with lighter colors are encoded with fewer bits (i.e., lower BPP). 
By incorporating the mutual attention blocks, our BiSIC method effectively identifies shared features between stereo views for redundancy reduction, thereby increasing the compression ratio.

\section{Extra Ablation Studies}
\label{sec:ab}
\subsection{Ablation Study on BiSIC-Fast}
\begin{figure}
    \centering
    \includegraphics[height=6cm]{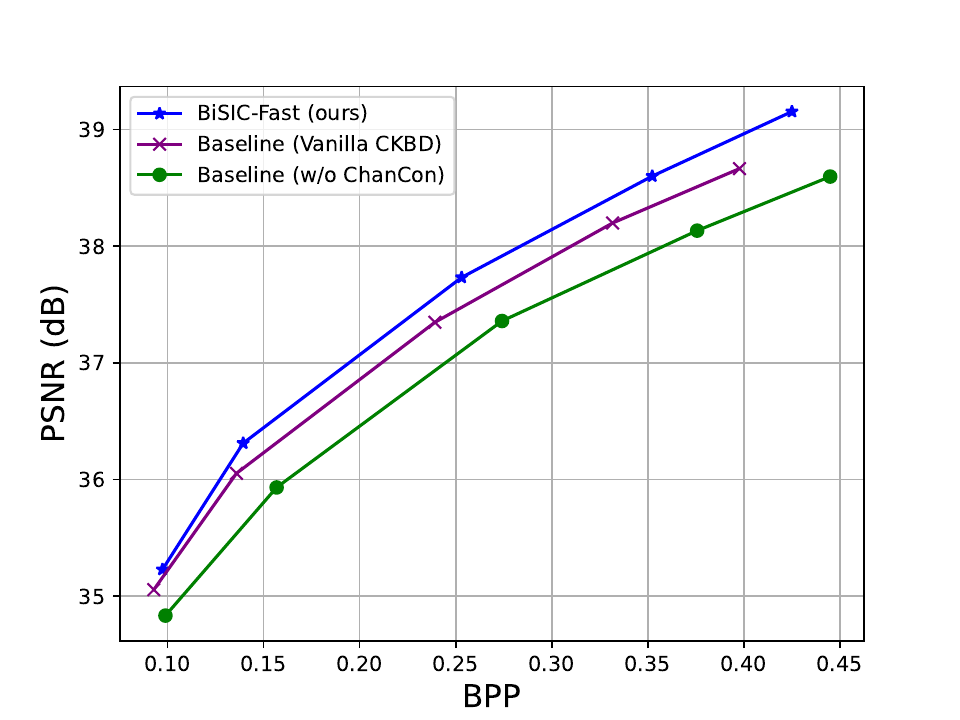}
    \caption{Ablation study of BiSIC-Fast on InStereo2K dataset. The Baseline (Vanilla CKBD) replaces the stereo-checkerboard structure in BiSIC-Fast with the vanilla checkerboard, while the Baseline (w/o ChanCon) removes the utilization of channel context in BiSIC-Fast.}
    \label{fig:FastAb}
\end{figure}

The ablation studies for the proposed BiSIC method are shown in  Sec. 4, which illustrates the impact of each proposed component, including the 3D convolution backbone, cross-dimensional entropy model, and mutual attention block. In this subsection, we provide ablation studies for our fast variant, BiSIC-Fast. Specifically, we investigate the effect of our designed stereo-checkerboard structure and evaluate the significance of channel context. The RD performance is illustrated in \cref{fig:FastAb}, and we also calculate the Bjøntegaard Delta PSNR (BD-PSNR) \cite{bjontegaard2001calculation} for comparison. 

\begin{figure}
    \centering
    \includegraphics[height=3.5cm]{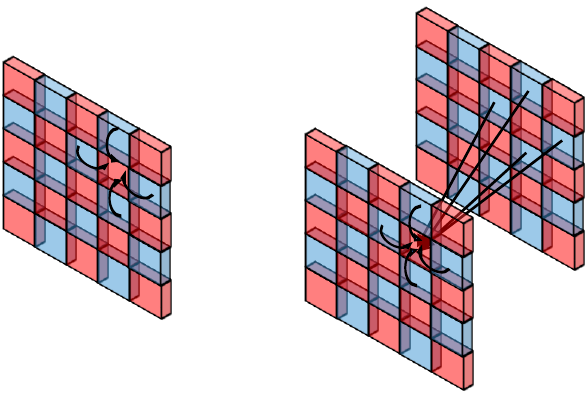}
    \caption{(Left) Illustration of dependencies learned in a vanilla checkerboard structure. (Right) Illustration of the dependency utilized by a stereo-checkerboard structure. For simplicity, in both examples, we only show the conditional effect of the four neighboring entries around one target entry.}
    \label{fig:StereoCKBD}
\end{figure}

\noindent \textbf{Effectiveness of Stereo-Checkerboard.} The stereo-checkerboard structure enables joint learning from both views, aided by 3D convolution. To illustrate the effect of the stereo-checkerboard structure, we replace it with the vanilla checkerboard in \cite{he2021checkerboard} and present the ablation results in \cref{fig:FastAb}. We observe an RD performance degradation in this baseline compared to our BiSIC-Fast, specifically, with a BD-PSNR of -0.213 dB. This is because the cooperation of the stereo-checkerboard and 3D convolution enables the utilization of references from both self-view and the other view, and thus extracts more information compared with the vanilla checkerboard method, as shown in \cref{fig:StereoCKBD}.

\noindent \textbf{Effectiveness of Channel Context.} To evaluate the contribution of channel context, we remove the slicing process on channel axis and the channel context model. As shown in \cref{fig:FastAb}, channel context contributes to a significant improvement in performance, quantified as an improvement of 0.6173 dB in BD-PSNR. Therefore, without channel context, the stereo anchor part is conditioned only on hyperprior, which is relatively insufficient. Consequently, an off-the-optimal stereo anchor progressively provides an inadequate condition for the stereo non-anchor part, resulting in unsatisfactory performance.

\subsection{Ablation Study on Number of Slices}
\begin{figure}[t]
    \begin{minipage}[b]{0.5\linewidth}
        \centering
        \includegraphics[height=5cm]{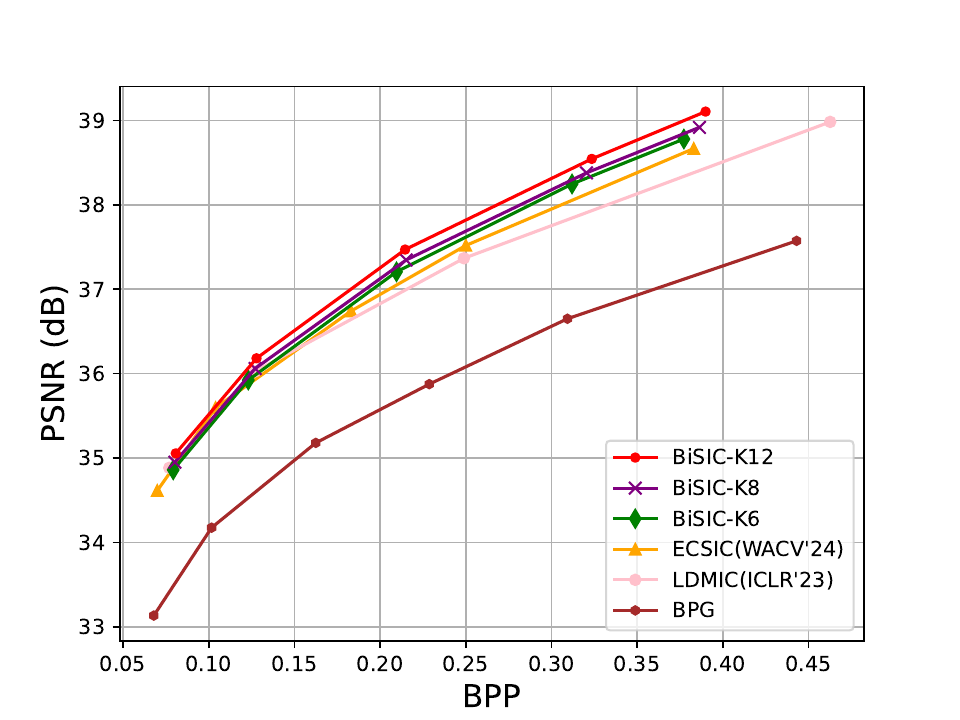}
        \caption{RD performance of our BiSIC model with various numbers of slices $K$.}
        \label{fig:adjustK}
    \end{minipage}
    \hspace{1 pt}
    \begin{minipage}[b]{0.44\linewidth}
        \centering
        \setlength{\tabcolsep}{3pt}
            \centering
            \begin{tabular}{ccc} 
            \toprule
                Method & Time & \multicolumn{1}{l}{BDBR} \\
                \midrule
                BiSIC-K12 & 167.25s & -48.07\% \\
                BiSIC-K8 & 116.41s & -45.95\% \\
                BiSIC-K6 & 96.84s & -44.88\% \\
                \bottomrule
            \end{tabular}%
            \vspace{+30pt}
            \captionof{table}{Runtime and BDBR results with various numbers of slices $K$.}
            \label{tab:adjustK}%
    \end{minipage}
\end{figure}

In the channel-wise auto-regressive entropy model, the previously decoded part serves as a condition for the later part. Thus, it is interesting to investigate the relationship between the number of slices, compression performance, and model efficiency. Note that a higher precision in slicing generates abundant conditions, but more slices directly increase the time consumption of compression. This forms a trade-off between compression performance and speed. In this subsection, we conduct an ablation study for our proposed BiSIC on the number of slices $K$ and investigate its effect on the trade-off between performance and efficiency. 
The RD performance on InStereo2K is shown in \cref{fig:adjustK}, along with several baselines for comparison. The Bjøntegaard Delta Bitrate (BDBR) \cite{bjontegaard2001calculation} results relative to BPG are shown in \cref{tab:adjustK}. As demonstrated, reducing the number of slices leads to a slight decrease in the RD performance and accelerates the encoding/decoding process. Specifically, when $K=6$, the time consumption is reduced by $42\%$, while the RD performance experiences a degradation of $3.19\%$. Nonetheless, it still outperforms other baselines, as shown in \cref{fig:adjustK}.

% {\color{blue}
\subsection{Ablation Study on ELIC Backbone}
\begin{figure}
    \centering
    \includegraphics[width=0.5\linewidth]{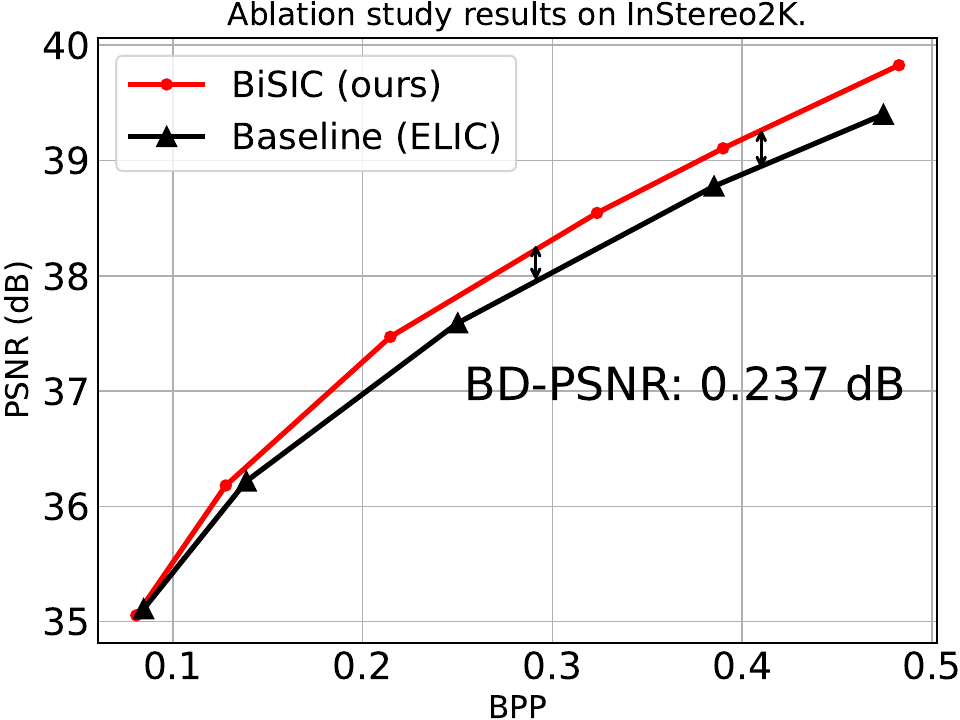}
    \caption{RD performance of our BiSIC model compared with Baseline (ELIC).}
    \label{fig:ELIC_base}
\end{figure}

\begin{figure}[t!]
\centering
    \begin{subfigure}{0.3\textwidth}
    \centering
    \includegraphics[width=\linewidth]{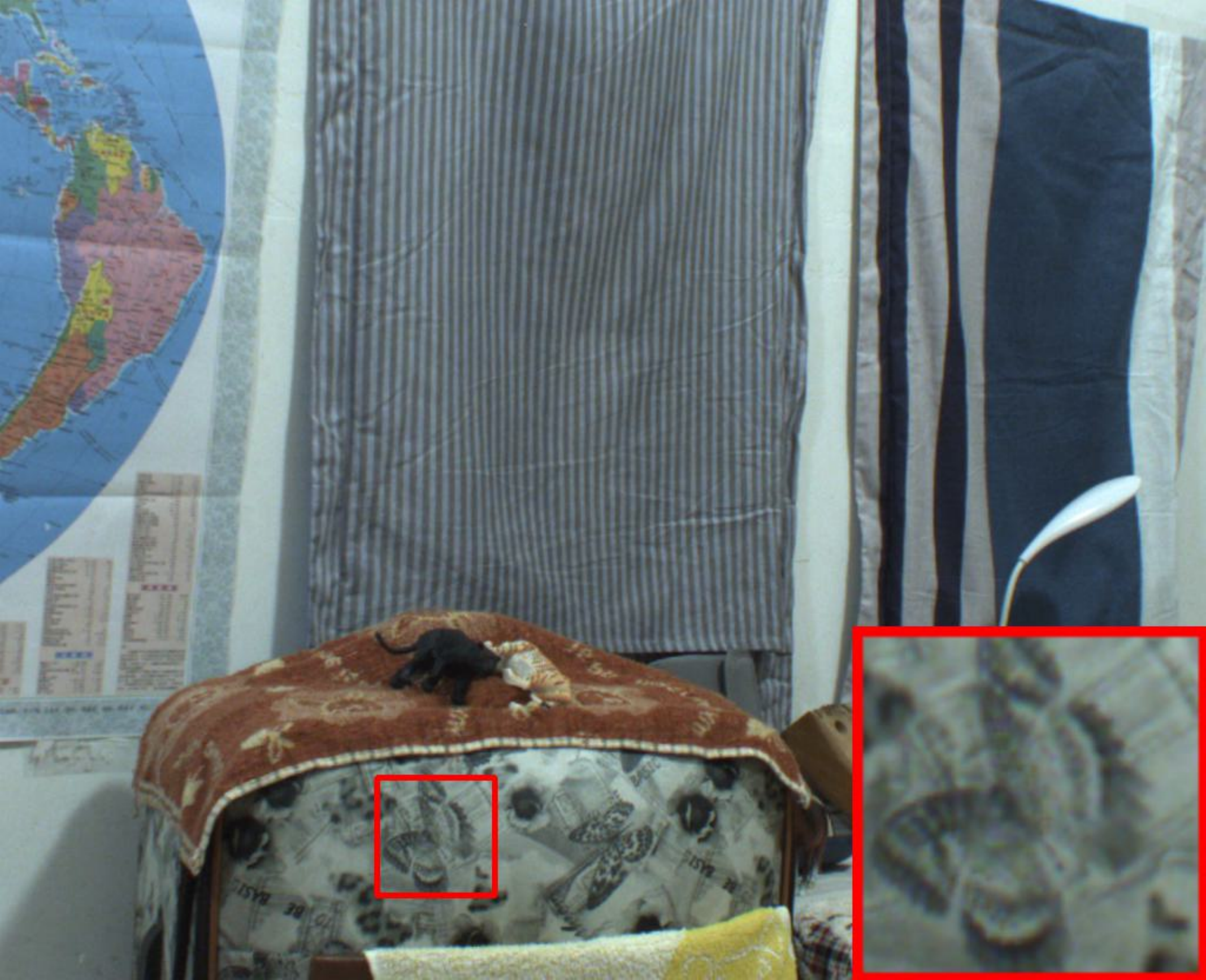}
    \caption{Ground truth left. \\ \quad}
    \label{fig:34LGT}
    \end{subfigure}
    ~
    \begin{subfigure}{0.3\textwidth}
    \centering
    \includegraphics[width=\linewidth]{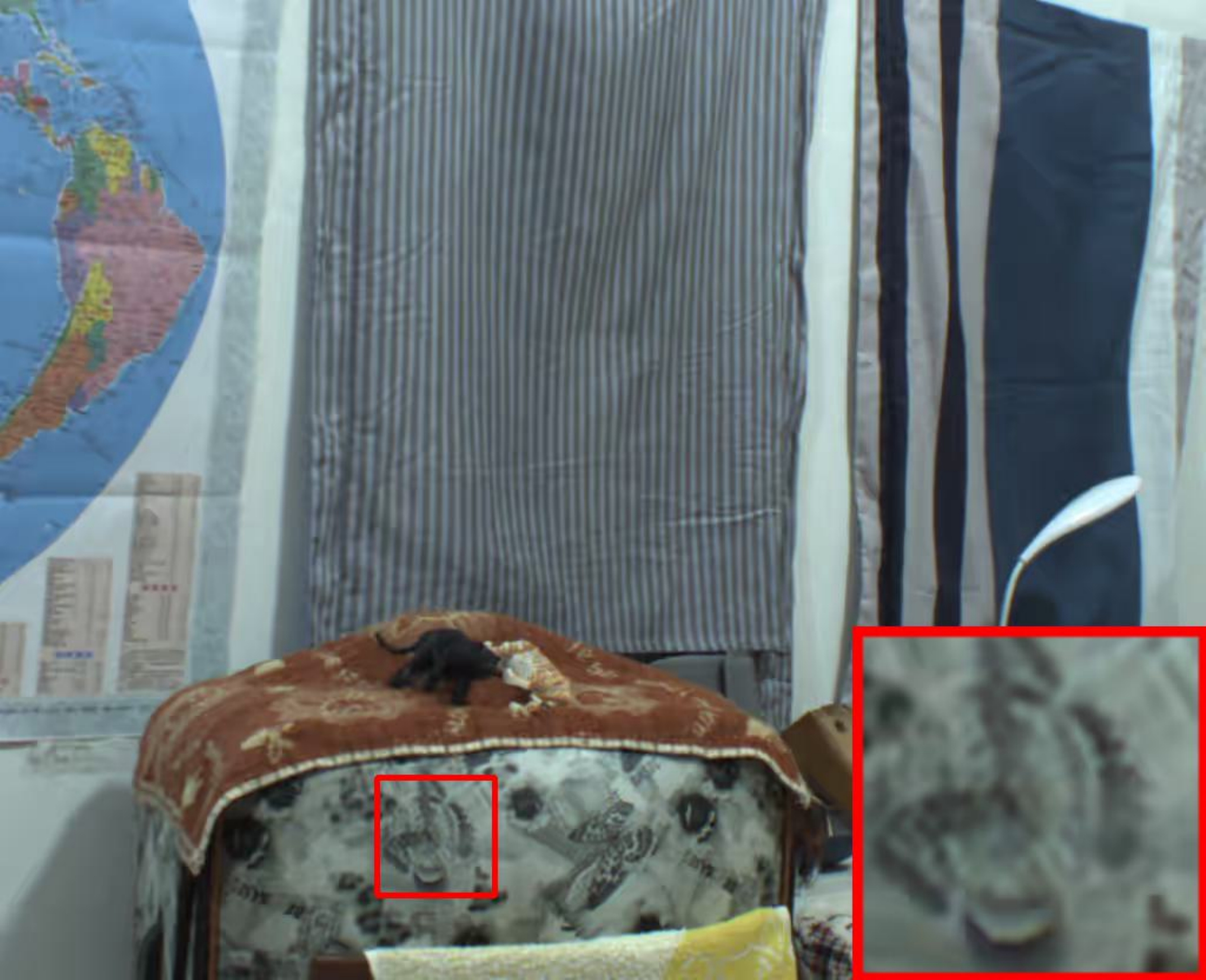}
    \caption{BPG left. \\ BPP: 0.1818  PSNR: 33.492 dB}
    \label{fig:34LBPG}
    \end{subfigure}
    ~
    \begin{subfigure}{0.3\textwidth}
    \centering
    \includegraphics[width=\linewidth]{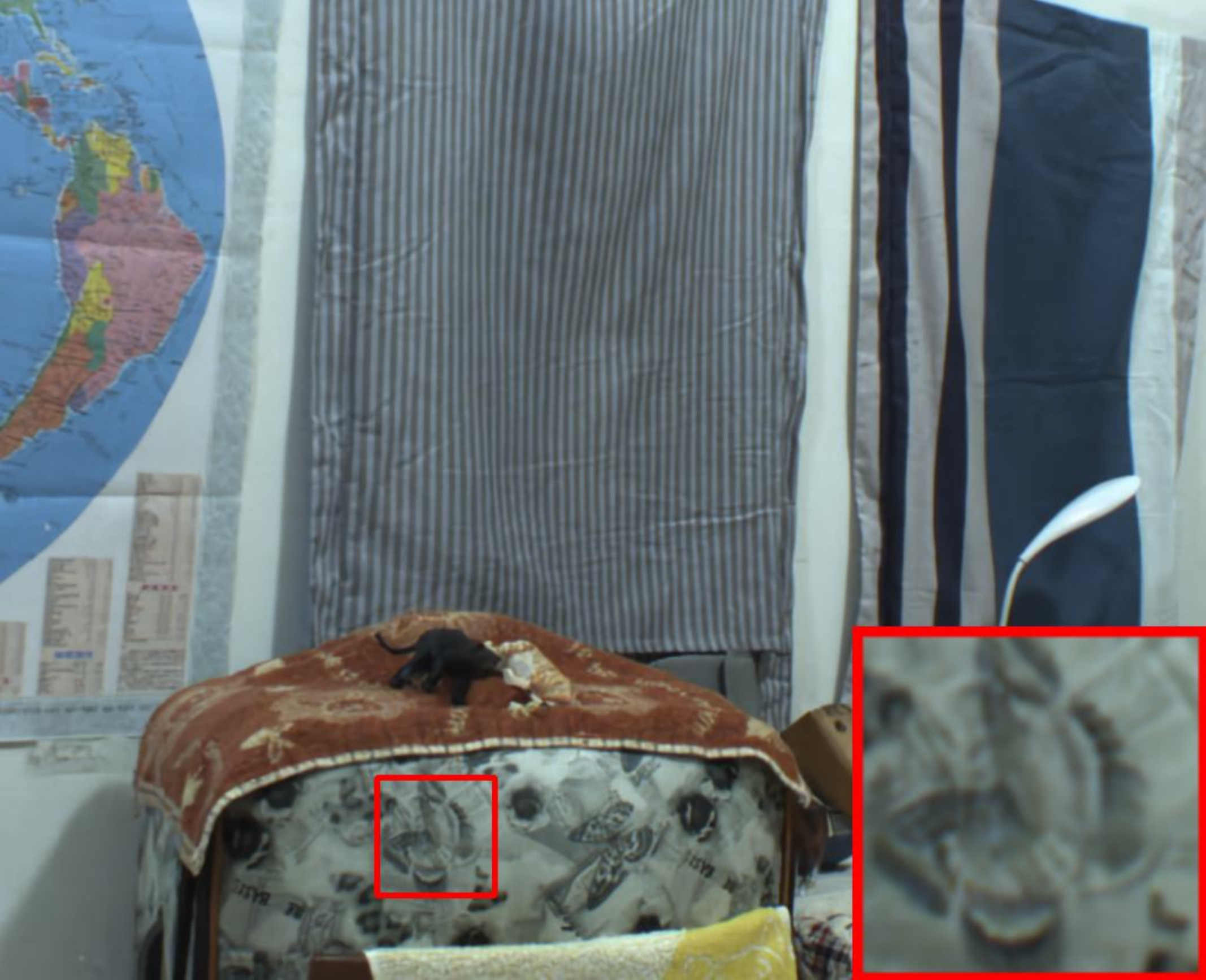}
    \caption{Proposed BiSIC left. \\ BPP: 0.1690 PSNR: 35.524 dB}
    \label{fig:34LMy}
    \end{subfigure}
    \\
    
    \begin{subfigure}{0.3\textwidth}
    \centering
    \includegraphics[width=\linewidth]{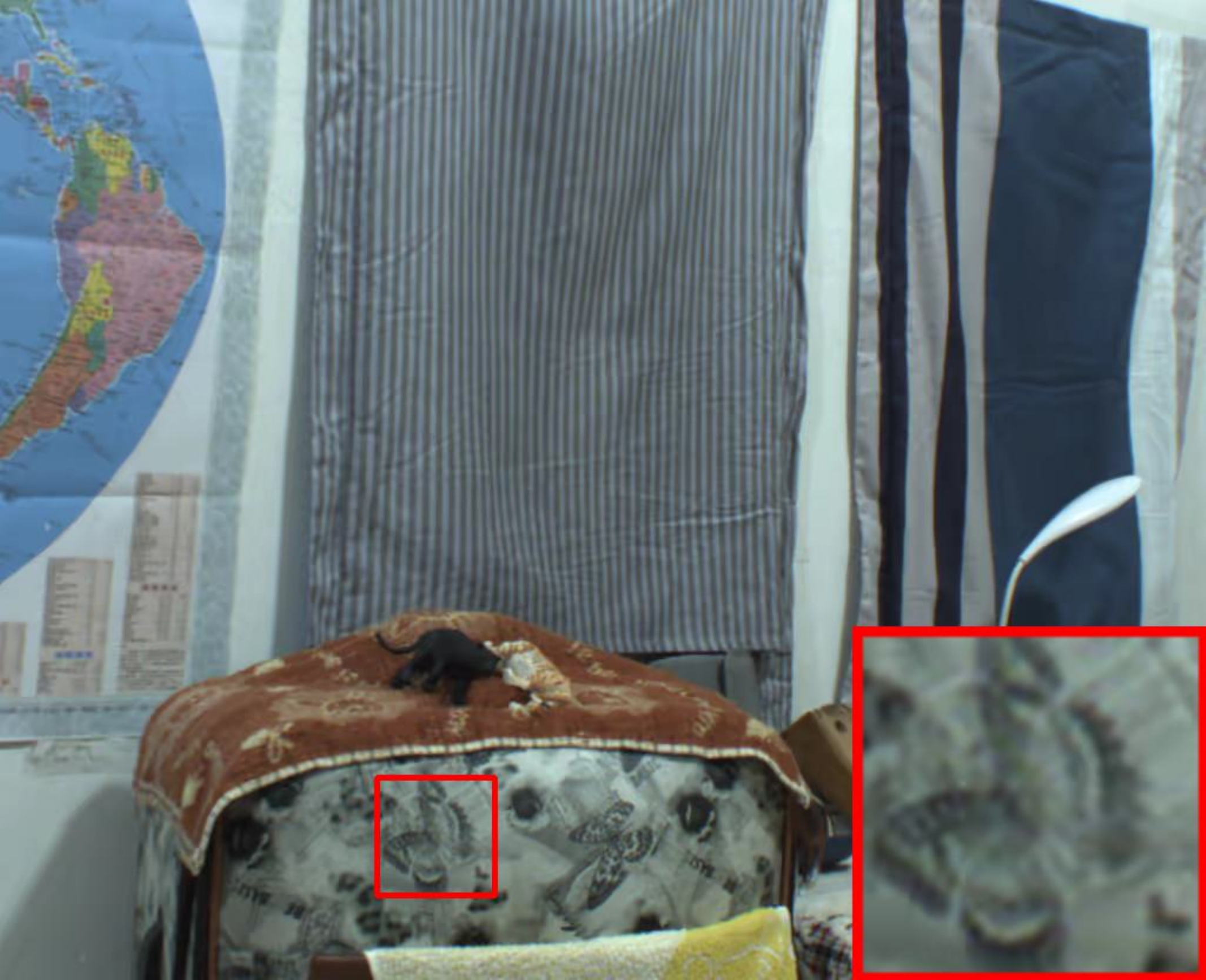}
    \caption{HEVC left. \\ BPP: 0.1467 PSNR: 34.702 dB}
    \label{fig:34LHEVC}
    \end{subfigure}
    ~
    \begin{subfigure}{0.3\textwidth}
    \centering
    \includegraphics[width=\linewidth]{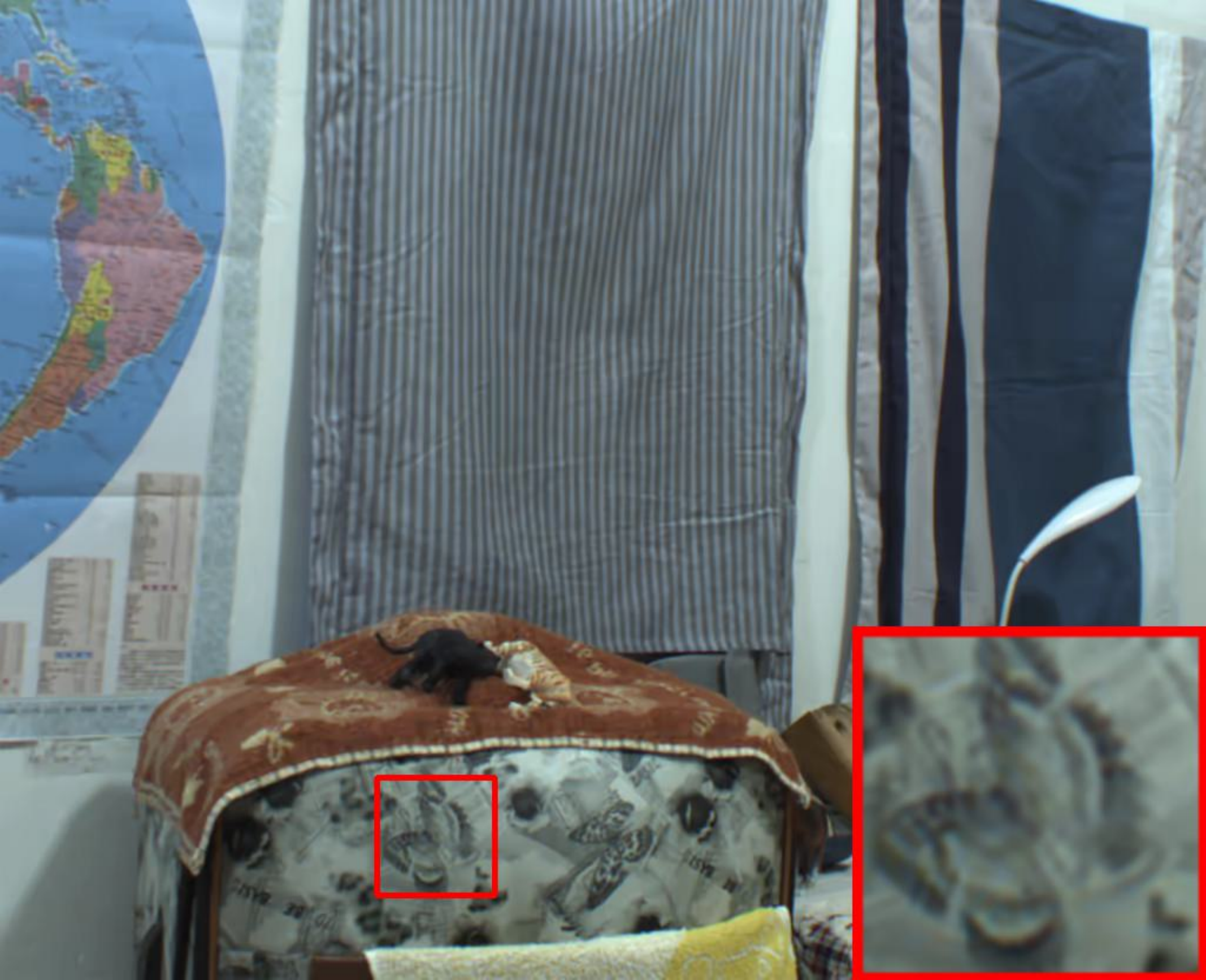}
    \caption{VVC left. \\ BPP: 0.1601 PSNR: 35.450 dB}
    \label{fig:34LVVC}
    \end{subfigure}
    ~
    \begin{subfigure}{0.3\textwidth}
    \centering
    \includegraphics[width=\linewidth]{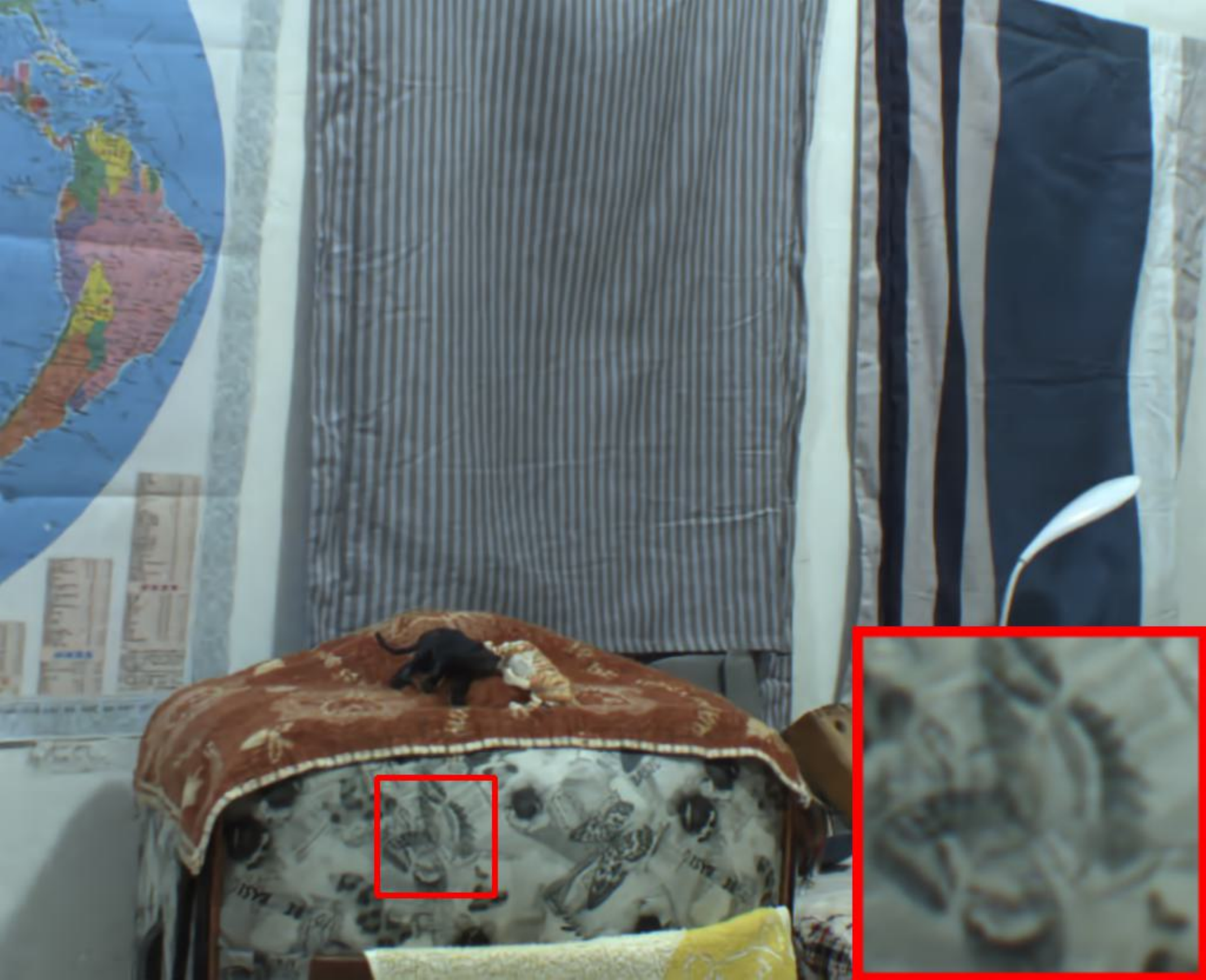}
    \caption{ECSIC left. \\ BPP: 0.1560 PSNR: 34.831 dB}
    \label{fig:34LEC}
    \end{subfigure}
    \\
    
    \begin{subfigure}{0.3\textwidth}
    \centering
    \includegraphics[width=\linewidth]{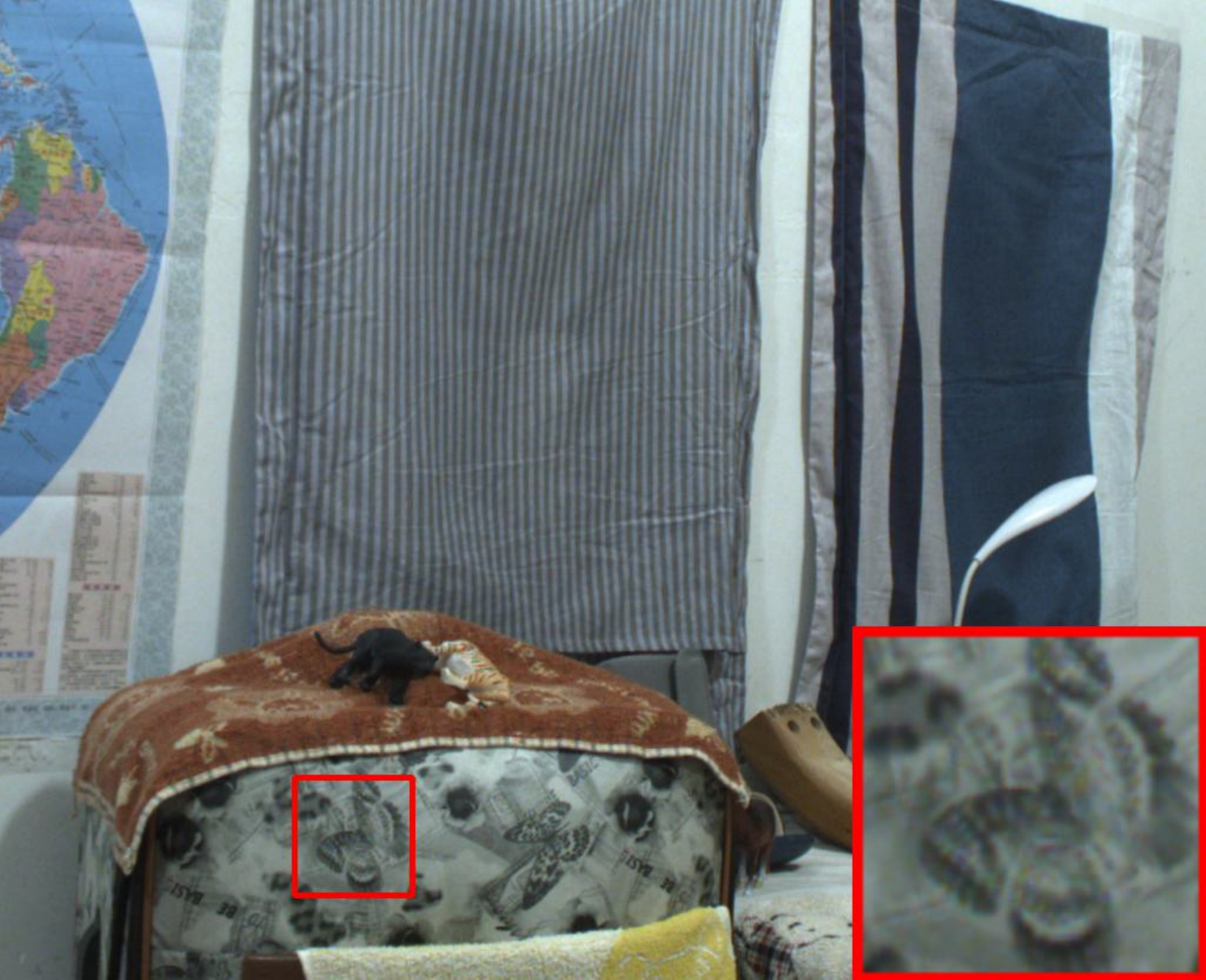}
    \caption{Ground truth right. \\ \quad}
    \label{fig:34RGT}
    \end{subfigure}
    ~
    \begin{subfigure}{0.3\textwidth}
    \centering
    \includegraphics[width=\linewidth]{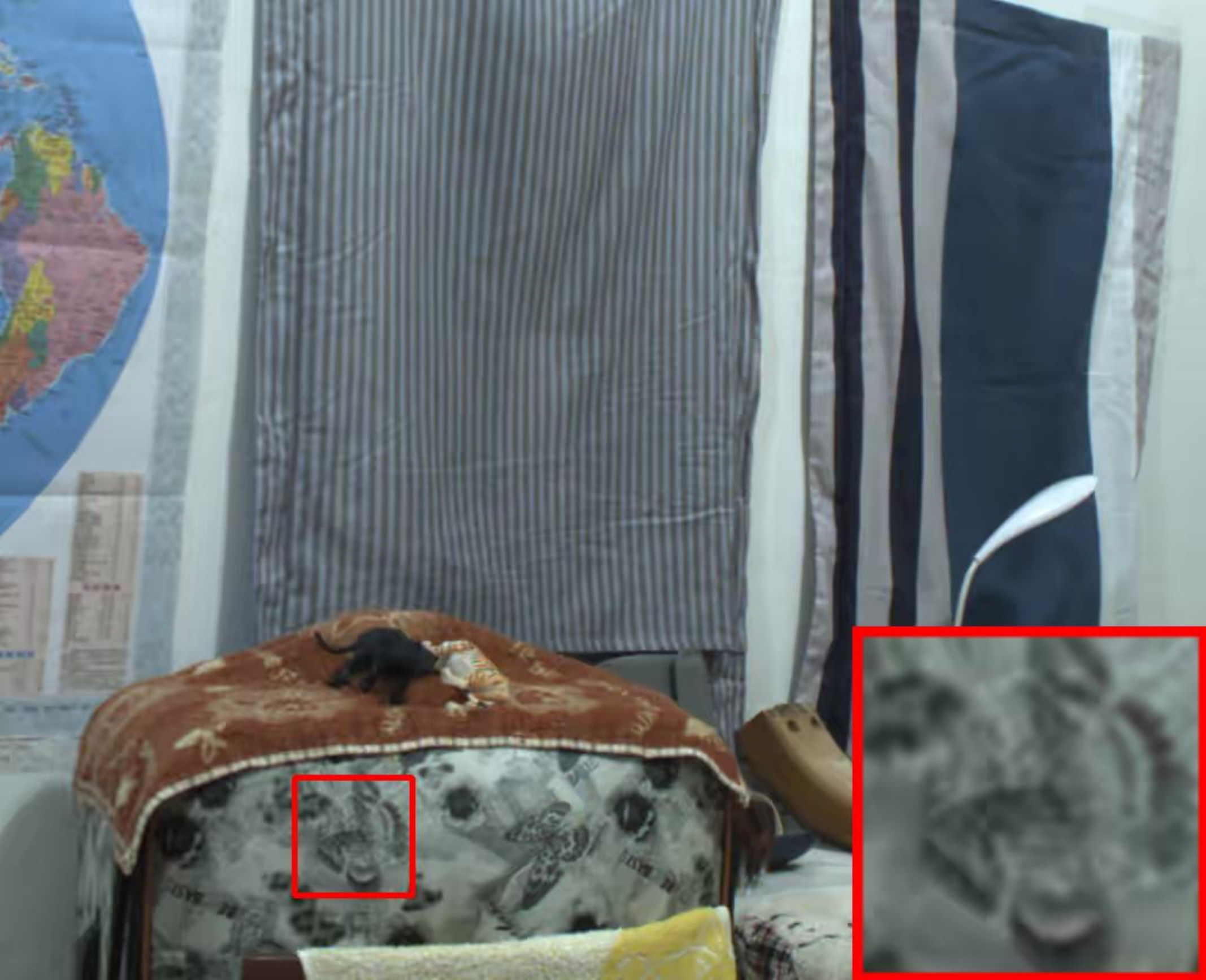}
    \caption{BPG right. \\ BPP: 0.1705 PSNR: 33.558 dB}
    \label{fig:34RBGP}
    \end{subfigure}
    ~
    \begin{subfigure}{0.3\textwidth}
    \centering
    \includegraphics[width=\linewidth]{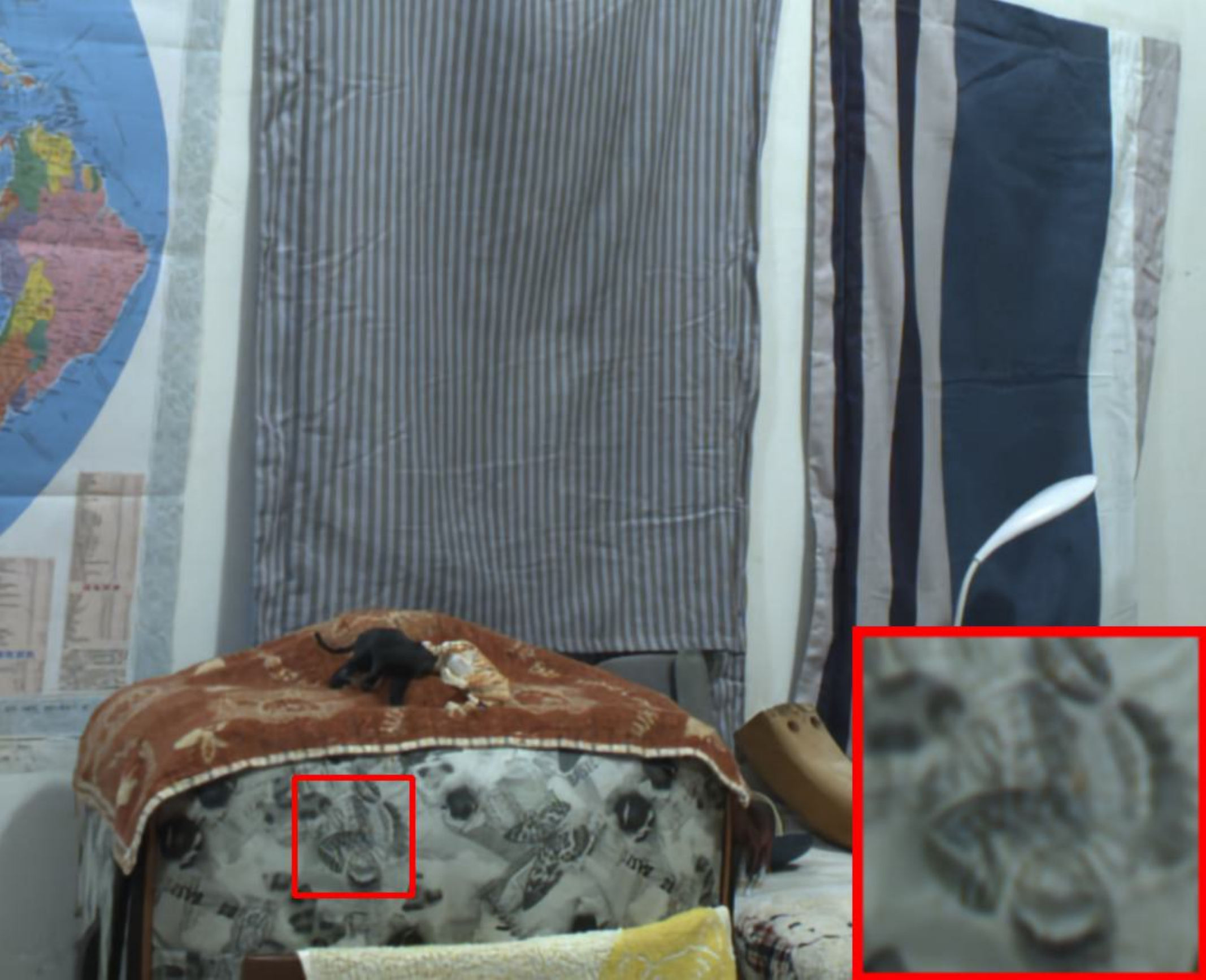}
    \caption{Proposed BiSIC right. \\ BPP: 0.1615 PSNR: 35.631 dB}
    \label{fig:34RMy}
    \end{subfigure}
    \\
    
    \begin{subfigure}{0.3\textwidth}
    \centering
    \includegraphics[width=\linewidth]{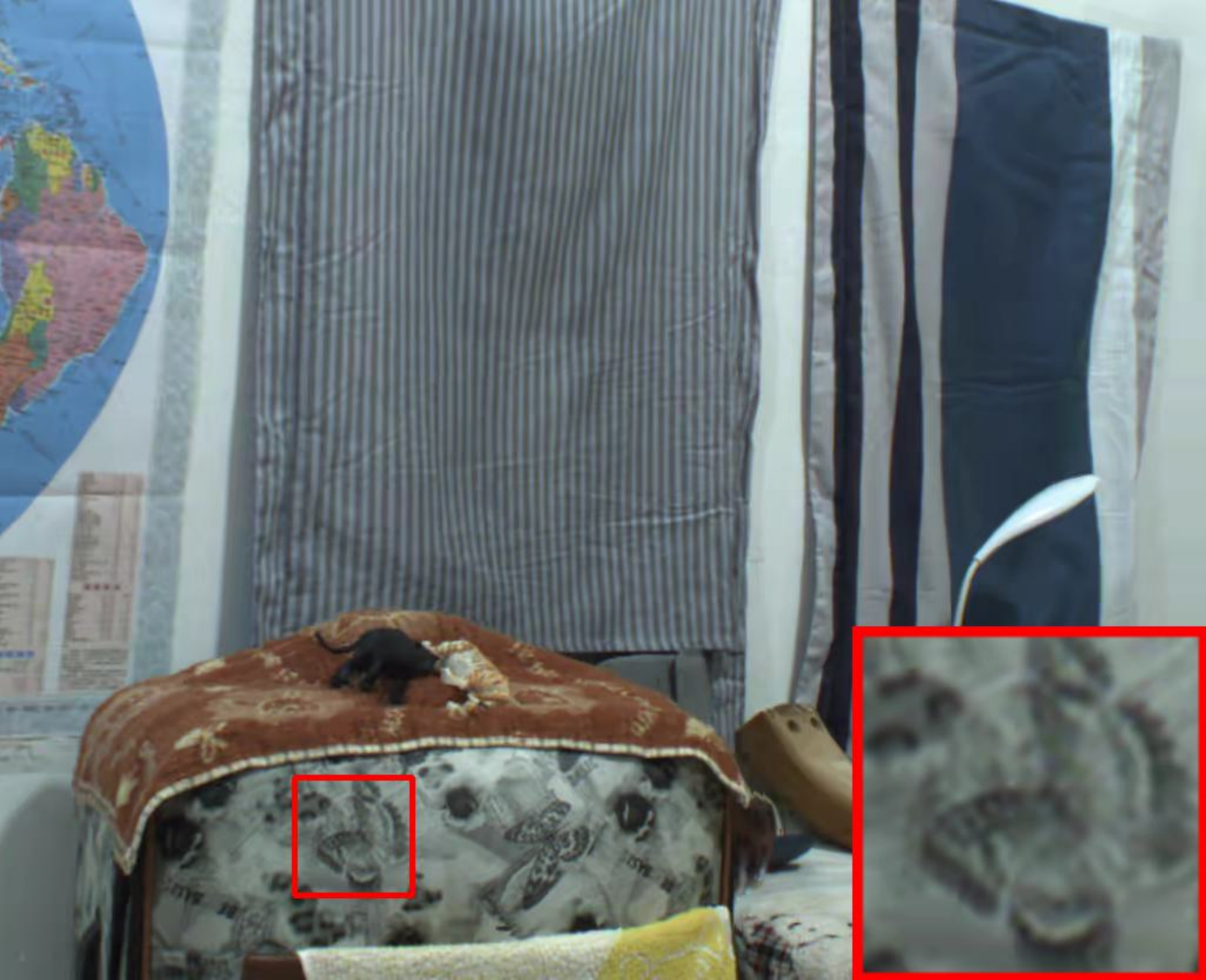}
    \caption{HEVC right. \\ BPP: 0.1467 PSNR: 32.558 dB}
    \label{fig:34RH}
    \end{subfigure}
    ~
    \begin{subfigure}{0.3\textwidth}
    \centering
    \includegraphics[width=\linewidth]{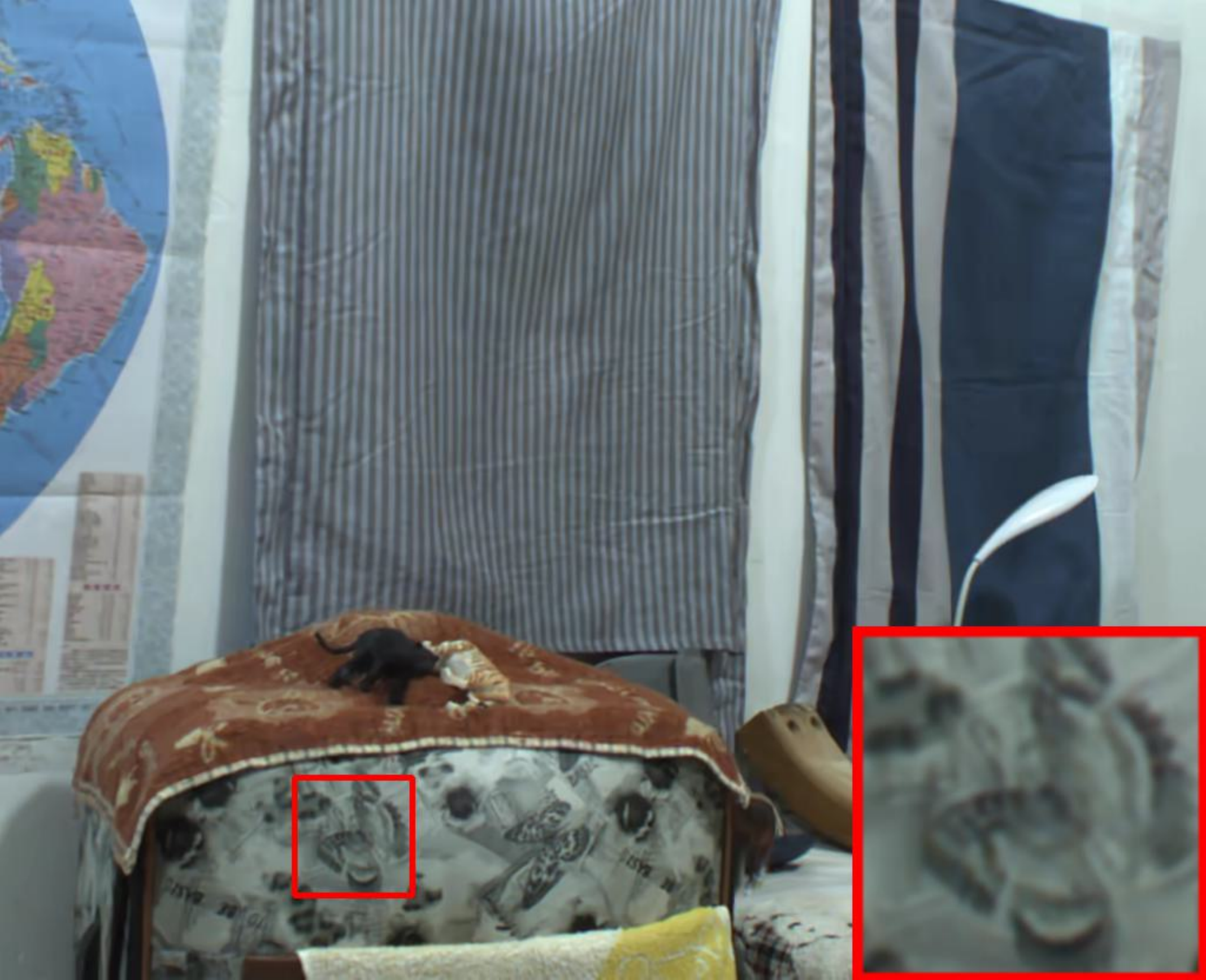}
    \caption{VVC right. \\ BPP: 0.1601 PSNR: 33.477 dB}
    \label{fig:34RVVC}
    \end{subfigure}
    ~
    \begin{subfigure}{0.3\textwidth}
    \centering
    \includegraphics[width=\linewidth]{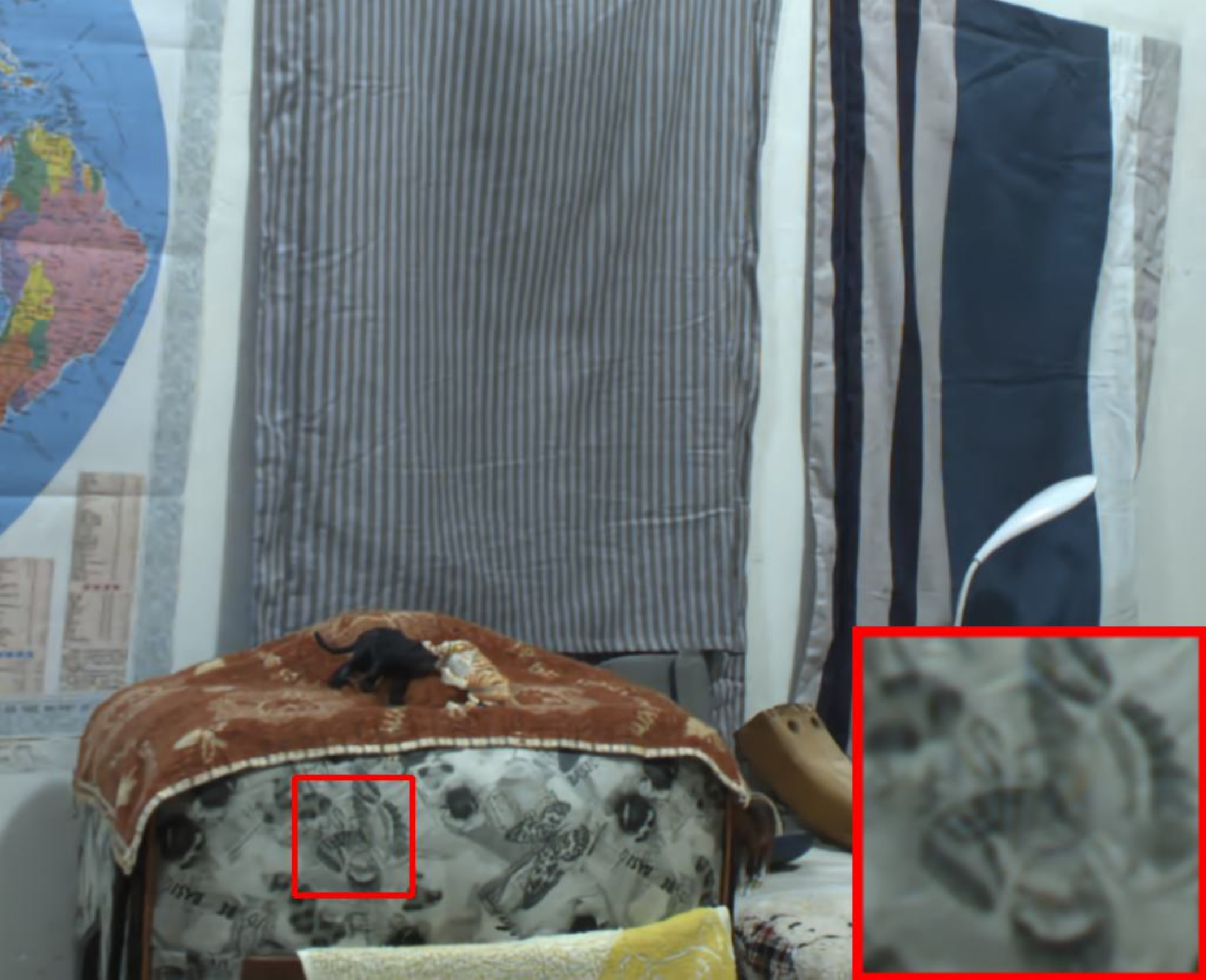}
    \caption{ECSIC right. \\ BPP: 0.1431 PSNR: 35.010 dB}
    \label{fig:34REC}
    \end{subfigure}

\caption{Visualization of the reconstructed images.
For classical video coding methods, such as HEVC and VVC, BPP is calculated as an average across two views.}
\label{fig:ins34}
\end{figure}

\begin{figure}[t!]
    \centering

    \begin{subfigure}{0.99\textwidth}
        \centering
        \includegraphics[width=\linewidth]{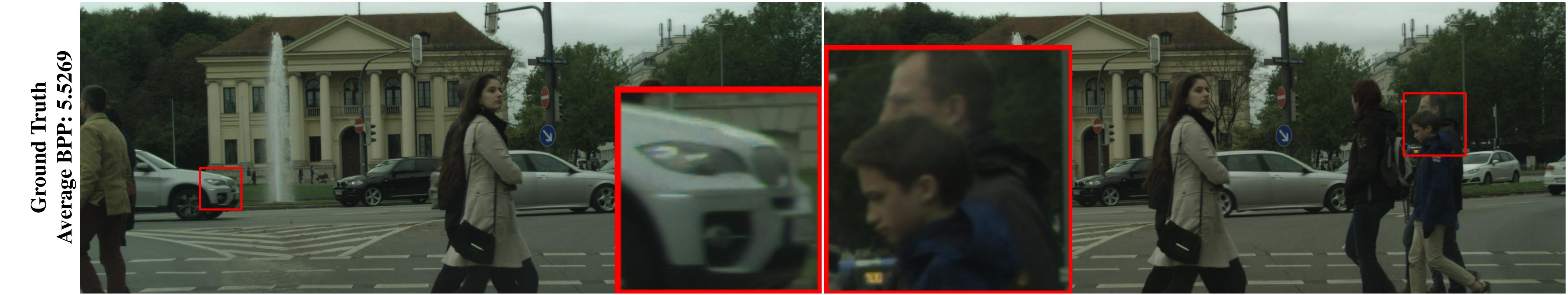}
        \caption{Ground truth images.}
    \end{subfigure}
    \\
    
    \begin{subfigure}{0.99\textwidth}
        \centering
        \includegraphics[width=\linewidth]{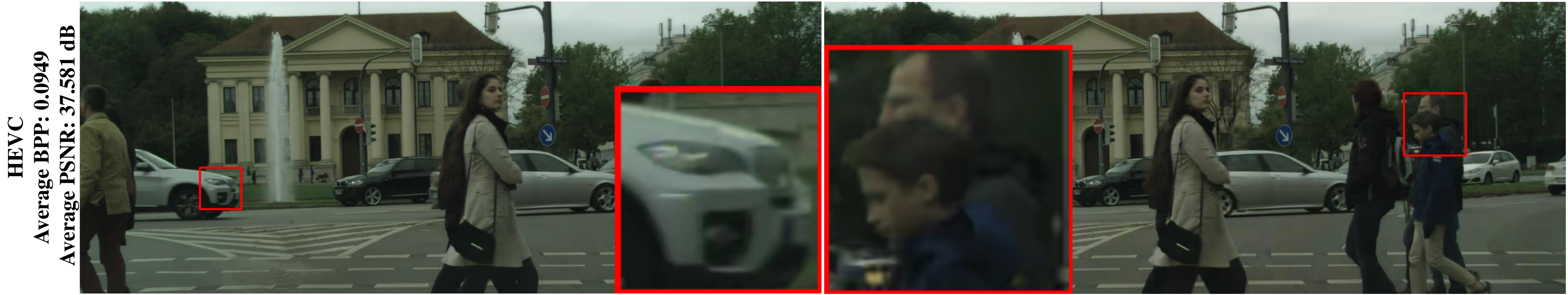}
        \caption{Results of HEVC. Left PSNR: 38.619 dB. Right PSNR: 37.871 dB.}
    \end{subfigure}
    \\

    \begin{subfigure}{0.99\textwidth}
        \centering
        \includegraphics[width=\linewidth]{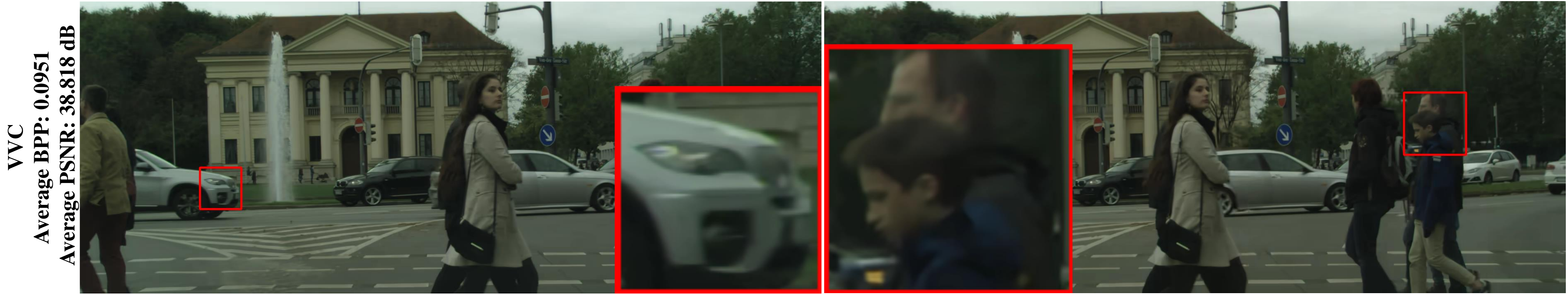}
        \caption{Results of VVC. Left PSNR: 39.665 dB. Right PSNR: 37.971 dB.}
    \end{subfigure}
    \\

    \begin{subfigure}{0.99\textwidth}
        \centering
        \includegraphics[width=\linewidth]{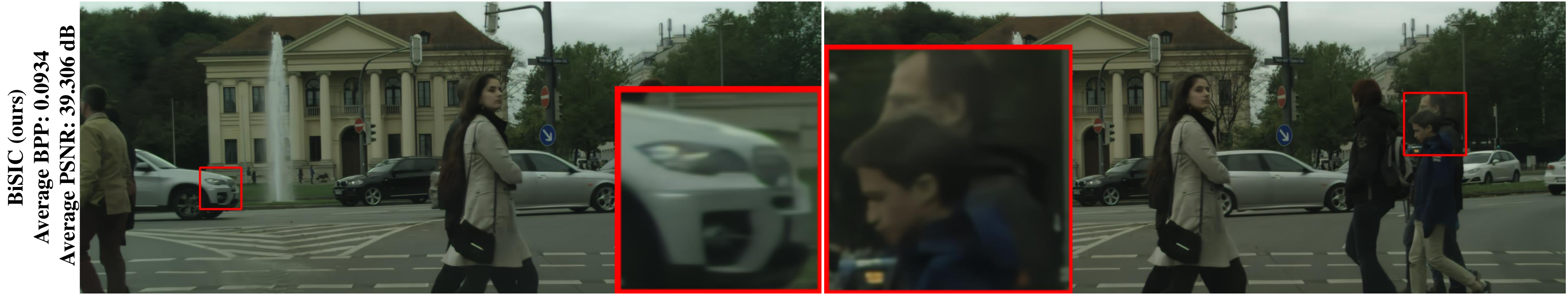}
        \caption{Results of the proposed BiSIC. Left PSNR: 39.139 dB. Right PSNR: 39.473 dB.}
    \end{subfigure}
    \\

    \begin{subfigure}{0.99\textwidth}
        \centering
        \includegraphics[width=\linewidth]{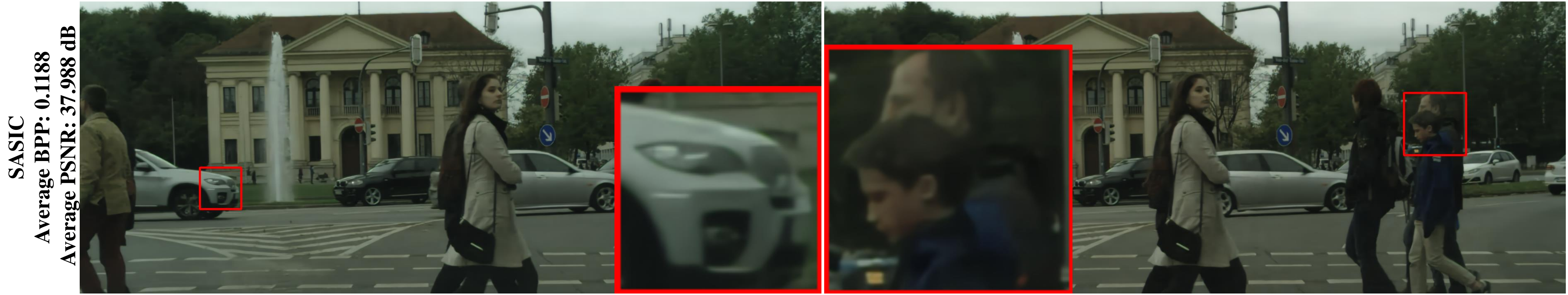}
        \caption{Results of SASIC. Left PSNR: 37.788 dB. Right PSNR: 38.189 dB.}
    \end{subfigure}
    \\

    \begin{subfigure}{0.99\textwidth}
        \centering
        \includegraphics[width=\linewidth]{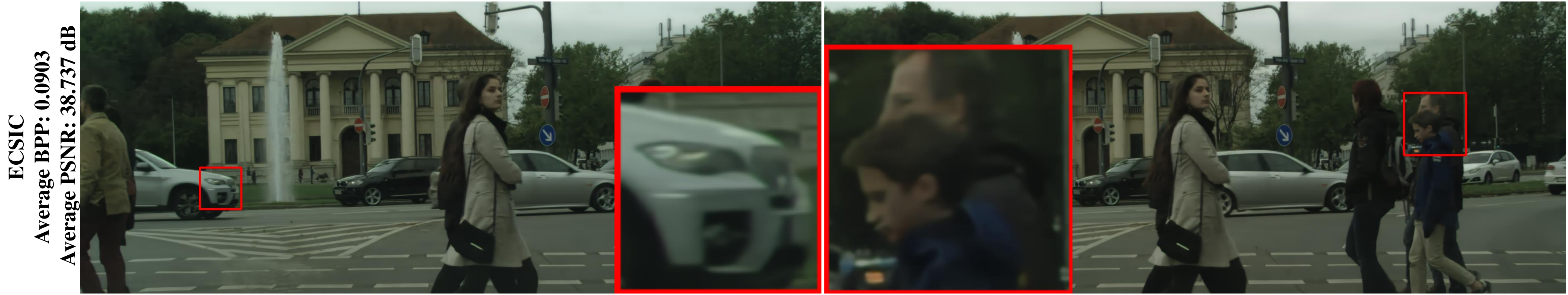}
        \caption{Results of ECSIC. Left PSNR: 38.402 dB. Right PSNR: 39.072 dB.}
    \end{subfigure}
    \caption{Visualization on the Cityscapes dataset. We compare our BiSIC with HEVC, VVC, SASIC, and ECSIC. BiSIC achieves the best PSNR performance with a relatively low BPP. Moreover, BiSIC maintains balanced qualities between stereo views.}
    \label{fig:City1459}
\end{figure}

In our work, we employ 3D convolutional layers as the backbone of the codec. In Section 4.4 of the main body of the paper, we have provided ablation study compared with plain 2D convolution backbone baseline. Here, we provide the comparison results with previous SOTA codec backbone of ELIC \cite{he2022elic}, to further validate the effectiveness of our 3D convolution based backbone in the codec for stereo image compression. Specifically, we maintain the other part of our model and replace our backbone with the residual block and attention based paradigm as in ELIC. We refer to this variant as Baseline (ELIC). Note that this baseline (43.1M parameters, 4353G FLOPs) has similar model size but higher computation cost compared with ours (49.3M parameters, 2978G FLOPs). \cref{fig:ELIC_base} shows that our method achieves a BD-PSNR gain of $0.237$ dB over this baseline. 
% }

\section{Experimental Details}
\label{sec:exp}
In this section, we provide more details about the neural network architectures and the process of model training.

\noindent \textbf{Details of Entropy Model.}
The proposed cross-dimensional entropy model aggregates the hyperprior, spatial context, channel context, and stereo dependency to estimate the probability distributions of the compact latents. 
Let $\mathbf{G}$ denote the network and $\theta$ denote its learnable parameters. The hyperprior, channel context model, and spatial context model are formulated as follows:
\begin{align} % \setcounter{equation}{10}
\label{eq:chan}
\tilde{\textbf{\textit{z}}}_l, \tilde{\textbf{\textit{z}}}_r&={h}_s(\hat{\textbf{\textit{z}}}_l, \hat{\textbf{\textit{z}}}_r), \\
\mathbf{\Theta}_{l},\mathbf{\Theta}_{r}&=\mathbf{G}_{\textrm{ch}}(\hat{\textbf{\textit{y}}}_{l}^{<k},\hat{\textbf{\textit{y}}}_{r}^{<k};\theta_{\textrm{ch}}), \\
\mathbf{\Upsilon}_{l},\mathbf{\Upsilon}_{r}&=\mathbf{G}_{\textrm{sp}}(\hat{\textbf{\textit{y}}}^k_{l,<i},\hat{\textbf{\textit{y}}}^k_{r,<i};\theta_{\textrm{sp}}).
\end{align}
The hyperprior dependency is obtained through a hyper decoder, which is also depicted in Fig. 1 and Eq. (4) in the main text.
The channel context model $\mathbf{G}_{\textrm{ch}}$ comprises four convolutional layers with $1\times 1$ kernel size, and it produces the channel dependency feature with 128 channels.
The spatial context model $\mathbf{G}_{\textrm{sp}}$ is obtained through one layer of masked 3D convolution, which is illustrated in Fig. 4(a) in the main body of the paper. The estimated mean and variance are produced with the aggregation model as follows:
\begin{align}
    \bm{{\mu}}_{l},\bm{{\sigma}}^2_{l}&=\mathbf{G}_{\textrm{ag}}(\tilde{\textbf{\textit{z}}}_l, \mathbf{\Theta}_{l},\mathbf{\Upsilon}_{l};\theta_{\textrm{ag}}),\\
    \bm{{\mu}}_{r},\bm{{\sigma}}^2_{r}&=\mathbf{G}_{\textrm{ag}}(\tilde{\textbf{\textit{z}}}_r, \mathbf{\Theta}_{r},\mathbf{\Upsilon}_{r};\theta_{\textrm{ag}}),
\end{align}
where $\mathbf{G}_{\textrm{ag}}$ represents the network that aggregates multiple references and provides estimations. This network consists of four convolutional layers with $1\times 1$  kernels.
The conditional estimations of the two views are as follows:
\begin{align}
    p_{\hat{\textbf{\textit{y}}}_{l}}(\hat{\textbf{\textit{y}}}_{l} | \hat{\textbf{\textit{z}}}_{l}; \theta_{\textrm{ag}}, \theta_{{h}_s}, \theta_{\textrm{ch}}, \theta_{\textrm{sp}})=\mathcal{N}&(\bm{{\mu}}_{l},\bm{{\sigma}}^2_{l}), \\
    p_{\hat{\textbf{\textit{y}}}_{r}}(\hat{\textbf{\textit{y}}}_{r} | \hat{\textbf{\textit{z}}}_{r}; \theta_{\textrm{ag}}, \theta_{{h}_s}, \theta_{\textrm{ch}}, \theta_{\textrm{sp}})=\mathcal{N}&(\bm{{\mu}}_{r},\bm{{\sigma}}^2_{r}).
\end{align}

\noindent \textbf{Details of Stereo-Checkerboard.}
The proposed fast variant relies on the stereo-checkerboard structure, which transforms the entry-by-entry auto-regressive process into a two-fold operation. Specifically, the stereo views are split into two parts: stereo anchor part $\hat{\textbf{\textit{y}}}_{\textrm{ach}}$ and stereo non-anchor part $\hat{\textbf{\textit{y}}}_{\textrm{nac}}$, as shown in Fig. 5 in the main body. 
The anchor part is encoded/decoded with a hyperprior and the channel-wise condition, where the estimations of mean $\bm{{\mu}}_{\textrm{ach}}$ and variance $\bm{{\sigma}}^2_{\textrm{ach}}$ for stereo anchor part are  formulated as:
\begin{align}
\bm{{\mu}}_{l,\textrm{ach}},\bm{{\sigma}}^2_{l,\textrm{ach}}&=\mathbf{G}_{\textrm{ag-ach}}(\tilde{\textbf{\textit{z}}}_l, \mathbf{\Theta}_{l};\theta_{\textrm{ag-ach}}),\\
    \bm{{\mu}}_{r,\textrm{anc}},\bm{{\sigma}}^2_{r,\textrm{anc}}&=\mathbf{G}_{\textrm{ag-ach}}(\tilde{\textbf{\textit{z}}}_r, \mathbf{\Theta}_{r};\theta_{\textrm{ag-ach}}).
\end{align}
Then, with the existing anchor part, we obtain the anchor context feature $\mathbf{\Upsilon}_{\textrm{ach}}$ using 3D convolution as:
\begin{equation}
    \mathbf{\Upsilon}_{\textrm{ach}}=\mathbf{G}_{\textrm{ach}}(\hat{\textbf{\textit{y}}}_{\textrm{ach}};\theta_{\textrm{ach}}).
\end{equation}
Notably, $\mathbf{G}_{\textrm{ach}}$ is an ordinary 3D convolutional layer with a kernel size of $(3, 5, 5)$, as the whole stereo anchor part has been obtained and the non-anchor entries are set to zero, eliminating the need for a mask. The anchor context feature $\mathbf{\Upsilon}_{\textrm{ach}}$ serves as a reference for the stereo non-anchor part. Thus, the mean $\bm{{\mu}}_{\textrm{nac}}$ and variance $\bm{{\sigma}}^2_{\textrm{nac}}$ of the stereo non-anchor part are estimated by:
\begin{align}
\bm{{\mu}}_{l,\textrm{nac}},\bm{{\sigma}}^2_{l,\textrm{nac}}&=\mathbf{G}_{\textrm{ag-nac}}(\tilde{\textbf{\textit{z}}}_l, \mathbf{\Theta}_{l},\mathbf{\Upsilon}_{l,\textrm{ach}};\theta_{\textrm{ag-nac}}),\\
\bm{{\mu}}_{r,\textrm{nac}},\bm{{\sigma}}^2_{r,\textrm{nac}}&=\mathbf{G}_{\textrm{ag-nac}}(\tilde{\textbf{\textit{z}}}_r, \mathbf{\Theta}_{r},\mathbf{\Upsilon}_{r,\textrm{ach}};\theta_{\textrm{ag-nac}}),
\end{align}
where the aggregation networks $\mathbf{G}_{\textrm{ag-ach}}$ and  $\mathbf{G}_{\textrm{ag-nac}}$ consist of four convolutional layers with $1\times 1$ kernels.

\noindent \textbf{Implementation Details.}
All training and testing settings on datasets follow previous works\cite{Wodlinger_2022_CVPR,zhang2023ldmic,wodlinger2024ecsic} to ensure a fair comparison. 
Specifically, each image in the InStereo2K dataset is pre-processed to ensure that its size is divisible by 64. 
For the Cityscapes dataset, rectification artifacts and the self-vehicle are removed, with 64, 256, and 128 pixels being cut off from the top, bottom, and sides, respectively, of every image.
In the test phase, we evaluate the performance using images of size $1,024 \times 832$ from the InStereo2K dataset and images of size $1,792 \times 704$ from the Cityscapes dataset.

For the traditional codec baselines, BPG \cite{BPGWeb} is implemented with YUV 4:4:4 to maintain its good performance. HEVC and VVC are implemented based on JVET\footnote{\url{https://vcgit.hhi.fraunhofer.de/jvet}}, where we first convert the stereo image pair into a YUV 4:4:4 video using ffmpeg\footnote{\url{https://ffmpeg.org/}}, followed by video compression. The left view is regarded as an I frame and the other one is regarded as a P frame during video compression. 
Notably, MV-HEVC only supports the 4:2:0 chroma mode, which results in suboptimal PSNR scores at higher bitrates \cite{wodlinger2024ecsic}.
In addition, we reproduce BCSIC \cite{Lei_2022_CVPR} and test it under the same test image settings as in \cite{Wodlinger_2022_CVPR,zhang2023ldmic,wodlinger2024ecsic}. This is because the original RD curves reported in the paper \cite{Lei_2022_CVPR} are tested on $512 \times 512$ images, which yields much lower values compared to other shown baselines. Therefore, we present the results under the same testing setup for a fair comparison.

\section{Extensions}
\label{sec:exten}
Based on our proposed bidirectional stereo image compression model, BiSIC, and its fast variant, several interesting follow-up directions are worth investigating. Firstly, the Vision Transformers \cite{dosovitskiy2020image,liu2021swin} has proven effective in single image compression \cite{zhu2021transformer,zou2022devil} due to its ability in feature learning and latent representation. Thus, it has the potential to further optimize the RD performance when used as the backbone of our model. Secondly, it is interesting to extend this work to multi-view video compression or immersive video compression pipelines, which further cater to the current boom in AR/VR technology.

\end{document}